\documentclass{article}

\usepackage{amsmath}
\usepackage{amssymb}
\usepackage{graphicx}
\usepackage{color}

\newcommand{\Slash}[1]{{\ooalign{\hfil/\hfil\crcr$#1$}}}

\newcommand{\ltsim}{\protect\raisebox{-0.5ex}{$\:\stackrel{\textstyle <}{\sim}\:$}}
\newcommand{\gtsim}{\protect\raisebox{-0.5ex}{$\:\stackrel{\textstyle >}{\sim}\:$}}
\newcommand{\wightman}{\tiny \raisebox{-0.5pt}[0pt][0pt]{$\stackrel{ \textstyle >}{\textstyle \rule{0pt}{1.5pt} \smash <}$}}
\newcommand{\wightmaninverse}{\tiny \raisebox{-0.5pt}[0pt][0pt]{$\stackrel{ \textstyle <}{\textstyle \rule{0pt}{1.5pt} \smash >}$}}

\newcommand{\be}{\begin{eqnarray}}
\newcommand{\ee}{\end{eqnarray}}
\newcommand{\n}{\nonumber \\}

\begin{document}

\begin{flushright}
{KEK-TH-1697}
\end{flushright}
\vskip 1cm

\begin{center}
{\LARGE{\bf 
Kadanoff-Baym approach \\ to the thermal resonant leptogenesis}
}
\vskip 1cm

\renewcommand{\thefootnote}{\fnsymbol{footnote}}
{\large 
Satoshi Iso$^a$\footnote[1]{e-mail address: satoshi.iso@kek.jp},
Kengo Shimada$^a$\footnote[2]{e-mail address: skengo@post.kek.jp},
and Masato Yamanaka$^b$\footnote[3]{e-mail address: yamanaka@eken.phys.nagoya-u.ac.jp}
}
\vskip 0.5cm

{\it 
$^a$High Energy Accelerator Research Organization (KEK)  \\
and  \\
The Graduate University for Advanced Studies (SOKENDAI), \\
Oho 1-1, Tsukuba, Ibaraki 305-0801, Japan
}
\vskip 0.3cm
{\it 
$^b$Department of Physics, Nagoya University, Nagoya 464-8602, Japan
}
\end{center}

\vskip 1cm
\begin{center}
\begin{bf}
Abstract
\end{bf}
\end{center}
Using  the non-equilibrium Green function method (Kadanoff-Baym equations) in the expanding universe,
we investigate evolution of the lepton number asymmetry when the right-handed (RH) neutrinos have almost degenerate masses $|M_i^2-M_j^2| \ll M_i^2$. 
The resonantly enhanced $CP$-violating parameter $\varepsilon_i$ associated with the decay of the RH neutrino $N_i$ is obtained.
It is proportional to an enhancement factor $(M_i^2-M_j^2) M_i \Gamma_j/ ((M_i^2-M_j^2)^2 +R_{ij}^2)$ 
with the regulator $R_{ij}=M_i \Gamma_i+M_j \Gamma_j$. 
The result is consistent with the previous result obtained by Garny {\it et al.}, in a constant background with an out-of-equilibrium initial state.
We discuss the origin of such a regulator,
and why it is not like $R_{ij}=M_i \Gamma_i-M_j \Gamma_j$.

\newpage
\tableofcontents
\newpage
\renewcommand{\thefootnote}{\arabic{footnote}}

\section{Introduction}
\setcounter{equation}{0}
Origin of the Baryon asymmetry in the universe 
is one of the unsolved issues in particle physics. 
Although the standard model (SM)  satisfies the Sakharov's three conditions \cite{Sakharov},
sufficient number of baryon asymmetry cannot be produced 
due to the smallness of the $CP$-asymmetry in the CKM matrices 
and the modest electroweek phase transition.
On the other hand, 
the neutrino oscillation which implies tiny neutrino masses demands that some extension of the SM
is necessary.
Introducing right-handed (RH) neutrinos $N_i$ with large Majorana masses $M_i$
gives a natural solution to explain the smallness of the neutrino masses via see-saw
mechanism, but it also naturally explain the Baryon number asymmetry in the universe
through the leptogenesis \cite{Fukugita:1986hr}.  
 (See e.g., a very nice recent review \cite{Fong:2013wr}.)
In this scenario, RH neutrinos are produced thermally by the reheating after inflation.
As temperature decreases with the expansion of the universe down  
to the Majorana mass scale, 
RH neutrinos become out of thermal equilibrium and 
 their $CP$-asymmetric decay into the SM leptons and the Higgs  produce  lepton number 
 asymmetry in the universe.
The lepton number asymmetry produced  is then 
converted into the baryon number asymmetry through the nonperturbative
$B+L$ -violating process of sphalerons in the SM \cite{Kuzmin:1985mm}.

If the Majorana masses of the RH neutrinos have a hierarchical structure,
the lightest Majorana mass must satisfy the  Davidson-Ibarra(DI) bound \cite{Davidson:2002qv},
$M\gtsim 10^9 $GeV in order to produce sufficient lepton number asymmetry.
When at least  two of the RH neutrinos are degenerate in their masses,
the DI bound can be evaded.
In this case,
quantum oscillation of almost degenerate RH neutrinos resonantly enhance the 
$CP$-violating decay  and hence 
lepton number asymmetry can be  produced sufficiently even
 for RH neutrino masses as light as TeV scale.
This scenario is known as the resonant leptogenesis \cite{Pilaftsis:1997jf} \cite{Pilaftsis:2003gt} \cite{Pilaftsis:2005rv}. Such light RH neutrinos might induce detectable non-unitarity of the mixing matrix of 
active neutrinos \cite{Antusch:2009gn} \cite{Antusch:2009pm} and have attracted much attention. 

TeV scale leptogenesis has attracted  enormous 
attention in light of the LHC experiment \cite{Hambye:2001eu}--\cite{Deppisch:2013jxa}.  
The scale can be made  even smaller if the leptotenesis occurs through  $CP$-violating oscillations
between RH neutrinos far away from the  thermal equilibrium. The mechanism plays an important
role in the model of $\nu$MSM \cite{Shaposhnikov:2006xi}--\cite{Canetti:2012kh}.

Furthermore, light RH neutrinos do not give large radiative corrections to the Higgs boson 
mass and are safe in view of the naturalness \cite{Boyarsky:2009ix}.
Related to the naturalness of the electroweak weak against higher physical scales,
 one of the authors proposed a classically conformal 
$U(1)_{B-L}$ extension of the SM \cite{Iso:2009ss} \cite{Iso:2009nw}.
In this model, $B-L$ gauge symmetry is spontaneously broken via the Coleman-Weinberg mechanism
which triggers the electroweak gauge symmetry. 
In  \cite{Iso:2012jn}, we further showed
that if the Higgs potential is flat at the Planck scale, the model  naturally predicts 
the Higgs boson mass at around 126 GeV and  TeV scale $B-L$ breaking (or
the leptogenesis).
This motivated us to investigate the TeV scale leptogenesis in the $U(1)_{B-L}$ model 
\cite{Iso:2010mv}.

\begin{figure}
\begin{center}
\includegraphics[width=0.8\textwidth ]{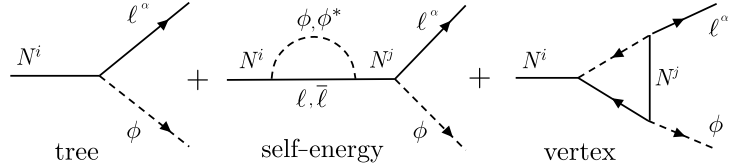}
\caption{Tree and one-loop diagrams of the RH neutrino decay.
In the resonant case, an interference of the tree and the self-energy diagram \cite{Flanz:1994yx}
\cite{Covi:1996wh}
gives a dominant contribution to the $CP$-violating parameter.
}
\label{FigCPDecay}
\end{center}
\end{figure}

In the resonant case, 
the $CP$-asymmetry in the decay of $N_i$ mainly comes from an interference
of the tree and the self-energy one-loop diagrams (See Fig.\ref{FigCPDecay}).
It is expressed by the $CP$-violating parameter
\begin{align}
\varepsilon_{i}&\equiv \frac{\Gamma_{N_i \to \ell \phi}-\Gamma_{N_i \to \overline{\ell} \overline{\phi}}}{\Gamma_{N_i \to \ell \phi}+\Gamma_{N_i \to \overline{\ell} \overline{\phi}}}= \sum_{j(\ne i)}\frac{\Im (h^{\dag}h)^{2}_{ij}}{(h^{\dag}h)_{ii}(h^{\dag}h)_{jj}} \frac{ (M^{2}_{i}-M^{2}_{j}) M_{i}\Gamma_{j}}{(M^{2}_{i}-M^{2}_{j})^2+ R_{ij}^2 } 
\label{CPVparameter}
\end{align}
where $h$ is the neutrino Yukawa coupling and $\Gamma_i \simeq (h^{\dag}h)_{ii}M_i /8\pi$ is the decay width of $N_i$.
The resonant enhancement of the $CP$-violating parameter was discussed in \cite{Flanz:1996fb}.
Systematic considerations were performed by Pilaftsis 
\cite{Pilaftsis:2003gt}\cite{Pilaftsis:1997dr}\cite{Pilaftsis:1998pd}, and he found that
the regulator in the denominator is given by $R_{ij}=M_i \Gamma_j$.
If the mass difference is larger than the decay width,
we have $|M^2_i -M^2_j| \gg R_{ij}$, 
and $\varepsilon_i$ is suppressed by $\Gamma_i/M \sim (h^{\dag}h)_{ii}$.
However, in the degenerate case, $|M_i -M_j|\sim \Gamma$ and 
$\varepsilon$ can be enhanced to  ${\mathcal O}((h^\dagger h)^0)\sim 1$. 
Hence the determination of the regulator $R_{ij}$ is essential for a precise prediction 
of the lepton number asymmetry in the resonant leptogenesis.
The authors \cite{Buchmuller:1997yu} calculated the
resummed propagator of the RH neutrinos and obtained a different regulator
$R_{ij}=|M_i \Gamma_i-M_j \Gamma_j|$. By using their result, the enhancement
factor becomes much larger. The origin of the difference of the regulators 
is discussed in \cite{Anisimov:2005hr} \cite{Rangarajan:1999kt}.
Since the lower scale of the leptogenesis is strongly
 sensitive to the form of the regulator, it is very important to systematically evaluate
 the precise form of the regulator. 
 
Conventionally, leptogenesis is often calculated based on the classical 
Boltzmann equation which describes the time evolution of the phase space distribution function
of on-shell particles \cite{Kolb:1990vq}.
In the Boltzmann equation, the interactions between particles are taken into account
through the collision terms
 that comprise the $S$-matrix elements calculated separately in the framework of quantum field theory.
The authors \cite{Buchmuller:2000nd}
 applied the non-equilibrium Green's function method  with the Kadanoff-Baym (KB) equations
developed in studies of the transport phenomena \cite{Baym:1961zz}\cite{Baym:1962sx} and derived the full-quantum evolution equation for the lepton number in the hierarchical mass case.
Using this method, one can systematically take  into account 
quantum interference, finite temperature and finite density effects.\footnote{
Quantum oscillations in the leptogenesis are also investigated
in \cite{Akhmedov:1998qx}\cite{Asaka:2005pn}\cite{Canetti:2010aw}
based on the density matrix formalism \cite{Sigi}\cite{Gagnon:2010kt}.
}
The method was intensively used in the leptogenesis in various papers  
 \cite{Anisimov:2010aq}--\cite{Frossard:2013bra}.
 In the resonant leptogenesis, since the quantum interference effect is crucial to the evaluation 
of the $CP$-violating parameter, we can expect importance of such a full-quantum mechanical 
formulation based on the KB equations. 
In \cite{De Simone:2007rw}, the authors used the method to obtain an oscillating
$CP$-violating parameter in the flat space-time.
Then applying it to the Boltzmann equation in the expanding universe, 
they calculated the lepton number asymmetry. 
In the strong washout regime,
 the oscillation is averaged out and the lepton number asymmetry is expressed
with an effective $CP$-violating parameter.
Then the maximal value agrees with the case of $R_{ij}= M_i\Gamma_j$\cite{De Simone:2007pa}.
The authors of \cite{Garbrecht:2011aw} also found  an oscillatory behavior by a different calculation,
and discussed an implication to the leptogenesis in the expanding universe.
The quantum oscillations in the flavored leptogenesis was also performed in \cite{Cirigliano:2007hb}\cite{Beneke:2010dz}\cite{Drewes:2012ma}.

Recently Garny {\it et al.} \cite{Garny:2011hg} systematically
investigated generation of the lepton asymmetry in the resonant leptogenesis using the 
formulas developed in \cite{Anisimov:2010aq}\cite{Anisimov:2010dk}.
In the investigation, they considered a non-equilibrium initial condition in a time-independent
background and calculated generation of the lepton number asymmetry. 
Starting from the vacuum initial state for the RH neutrinos,
they read off the $CP$-violating parameter from the generated lepton number asymmetry.
The effective regulator they derived is  $R_{ij}=M_i \Gamma_i+M_j \Gamma_j$, which differs from 
the previous results,
$R_{ij}= M_i \Gamma_j$ by \cite{Pilaftsis:2003gt} or 
$R_{ij}=|M_i \Gamma_i-M_j \Gamma_j|$ by \cite{Buchmuller:1997yu}.

The purpose of the present paper is to perform  systematic 
investigations of the thermal resonant leptogenesis 
 based on the KB equations. 
We scrutinize various properties of the Green functions of the RH neutrinos, and
directly extract the $CP$-violating parameter $\varepsilon_i$  
 from the evolution equation for the lepton number 
  in the expanding universe, with an emphasis on the quantum flavor oscillations.

The paper is organized as follows.
In section 2.1 and 2.2, we first summarize the basic properties of various Green functions
and the Kadanoff-Baym (KB) equations that these Green functions must satisfy.
Then we derive the evolution equation of the lepton number in the expanding universe in section 2.3.
The evolution equation is written in terms of the propagators of the RH neutrinos,
the SM leptons and the Higgs. In section 2.4 we explain how the KB equation is reduced
to the ordinary Boltzmann equation. The most important ingredient necessary to solve the 
evolution equation for the lepton number is the Wightman functions of the RH neutrinos.
The flavor diagonal component is directly related to the distribution function, but more important
for the lepton asymmetry is its off-diagonal component.

In section 3, we investigate how the flavor oscillation affects the off-diagonal component of 
the propagators. In the section, we focus on the resonant oscillations in the thermal equilibrium.
In section 3.1, 3.2 and 3.3, we study the properties of  the retarded and advanced propagators
in which  information of the spectrum is encoded.
Then we study the Wightman functions with information of the distribution functions.

In section 4, we scrutinize the behavior of Green functions out of equilibrium.
In the expanding universe, Green functions are approximated in the leading order approximation
by the thermal values at the local temperature. But in order to calculate the lepton asymmetry,
deviations from the thermal values are important. We show in section 4.3 that the deviations of the 
flavor off-diagonal Wightman functions from the thermal values behave quite differently
from behaviors of other Green functions. 

In section 5, we apply the calculated deviations of the Wightman functions of the RH neutrinos into the 
evolution equation derived in section 2, and obtain the quantum Boltzmann equation for the 
lepton number asymmetry. The deviations are classified into 3 types.
One of them generate the lepton number asymmetry while the other two wash out the 
generated  asymmetry. In section 5.4, we read off the $CP$-violating parameter $\varepsilon$
 and show that the regulator is given by $R_{ij}=M_i \Gamma_i+M_j \Gamma_j$.

 In section 6, we give a physical interpretation why the regulator
 $R_{ij}=M_i \Gamma_i+M_j \Gamma_j$ appears
 instead of $R_{ij}=M_i \Gamma_i-M_j \Gamma_j$.
In particular, we show that if we {\it neglect}  what we call the off-shell contributions
the regulator is erroneously given by $R_{ij}=M_i \Gamma_i-M_j \Gamma_j$.

In section 7, we summarize our results.

In appendix A and B, we give a brief introduction to the closed time path (CTP) formalism and
the KB equations. In appendix C, we introduce the 2PI formalism and then in appendix D
we derive the self-energies for the RH neutrinos and the SM leptons based on the 2PI formalism.
In appendix E and F, some useful identities in calculating convolutions are given.
From appendix G to J, we give details of the calculations of various Green functions.
In appendix K, we give anther derivation of the off-diagonal component of the 
Wightman functions out of equilibrium.  The calculation explains why the 
regulator $R_{ij}=M_i \Gamma_i + M_j \Gamma_j$ naturally appears.
Appendices L and M are calculations of some equations in the paper.

\section{Evolution equations of lepton numbers}
\setcounter{equation}{0}

A systematic method to investigate the evolution of lepton asymmetry is the Kadanoff-Baym (KB) equations.
The advantage of the KB equation to the Boltzmann
equation is that it gives a quantum evolution equation of various correlation functions
which does not distinguish on-shell and off-shell states.
Accordingly it can take into account quantum coherence of the system and memory effects.
Also the doubling problem in the scattering processes with on-shell internal lines
can be systematically resolved (see \cite{Frossard:2012pc} and references therein).
The KB equation can be reduced to the classical 
Boltzmann equation only in  special cases where the memory effects can be neglected.

Time-evolution of a quantum system is determined by the Hamiltonian of the system
and the initial wave function at the initial time $t=t_{i}$.
Such time-evolution is described by the wave function at later times, or instead,
a set of all $n$-point Green functions.  
Of course, it is practically impossible to study the evolution equations containing all the 
$n$-point functions and we need to select  an important set
of observables. In the classical approach based on the Boltzmann equation,
one-particle distribution function on the phase space is selected.
In the quantum Boltzmann approach, two-point Green functions are selected.

In this section, we summarize  notations of various Green functions and their basic properties
in the thermal equilibrium. We also summarize the non-equilibrium evolution equation (KB equation)
for the Green functions. More details are given in  Appendices.
After brief reviews in section \ref{Green functions and KMS relations} and \ref{Kadanoff-Baym equations},
we derive the evolution equation of the lepton number in section \ref{Evolution of lepton number in the expanding universe} and \ref{Boltzmann equation for the lepton number}.

\subsection{Our model}
The model we consider is an extension of the SM with RH neutrinos $\nu _{ R,i}$.
$i$ is the flavor index, $i=1,2,3.$
We set $N_i=\nu_{R,i}+\nu_{R,i}^{c}$. 
 The Lagrangian is given by
 \begin{align}
{\mathcal L}={\mathcal L}_{SM}+\frac{1}{2}\overline{N}^{i}(i\Slash{\nabla}-M_i)N^{i}+{\mathcal L}_{int}\ , \label{L}
\end{align}
\begin{align}
{\mathcal L}_{int} \equiv -h_{\alpha i}(\overline{\ell}^{\alpha}_{a} \epsilon_{ab} \phi^{*}_{b})P_{R}N^{i}+h^{\dag}_{i\alpha}\overline{N}^{i}P_{L}(\phi_{b} \epsilon_{ba} \ell ^{\alpha}_{a}) \label{Lint}
\end{align}
where  $\alpha ,\beta =1,2,3$ and $a,b=1,2$ are flavor indices of the SM leptons $\ell_a^\alpha$ and
isospin $SU(2)_L$ indices respectively.
$M_i$ is the Majorana mass  of $N_i$ and $h_{i \alpha}$ is the Yukawa coupling of $N^i, \ell_a^\alpha$
and the Higgs $\phi_a$ doublet.
$P_{R(L)}$  are chiral projections on right(left)-handed fermions.
In the present paper, we consider the case of almost degenerate Majorana masses at TeV scale.
Then the Yukawa couplings become very small $h_{i \alpha} \ll 1$ so as to generate
tiny neutrino masses through the see-saw mechanism.
Hence the decay width $\Gamma_i \simeq (h^{\dag}h)_{ii}M_i /8\pi$
 is much smaller than the mass $M_i$.

\subsection{Green functions and KMS relations}
\label{Green functions and KMS relations}
Various Green functions are introduced in field theories.
Consider a fermion field $\psi$. The statistical propagator $G_{F}$ and the spectral density
$G_{\rho}$ are defined as
\begin{align}
G_{F}(x,y)=&\frac{1}{2}\langle [ \hat{\psi}(x) , \overline{\hat{\psi}}(y) ] \rangle \ , \label{defG_{F}} \\
G_{\rho}(x,y)=&i \langle \{ \hat{\psi}(x) , \overline{\hat{\psi}}(y) \} \rangle \ .\label{defG_{rho}}
\end{align}
The statistical propagator $G_{F}$ contains  information of the particle density of the state
on which  operators are evaluated.
On the other hand, the spectral density $G_{\rho}$ gives information of the particle's mass and
decay width. Because of the anti-commutator, $\gamma^0 G_\rho(x^0,y^0)$ becomes proportional to 
the spatial delta function $\delta^3 ({\bf x}-{\bf y})$ at the equal time $x^0=y^0$:
\be
\gamma_0 G_\rho(x,y) = i \delta^3({\bf x} -{\bf y}) {\bf 1}
\ee
where ${\bf 1}$ is an identity matrix in the flavor and the spinor indices.

Other useful Green functions are the Wightman functions
\begin{align}
G_{>}(x,y)=& G_{F}(x,y) -\frac{i}{2}G_{\rho}(x,y) = \langle \hat{\psi}(x) \overline{\hat{\psi}}(y) \rangle \ , \label{defG_{g}}\\
G_{<}(x,y)=& G_{F}(x,y) +\frac{i}{2}G_{\rho}(x,y) = - \langle \overline{\hat{\psi}}(y) \hat{\psi}(x) \rangle \label{defG_{l}}
\end{align}
and the retarded and advanced Green functions are given by
\begin{align}
G_{R/A}(x,y)=&\pm {\rm \Theta}(\pm (x^0 -y^0))G_{\rho}(x,y)\ .\label{defG_{RA}} 
\end{align}
The spectral function can be written as $G_\rho=G_R-G_A=i(G_>-G_<)$.

In the present paper, we assume homogeneity along the spatial directions so that
we can always use the Fourier transform in the 3-dimensional space.
If the state  is described by the thermal equilibrium state,
we can further Fourier transform in the time direction.\footnote{In the present paper, we
often use the Fourier transform in the time direction when the system is in the thermal
equilibrium at the local temperature $T(t)$ at time $t$. Then the Green functions in the 
four-momentum representation depends on time $t$ through the local temperature.} 
In the thermal equilibrium at temperature $T$, the Green functions $G(x,y)$ are
anti-periodic in the time direction  with an imaginary period $i\beta=i/T$ and 
their Fourier transforms  satisfy
the KMS (Kubo Martin Schwinger) relation
\begin{align}
G^{(eq)}_{\wightman}(q)=-i\left\{ \begin{matrix} 1-f(q_0) \\
-f(q_0) \end{matrix} \right\} G^{(eq)}_{\rho}(q) \ , \ \ 
G^{(eq)}_F(q)= -i \left( \frac{1}{2}-f(q_0) \right)  G^{(eq)}_{\rho}(q)\ .
 \label{KMSforG0}
\end{align}
Here $f(q_0)$ is the Fermi-Dirac distribution function 
$
f(q_0)=1/(e^{q_0 /T}+1 ).
$
In presence of the  chemical potential $\mu$, $q_0$ is replaced by $q_0 -\mu$.
Since the relation relates the fluctuation described by the Wightman function to the dissipation 
described by the retarded Green function, it is also called the fluctuation-dissipation relation.
By this relation, the spectrum of the system determines all the Green functions.
When the system becomes out of equilibrium, the KMS relation is violated. 
The violation plays an important role in the leptogenesis.

As a final remark in this section, let us recall that the explicit forms of the Wightman functions
of free  charged fermions (bosons)  are given by
\be
&& G_{\wightman}^{{\rm free}}(x,y) =
\int \frac{d^3 q}{(2\pi)^3}e^{+i{\bold q}\cdot({\bold x} -{\bold y})}\frac{1 }{2\omega_{q}}
\n 
&& \times 
{\Big [ } e^{-i\omega _{q}(x^0 -y^0)} \left\{ \begin{matrix} 1 - \eta f_{\bold q} \\  -\eta f_{\bold q}
 \end{matrix} \right\} 
\hat{g}_+
+e^{+i\omega _{q}(x^0 -y^0)} \left\{ \begin{matrix} -\eta \bar{f}_{-{\bold q}} \\ 1 - \eta \bar{f}_{-{\bold q}}
\end{matrix} \right\} 
\hat{g}_-  {\Big ] }
\label{WightmanFree}
\ee
where $\omega_q$ is the energy of the on-shell particle, and $\hat{g}_\pm= (\pm\omega_{q}\gamma^0 -{\bold q}\cdot{\boldsymbol \gamma} + m) $, $\eta=+1$ for Dirac fermions with their mass $m$ and $\hat{g}_\pm=1$, $\eta =-1$ for bosons.
$f_{\bold q}$ and $\bar{f}_{\bold q}$ are distribution functions of on-shell particles and anti-particles with their spatial momentum ${\bold q}$ respectively. 
They are not necessarily the equilibrium distribution functions.
\subsection{Kadanoff-Baym equations}
\label{Kadanoff-Baym equations}
If the system is out of equilibrium and the state is time-dependent,
we cannot use the ordinary perturbative method based on the Feynmann propagators.
A general formalism  is given by the closed-time-path (CTP) formalism in which 
perturbative vertices are inserted on the closed-time-path ${\mathcal C}={\mathcal C}_+ + {\mathcal C}_-.$
See Appendix \ref{AppCTP} for brief review and Figure \ref{FigCTP} there.
One of the self-consistent approximation of the Schwinger-Dyson equations in the CTP formalism is called Kadanoff-Baym (KB) equation.
Derivation of the KB equations is given in Appendix \ref{AppKB} and \ref{App2PI}.

The equations for the retarded and advanced Green functions are
\begin{align}
 iG_{0(x)} ^{-1}G_{R/A}(x,y)-\int^{\infty}_{t_{int}}d^{4}z_g 
\ \Pi _{R/A}(x,z)G_{R/A}(z,y) =-\delta ^{g}(x-y) \label{KBeqG_{RA}} \ .
\end{align}
Here $d^4 z_g$ is an abbreviation of $d^4 z \sqrt{-g(z)}$ and $\delta^g(x-y)=\delta^4(x-y)/\sqrt{-g}$. $G_{0(x)}^{-1}$ is the free kinetic operator
whose derivatives act on a field at $x$. 
$\Pi_{R/A}$ is the self-energy and defined in eq.(\ref{PiRAApp}).
They have the same  properties as  $G_{R/A}$, e.g.,  $\Pi_R(x,y)=0$ for $x^0<y^0$ is satisfied.
Note that the integration range in (\ref{KBeqG_{RA}}) is constrained between $x_0$ and $y_0$:
\be
\int^{x^0  (y^0)}_{y^0(x^0)}d^{4}z_g 
\ \Pi _{R/A}(x,z)G_{R/A}(z,y) 
\label{nomemoryRA}
\ee
because of the step functions in $\Pi_{R/A}$ and $G_{R/A}$.
Therefore $G_{R/A}(x,y)$ is determined by the local information between $x_0$ and $y_0$.
Namely $G_{R/A}$ does not depend on the information of the system
in the past:
there is no memory effect for $G_{R/A}$. 

Other Green functions $G_*$  ($*=F, \rho, \lessgtr$)  satisfy
\begin{align}
 iG_{0(x)} ^{-1}G_{*}(x,y) &=& \int^{\infty}_{t_{int}}d^{4}z_g  \ \Pi _{R}(x,z)G_{*}(z,y) 
+ \int^{\infty}_{t_{int}}d^{4}z_g  \ \Pi _{*}(x,z)G_{A}(z,y) \n
&=& \int^{x^0}_{t_{int}}d^{4}z_g  \ \Pi _{R}(x,z)G_{*}(z,y)
+ \int^{y^0}_{t_{int}}d^{4}z_g  \ \Pi _{*}(x,z)G_{A}(z,y) 
\ . \label{KBeqG_{*}}  
\end{align}
In the second equality, we have used the properties of $R/A$ functions.
By using eq.(\ref{KBeqG_{RA}}), 
this equation can be solved formally in terms of the self-energy function
and the $R/A$ Green functions as
\be
G_* (x,y) &=& - \int_{t_{int}}^\infty d^4 z_g d^4 w_g G_R(x,z) \Pi_*(z,w) G_A(w,y) \n
 & \equiv&  - (G_R * \Pi_{*} * G_A) (x,y) \ .
 \label{GformalSol}
\ee
In the last line $*$-operation denotes the convolution operation.

Let us see the memory effect of $G_*$.
Generally speaking, the integrals in (\ref{KBeqG_{*}}) over $z$ are performed from the past 
at the initial time $t_{int}$ to $x^0$ or $y^0$.
This makes  Green functions
dependent on the state of the system in the past before $x^0$ or $y^0$. 
This is indeed the case for $G_F$ and $G_\lessgtr$, but for the spectral density $G_\rho$,
there is no memory effect. It can be seen by using  $\Pi_\rho=\Pi_R-\Pi_A$.
Then  the integral of  (\ref{KBeqG_{*}}) can be rewritten as
\begin{align}
 iG_{0(x)} ^{-1}G_{\rho}(x,y) 
= \int^{x^0}_{y^0}d^{4}z_g  \ \Pi _{\rho}(x,z)G_{\rho}(z,y) \ .
\end{align}
Or it can be directly seen from  
the relation $G_\rho=G_R-G_A$.
The relation $G_\rho=-G_R \Pi_\rho G_A=G_R-G_A$ is equivalent to
$\Pi_\rho=\Pi_R-\Pi_A=G_R^{-1}-G_A^{-1}.$

In the thermal equilibrium, since the system is translationally invariant,
(\ref{GformalSol}) can be Fourier transformed and 
\be
G_*^{(eq)}(p)= - G_R^{(eq)}(p)\Pi_*^{(eq)}(p)G_A^{(eq)}(p) \ .
\ee

These equations  (\ref{KBeqG_{RA}}), (\ref{KBeqG_{*}}) 
are not closed within the two-point Green functions because
the self-energy $\Pi$ contains $n(>2)$-point functions.
Hence, in order to solve them explicitly, 
we need to make an approximation to express $n(>2)$-point functions
in terms of the two-point functions.
2PI effective action method is one of the simplest and self-consistent methods.
(See Appendix \ref{App2PI} for  brief explanation.)
By using it, the self-energies $\Pi$ in the above equations (\ref{KBeqG_{RA}}) (\ref{KBeqG_{*}})
are represented as  a sum of 1PI diagrams made of full propagators, and 
consequently these equations can be interpreted as simultaneous 
equations for various propagators in the system. 
These self-consistent equations among the propagators are especially called
the Kadanoff-Baym equations.

\subsection{Evolution of lepton number in the expanding universe}
\label{Evolution of lepton number in the expanding universe}

Now we investigate the KB equations of lepton numbers in the expanding universe.
We first define Green functions, $G$, $S$ and $\Delta$ for 
the RH neutrinos, the SM lepton doublet and the Higgs doublet respectively:
\begin{align}
G_{>}^{ij}(x,y)= \langle \hat{N}^{i}(x) \overline{\hat{N}}\hspace{0pt}^{j}(y) \rangle \ ,& \ \ G^{ij}_{<}(x,y)= -\langle \overline{\hat{N}}\hspace{0pt}^{j}(y) \hat{N}^{i}(x)  \rangle \ ,\label{defG_{W}} \\
S^{\alpha \beta}_{ab >}(x,y)= \langle \hat{\ell} ^{\alpha}_{a}(x) \overline{\hat{\ell}}\hspace{0pt}^{\beta}_{b}(y) \rangle \ ,& \ \ S^{\alpha \beta}_{ab <}(x,y)= -\langle  \overline{\hat{\ell}}\hspace{0pt}^{\beta}_{b}(y) \hat{\ell} ^{\alpha}_{a}(x)  \rangle \ ,\label{defS_{W}} \\
\Delta_{ab>}(x,y)= \langle \hat{\phi}_{a}(x) \hat{\phi}^{\dag}_{b}(y) \rangle \ ,& \ \ \Delta_{ab<}(x,y)= +\langle \hat{\phi}^{\dag}_{b}(y) \hat{\phi}_{a}(x) \rangle \label{defDel_{W}} \ .
\end{align} 
The classical inverse propagators are given by
\begin{align}
iG_{0}^{-1\ ij}(x,y)=&(i\Slash{\nabla}_{x}-M_i) \delta^{ij}\delta^g(x-y)\ ,  \label{G_{0}^(-1)}\\ 
iS_{0 \ \ \ ab}^{-1\ \alpha \beta}(x,y)=&i\Slash{\nabla}_x P_{L}  \delta ^{\alpha \beta} \delta_{ab} 
\delta^g(x-y)\ ,\label{S_{0}^(-1)}   \\
i\Delta _{0 \ \ ab}^{-1}(x,y)=&-\nabla^{2}_{x} \delta_{ab} \delta^g(x-y) \label{Del_{0}^(-1)} \ .
\end{align}

In this paper, we consider the spatially flat space-time with the scale factor $a(t)$:
\begin{align}
ds^2 = dt^2 -a^2(t) d{\bold x}\cdot d{\bold x} \ .
\end{align}
We use $\tilde{\mu},\tilde{\nu},\dots$ as the space-time indices and
$\mu, \nu,\dots$ as the local Lorentz indices.
$\gamma$ matrices are written as
$\gamma^{\tilde{\mu}}(t)=\gamma^{\mu}e_{\mu}^{\ \tilde{\mu}}$
where the vier-bein field $e_\mu^{\ \tilde{\mu}}$ satisfies 
$e_{\mu}^{\ \tilde{\mu}}e_{\nu}^{\ \tilde{\nu}}g_{\tilde{\mu} \tilde{\nu}}=\eta_{\mu \nu}$.
In the following we mainly use $t$-independent
$\gamma^{\mu}=(\gamma^0 ,{\boldsymbol \gamma})$ instead of 
$t$-dependent $\gamma^{\tilde{\mu}}(t)$. 
The delta-function becomes $\delta^g(x-y)=\delta^4(x-y)/a^{3}(x^0)$.

In the background, the covariant derivative (\ref{Del_{0}^(-1)}) becomes
\begin{align}
\nabla_{x^{\tilde{\mu}}}=\partial _{\tilde{\mu}} +3H(x^0)\delta _{\tilde{\mu}}^{0} \ .
\end{align}
Since the spin connection is given by 
$\Omega _{\tilde{\mu}}=aH[\gamma_{\mu},\gamma_{0}]/4$, the covariant derivative
for spinors in (\ref{G_{0}^(-1)}), (\ref{S_{0}^(-1)}) is given by
\begin{align}
\Slash{\nabla}_{x} =&\gamma^{\tilde{\mu}}(x)(\partial_{\tilde{\mu}}+\Omega_{\tilde{\mu}}) 
=\gamma^{0}\left( \partial_{x^0}+\frac{3}{2} H(x^{0})\right)-\frac{{\boldsymbol \gamma}\cdot \partial_{\bold x}}{a(x^0)} \ .
\label{covderspinor}
\end{align}
Here the Hubble parameter is defined by 
$H(t)=\dot{a}/a$. In the radiation dominant universe, it is given by
\begin{align}
H(t)=1.66 \sqrt{g_*}\frac{T^{2}}{M_{pl}} \sim \frac{T^2}{10^{18}{\rm GeV}} \ . 
\label{HubbleParameter}
\end{align}

Lepton number density $n_{L}$ is a matrix with  flavor indices $\alpha,\beta$ and 
isospin indices $a,b$. It is given by the $\tilde{\mu}=0$ component of 
the  lepton number current 
\begin{align}
j^{\tilde{\mu} \beta \alpha}_{L ba}(x)\equiv &\langle \overline{\hat{\ell}}\hspace{0pt}^{\beta}_{b}(x)\gamma^{\tilde{\mu}}(x) \hat{\ell}^{\alpha}_{a}(x)\rangle \notag \\
=&-\left. {\rm tr}\{ \gamma^{\tilde{\mu}}(x)S^{\alpha \beta}_{ab >}(x,y) \} \right|_{y=x}  \n
=&-\left. {\rm tr}\{ \gamma^{\tilde{\mu}}(x)S^{\alpha \beta}_{ab <}(x,y) \} \right|_{y=x} \ . \label{leptoncurrent}
\end{align}
Here ${\rm tr}\{ \cdots\}$ is the trace of the spinors.
Because of the spatial homogeneity, divergence of the current $j_{L}$ is equal  to
\begin{align}
\nabla_{\tilde{\mu}}\ j^{\tilde{\mu} }_{L}(x) =&\frac{d n_{L}}{dt}+3H(t)n_{L} \ . \label{divleptoncurrent} 
\end{align}
On the other hand, it can be rewritten as\footnote{ Here, we have defined the derivative operator $\overleftarrow{\Slash{\nabla}}_{y}$ as
\[ S(x,y) \overleftarrow{\Slash{\nabla}}_{y}  \equiv -\nabla_{\tilde{\mu}}\left[  S(x,y) \gamma^{\tilde{\mu}}(y) \right] + S(x,y) \gamma^{\tilde{\mu}}(y) \Omega_{\tilde{\mu}} = \left(-\partial_{y^{\mu}} -\frac{3}{2} H(y^0)\right)S(x,y)\gamma^{\mu} . \]}
\begin{align}
\nabla_{\tilde{\mu}}\ j^{\tilde{\mu} }_{L}(x) 
=&- \ {\rm tr}\{ \Slash{\nabla}_{x}S_{\wightman}(x,y) - S_{\wightman}(x,y)\overleftarrow{\Slash{\nabla}}_{y} \}
 {\Big |}_{y=x} \n
=&\ i \int d^4 z_g \ {\rm tr}\{ iS_{0}^{-1}(x,z)S_{\wightman}(z,x) - S_{\wightman}(x,z) iS_{0}^{-1}(z,x)\} \ . 
\end{align}
In the second equality, we have used the definition of $S_0^{-1}(x,z)$ in (\ref{S_{0}^(-1)}).

By using the KB equation of (\ref{KBeqG_{*}}) for the SM lepton Green function $S_{\wightman}$,
we have
\begin{align}
\int d^4 z_g &\ iS_{0}^{-1}(x,z)S_{\wightman}(z,x) 
=  \int^{x_0}_{t_{int}}d^{4}z_g  \ {\Big (} \Sigma _{R}(x,z)S_{\wightman}(z,x) +
\Sigma _{\wightman}(x,z)S_{A}(z,x){\Big )} 
 \notag \\
&= - i \int^{x^0}_{t_{int}}d^{4}z_g \ {\Big (} \Sigma _{<}(x,z)S_{>}(z,x) - \Sigma _{>}(x,z)S_{<}(z,x){\Big )} 
\end{align}
where $\Sigma$ is the self-energy of the SM lepton.
The second equality is obtained by using the relations (\ref{Grho=Gwightman}) and (\ref{GRA=Grho}).

Acting $i S_{0}^{-1}$ from the right, a similar equation can be derived:
\begin{align}
\int d^4 z_g \ S_{\wightman}(x,z)iS_{0}^{-1}(z,x) 
&= - i \int^{x^0}_{t_{int}}d^{4}z_g \ {\Big (} S_{<}(x,z)\Sigma _{>}(z,x) - S_{>}(x,z)\Sigma _{<}(z,x) {\Big )} \ .
\end{align}
By using these equation, 
(\ref{divleptoncurrent}) becomes 
\begin{align}
\frac{d n_{L}}{dt}+3H(t)n_{L} 
=\ \int^{x^0}_{t_{int}}d^{4}z_g  \ {\rm tr}{\Big \{ }& \Sigma _{<}(x,z)S_{>}(z,x)- \Sigma _{>}(x,z)S_{<}(z,x) \notag \\
-&S_{<}(x,z)\Sigma_{>}(z,x) + S_{>}(x,z)\Sigma_{<}(z,x) {\Big \} } \ .
\label{KBeqforLeptonNumber}
\end{align}
This is the evolution equation for the lepton numbers in the expanding universe.

The right hand side (r.h.s.) is written as an integral of the full propagator $S$ of the SM lepton
and its self-energy $\Sigma$.  Since the self-energy $\Sigma$ contains various diagrams,
some systematic simplification of $\Sigma$ is necessary for practical calculations.
 A well-known approach is to use
the 2-particles-irreducible (2PI) formalism briefly reviewed in Appendix B. 
In the 2PI formalism, the self-energy diagrams are obtained 
by taking a variation of  2PI diagrams made of 
full propagators with respect to the full propagator.

\begin{figure}
\begin{center}
\includegraphics[width=0.3\textwidth ]{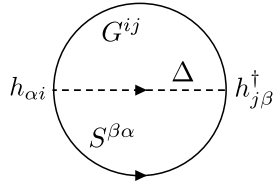}
\caption{
An example of 2PI diagrams for the Lagrangian (\ref{Lint}) with Yukawa interactions.
Each line represents a full propagator of the SM lepton, Higgs and the RH neutrino.
By taking a functional derivative with respect to each propagator, we can obtain 
the self-energy for the corresponding particle.
}
\label{Fig2PI-1}
\end{center}
\end{figure}

In the leading approximation, 
 the self-energy $\Sigma$ is obtained from
 the simplest 2PI diagram of Figure \ref{Fig2PI-1}.
Note that each propagator represents a full propagator, and the self-energy
of the SM lepton is obtained by cutting the propagator $\ell$. 
The next simplest 2PI diagram is given by Figure \ref{Fig2PI-2} in Appendix \ref{AppSE}, but
 in most of the present analysis, we consider only the contribution from  Figure \ref{Fig2PI-1}. 
It gives a good approximation if the RH neutrinos have almost degenerate masses.

The contribution to the lepton self-energy $\Sigma$ from Figure \ref{Fig2PI-1} 
is written in terms of the full propagators: 
\begin{align}
\Sigma^{\alpha \beta}_{ab\ \wightman}(x,y)=&-\delta_{ab} h_{\alpha i}h^{\dag}_{j\beta}P_{\rm R} G_{\wightman}^{ij}(x,y)P_{\rm L}\Delta_{\wightmaninverse} (y,x)
\equiv \ \delta_{ab} \Sigma^{\alpha \beta}_{\wightman}(x,y) \ .
\end{align}
Recall that $(i,j)$ are flavor indices of the RH neutrinos.
Then, summing the lepton flavor $\alpha, \beta$ and $SU(2)_L$ isospin
$a,b$ indices, we have
\begin{align}
\frac{d n_{L}}{dt} +3H n_{L}
= -g_w h_{\alpha i}h^{\dag}_{j\beta} \int^{x^0}_{t_{int}}d^{4}z_g \ {\bigg [}&{\rm tr}{\Big \{ }P_{\rm R} G_{<}^{ij}(z,x)P_{\rm L}S^{\beta \alpha}_{>}(x,z){\Big \} }\Delta_{>} (x,z) \n
 -&{\rm tr}{\Big \{ }P_{\rm R} G_{>}^{ij}(z,x)P_{\rm L}S^{\beta \alpha}_{<}(x,z){\Big \} }\Delta_{<} (x,z)  \n
-&{\rm tr}{\Big \{ }P_{\rm R}G_{>}^{ij}(x,z)P_{\rm L}S^{\beta \alpha}_{<}(z,x){\Big \} }\Delta_{<} (z,x) \n
+&{\rm tr}{\Big \{ }P_{\rm R} G_{<}^{ij}(x,z)P_{\rm L}S^{\beta \alpha}_{>}(z,x){\Big \} }\Delta_{>} (z,x){\bigg ]} \ .
\end{align}
Here we used the fact that the electroweak symmetry is restored at the temperature
$T \gtrsim {\rm TeV}$ 
we are in mind and hence the propagators are written in
$SU(2)$ symmetric forms:
$S^{\alpha \beta}_{ab}=S^{\alpha \beta}\delta_{ab}$, $\Delta_{ab}=\Delta \delta_{ab}$.
$g_w =2$ is the number of d.o.f. of $SU(2)_L$ doublets.
Since the third and the fourth terms are complex conjugate to the second and the first terms,
we can simplify the above equation as 
\begin{align}
\frac{d n_{L}}{dt} +3H n_{L} =
 2  \ \Re  
\int^{x^0}_{t_{int}} d\tau d^{3}{\bf z}_g \  h_{\alpha i}h^{\dag}_{j\beta} {\Big [}&
 {\rm tr}{\Big \{ }  
P_{\rm R} G_{<}^{ij}(x,z)P_{\rm L} \widetilde{\pi}_{>}^{\beta \alpha} (z,x) {\Big \} } \n
 -& {\rm tr}{\Big \{ }P_{\rm R} G_{>}^{ij}(x,z)P_{\rm L} \widetilde{\pi}_{<}^{\beta \alpha} (z,x)
 {\Big \} }
{\Big ]}
\label{dndt0} 
\end{align}
where we have defined $\tau=z^0$ and
\be
\widetilde{\pi}_{\gtrless}^{\beta \alpha } (z,x)= - g_w 
S^{ \beta  \alpha}_{\gtrless}(z,x) \Delta_{\gtrless} (z,x) \ . 
\ee
This is the equation we evaluate in the following investigations. 
As we mentioned above, the r.h.s. contains only the contribution from the simplest 2PI
diagram of Figure \ref{Fig2PI-1}. This corresponds to taking the processes of decay
and inverse-decay of the RH neutrinos. 
The effects of scattering can be taken into account\footnote{
A part of the scattering diagram
in which the internal particles are close to on-shell is taken into account
by considering the diagram of Figure \ref{Fig2PI-1}.  In the resonant case, it
 gives a dominant contribution to the scattering process and hence
it is sufficient to consider only the 2PI diagram of Figure \ref{Fig2PI-1}.  }
by considering the next simplest diagram of Figure \ref{Fig2PI-2}.
A systematical study of the KB equation 
including the scattering effects is given in \cite{Frossard:2012pc}\cite{Frossard:2013bra}.

\subsection{Boltzmann equation for the lepton number}
\label{Boltzmann equation for the lepton number}
The evolution equation (\ref{dndt0}) of the lepton number is determined 
by the behavior of 
full propagators of the RH neutrinos $G$, the SM leptons $S$ and the Higgs $\Delta$.
In  sections \ref{Sec-RORN} and \ref{Sec-POEQ}, 
we investigate detailed properties of the propagator $G$ of the RH neutrinos.
In this section, we will see how an ordinary Boltzmann-type equation can be 
derived from eq.(\ref{dndt0}) by using the quasi-particle approximation for the SM particles
described by $S$ and $\Delta.$

The quasi-particle approximation is an approximation to express
the Green functions in terms of distribution functions of 
quasi-particles with a mass  $m$ and a width $\Gamma$.
Hence the propagators in this approximation are obtained from the free Wightman function of
eq.(\ref{WightmanFree}) by  introducing the decay width $\Gamma$.
For a moment, we neglect the time-dependence of the background.
For the SM leptons, we have
\begin{align}
 S^{\beta \alpha }_{\wightman}(x,y)=\ &\delta^{\alpha \beta}
\int \frac{d^3 p}{(2\pi)^3} \frac{1}{2\omega_{p}}
e^{+i{\bold p}\cdot({\bold x} -{\bold y})} \ e^{-|x^0 -y^0|\Gamma_{\ell}/2} \notag \\
\times&{\Big [ } e^{-i\omega _{p}(x^0 -y^0)} \left\{ \begin{matrix} 1-f_{\ell p} \\ -f_{\ell p} \end{matrix} \right\} P_{\rm L}\Slash{p}_{+}P_{\rm R} 
+e^{+i\omega _{p}(x^0 -y^0)} \left\{ \begin{matrix} -f_{\overline{\ell} p} \\ 1-f_{\overline{\ell} p} \end{matrix} \right\} P_{\rm L}\Slash{p}_{-}P_{\rm R}{\Big ] }   \n
=\ &\delta^{\alpha \beta}
\sum_{\epsilon_{\ell}=\pm} \int \frac{d^3 p}{(2\pi)^3} \frac{1}{2\omega_{p}} 
e^{+i{\bold p}\cdot({\bold x} -{\bold y})} 
e^{-i\epsilon_{\ell}\omega _{p}(x^0 -y^0)-|x^0 -y^0|\Gamma_{\ell}/2} \n 
& \hspace{85pt}  \times (-1)^{\epsilon_{\ell}}
\left\{ \begin{matrix} 1-f^{\epsilon_{\ell}}_{\ell p} \\ -f^{\epsilon_{\ell}}_{\ell p} \end{matrix} \right\} 
P_{\rm L}\Slash{p}_{\epsilon_{\ell}}P_{\rm R} 
\label{S_{W}QP}
\end{align}
where 
$ \omega_{p}=\sqrt{m_{\ell}^{2}+|{\bold p}|^2 /a^2}$ and 
$\Slash{p}_{\pm}
\equiv \pm\omega_{p}\gamma^0 -{\bold p}\cdot{\boldsymbol \gamma}/a $.
Here we assumed the flavor independence  of the 
lepton propagators, $S^{\alpha \beta} \propto \delta^{\alpha \beta}$,
for simplicity.\footnote{
Generally, flavor structure plays an important role in the flavored leptogenesis
\cite{Abada:2006fw}\cite{Nardi:2006fx}.} 
Similarly the Wightman functions of the Higgs boson becomes
\begin{align}
\Delta_{\wightman}(x ,y)=&\int \frac{d^3 k}{(2\pi)^3} \frac{1}{2\omega_{k}}
e^{+i{\bold k}\cdot({\bold x} -{\bold y})}
e^{-|x^0 -y^0|\Gamma_{\phi}/2} \notag \\
& \times {\Big [ } e^{-i\omega _{k}(x^0 -y^0)} \left\{ \begin{matrix} 1+f_{\phi k} \\ +f_{\phi k} \end{matrix} \right\} 
+e^{+i\omega _{k}(x^0 -y^0)} \left\{ \begin{matrix} +f_{\overline{\phi} k} \\ 1+f_{\overline{\phi}k} \end{matrix} \right\} {\Big ] } \n
=&\sum_{\epsilon_{\phi}=\pm} \int \frac{d^3 k}{(2\pi)^3}
\frac{1}{2\omega_{k}} e^{+i{\bold k}\cdot({\bold x} -{\bold y})}
 e^{-i\epsilon_{\phi}\omega _{k}(x^0 -y^0)-|x^0 -y^0|\Gamma_{\phi}/2} \n
& \hspace{70pt}  \times 
(-1)^{\epsilon_{\phi}}
 \left\{ \begin{matrix} 1+f^{\epsilon_{\phi}}_{\phi k} \\ +f^{\epsilon_{\phi}}_{\phi k} \end{matrix} \right\}  \ 
\label{Del_{W}QP}
\end{align}
where $\omega_{k}=\sqrt{m_{\phi}^{2}+|{\bold k}|^2 /a^2}$ .

The thermal mass and width are given by
 $m_{\ell,\phi}\sim gT\ ,\ \Gamma_{\ell,\phi}\sim g^2 T$
where $g$ is the SM gauge coupling $g$.
The effects of the thermal plasma play very important roles, and are systematically investigated in 
\cite{Frossard:2012pc}\cite{Frossard:2013bra}. For example, the thermal
mass of the Higgs becomes larger than the RH neutrino masses at very high temperature.
In the present paper, we focus on the   largeness of $\Gamma_{\ell, \phi}$ as an important
thermal effect and do not consider other effects.

In these expressions we defined
 $(-1)^{\epsilon}=\pm 1$ for  $\epsilon=\pm$ respectively.
The distribution functions are assumed to be in the kinematical equilibrium 
and given by 
the Fermi-Dirac or the Bose-Einstein distributions at temperature $T$
with a chemical potential:
\begin{align}
f_{\ell p}=\frac{1}{e^{(\omega_{p}-\mu_{\ell})/T}+1}\ ,\ 
f_{\phi k}=\frac{1}{e^{(\omega_{k}-\mu_{\phi})/T}-1}\ .\  \label{BEdist}
\end{align}
For anti-particles, the signs of the chemical potentials are reversed and 
their distributions are given by
\be
 f_{\overline{\ell} p}=\frac{1}{e^{(\omega_{p}+\mu_{\ell})/T}+1}\ , \ \ f_{\overline{\phi} k}=\frac{1}{e^{(\omega_{k}+\mu_{\phi})/T}-1} \ .
\ee
In the second equalities of eq.
(\ref{S_{W}QP}) and (\ref{Del_{W}QP}), we have defined 
\begin{align}
f^{\epsilon}_{\ell p} \equiv \frac{1}{e^{(\epsilon \omega_{p}-\mu_{\ell})/T}+1}\ ,\ \ 
f^{\epsilon}_{\phi k} \equiv  \frac{1}{e^{(\epsilon \omega_{k}-\mu_{\phi})/T}-1}  
\end{align}
which satisfy
\be
f_{\ell p}=f^{+}_{\ell p}\ , \ \  f_{\overline{\ell} p}=(1-f^{-}_{\ell p})\ , \ \ 
 f_{\phi k}=f^{+}_{\phi k}\ , \ \ f_{\overline{\phi} k}=-(1+f^{-}_{\phi k})\ .
\ee

Now we come back to the time-dependence of the background.
Since the scale factor $a(t)$ is time-dependent,  temperature $T$, 
thermal mass and  width are  dependent on the time $t$ and
 we need to specify at which time these quantities in 
the quasi-particle approximation of eq. 
 (\ref{S_{W}QP}) and (\ref{Del_{W}QP}) are defined.
If the temperature of the universe is sufficiently low (e.g., $\sim$ 10 TeV),
the decay width is much larger than the Hubble expansion rate:
\begin{align}
\Gamma_{\ell ,\phi} \sim g^2 T \gg  H \sim \frac{T^2}{10^{18}{\rm GeV}}
 \label{GammaSMggH}
\end{align}
and the propagators damp quickly at $|x^0 -y^0| \gg 1/\Gamma_{l,\phi}$.
For such short period, time-dependence of the 
physical quantities such as the scale factor
 in the propagators (\ref{S_{W}QP}) and (\ref{Del_{W}QP}) 
 are suppressed by  $H/\Gamma_{\ell,\phi}$, 
and we can 
 approximate these quantities as being constant in the integration 
 of $\tau$ in (\ref{dndt0}). 
Then  the physical quantities can be evaluated at time $t=X_{xy}=(x^0 +y^0)/2$ as we see in (\ref{SE-out-of-eq}).

By Fourier transforming in the spatial direction and using
the above approximation,
 (\ref{dndt0}) becomes
\begin{align}
\frac{d n_{L}}{dt} +3H n_{L} = 2\Re \int\frac{d^3 q}{(2\pi)^3}   \int^{t}_{-\infty}d\tau\ 
 (h^{\dag}h)_{ji}
{\bigg [}&{\rm tr}{\Big \{ }P_{\rm R} G_{<}^{ij}(t,\tau ;
   {\bold q}  
)P_{\rm L}\widetilde{\pi}_{>}(\tau,t ;
{\bold q}  
) {\Big \} } \n
-&{\rm tr}{\Big \{ }P_{\rm R} G_{>}^{ij}(t,\tau ;{\bold q})P_{\rm L}\widetilde{\pi}_{<}
(\tau,t ; {\bold q}
) {\Big \} }{\bigg ]}
\label{dndt1}
\end{align}
where $t=x^0$ and $d^3 q=d^3 {\bf q}/a^3(t)$. 
Using the quasi-particle approximations (\ref{S_{W}QP}) and (\ref{Del_{W}QP}),
$\widetilde{\pi}_{\wightman} (\tau,t;{\bf q})$ are given by
\be
\widetilde{\pi}_{\wightman} (\tau,t;{\bf q})=P_{\rm L}\pi_{\wightman}(\tau,t;{\bf q})P_{\rm R} \ , \label{defpitilde}
\ee
\begin{align}
\pi_{\wightman}(\tau,t;{\bf q}) \equiv
(-g_w) \sum_{\epsilon_{\ell},\epsilon_{\phi}}&\int \frac{d^3 p}{(2\pi)^3 2\omega_{ p}}
\frac{d^3 k}{(2\pi)^3 2\omega_{ k}}\  (2\pi)^3 \delta ^3(q-p-k) 
\n
& \times \  {\mathcal D}_{\wightman (p,k)}^{\epsilon_{\ell} \epsilon_{\phi}}  \ \Slash{p}_{\epsilon _{\ell}} \ e^{-i(\epsilon_{\ell}\omega _{p} +\epsilon_{\phi}\omega _{k})(\tau -t) - \Gamma_{\ell \phi} /2 |\tau -t|} \label{defPiq}
\end{align}
where  $\Gamma_{\ell \phi} \equiv \Gamma_{\ell}+\Gamma_{\phi}$ and 
${\mathcal D}_{\wightman (p,k)}^{\epsilon_{\ell} \epsilon_{\phi}}$
 is defined as
\begin{align}
{\mathcal D}_{\wightman (p,k)}^{\epsilon_{\ell} \epsilon_{\phi}} \equiv 
(-1)^{\epsilon_{\ell}} (-1)^{\epsilon_{\phi}} 
\left\{ \begin{matrix} 
(1-f_{\ell p}^{\epsilon_{\ell}})(1+f_{\phi k}^{\epsilon_{\phi}})  
\\ (-f_{\ell p}^{\epsilon_{\ell}})(+f_{\phi k}^{\epsilon_{\phi}})  \end{matrix} \right\} \ .
\label{dndtD}
\end{align}
From (\ref{defPiq}) and (\ref{dndtD}), we can 
see that the term with $\widetilde{\pi}_>$ in (\ref{dndt1}) contains a factor $(1-f_{\ell})$
or $(1-f_{\ell}^-) =f_{\bar{\ell}}$ and corresponds to  gain in the lepton number
while  the other term with $\widetilde{\pi}_<$ contains a factor $f_{\ell}$
or $f_{\ell}^- =(1-f_{\bar{\ell}})$ and corresponds to  loss.
Hence the evolution equation (\ref{dndt1}) can be interpreted as the Boltzmann-like
equation for the lepton number.

In order to solve the evolution equation (\ref{dndt1}), we need detailed information
of the Wightman function  $G_\gtrless^{ij}(x,y)$ of the RH neutrinos.
In the following sections, we obtain behaviors of  the Wightman Green functions,
especially deviations from the thermal equilibrium values in the expanding universe.

Here we briefly comment on the basic structures of the r.h.s.
First, contributions from the flavor diagonal part $i=j$ are evaluated by
the quasi-particle approximation. $G_\lessgtr^{ii}$ is proportional to
$(f_{N_i}^{\epsilon})$ or $(1-f_{N_i}^{\epsilon})$ respectively where $f_{N_i}$ is the 
distribution function of the RH neutrino $N_i$. 
Therefore, combined with the distribution
functions from $\widetilde{\pi}_\gtrless$, flavor diagonal term gives
 the tree-level decay or inverse-decay
of Figure \ref{FigCPDecay}, and  wash out the generated lepton number asymmetry\footnote{ Using the so-called extended quasi-particle approximation\cite{Kohler:1993zz}\cite{Spicka:1994zz}\cite{MorozovRopke}, we can take into account the finite decay width of the RH neutrinos as the real intermediate state (RIS) subtracted scattering processes of lepton and Higgs \cite{Frossard:2012pc}.
These processes are mediated by off-shell RH neutrinos and also contribute to the washout of the lepton asymmetry.}.  

On the other hand, by using
the formal solution of  the Wightman function in (\ref{GformalSol}),
contributions from the flavor off-diagonal part $i \neq j$ are interpreted 
as an interference effect between the tree and 1-loop diagrams as follows.
The formal solution  is  written in terms of the self-energy $\Pi_\gtrless$ as
\begin{align}
G_{\wightman}^{ij}(x,y)=& - \sum_{k,l} \int_{-\infty}^{x^0} d u^0  d^3 {\bf u}_g \int_{-\infty}^{y^0} 
 dv^0  d^3 {\bf v}_g \ G_{R}^{ik}(x,u)
\Pi_{\wightman}^{kl} (u,v)G^{lj}_{A }(v,y) \ .
\label{GwightmanSec2}
\end{align}
In the leading order approximation with the 2PI diagram of Figure \ref{Fig2PI-1},
the self-energy $\Pi_\gtrless$ is written as a functional of the full propagators
of the SM lepton and the Higgs as in eq.(\ref{Pi_{W}}).
Hence it can be interpreted as an interaction vertex of $\ell \phi \leftrightarrow N$
at $u \sim v$. The  RH neutrino propagates from $u \sim v$ to another interaction vertex at $x \sim y$.
By inserting this expression into eq.(\ref{dndt1}) and taking the 
on-shell limit of the RH neutrinos, 
$CP$-asymmetric interference between the
tree and the one-loop self-energy diagram can be obtained.
If the RH neutrinos propagating between these vertices are off-shell,
the contribution is interpreted as s-channel  scattering processes. 
Hence flavor off-diagonal terms in the r.h.s. of (\ref{dndt1}) give
both of the $CP$-asymmetric decay of the RH neutrinos and the washout of the lepton
numbers via s-channel scattering of leptons and Higgs.

In the resonant leptogenesis where the RH neutrinos have almost degenerate masses,
however, 
it is not legitimate to separate the on-shell and off-shell contributions as above
since the RH neutrinos are coherently mixed between different flavors,
as has been mentioned in \cite{Garny:2009qn}\cite{Frossard:2012pc}.
Therefore we need to scrutinize the behavior of 
the Wightman functions $G_\gtrless^{ij}(t,\tau;{\bf q})$ 
in the expanding universe.
\subsection{Summary of this section }
The evolution of the lepton number is given by (\ref{dndt0}) or its Fourier transform (\ref{dndt1}).
They are the basic equations we evaluate in the following sections.
If we adopt the quasi-particle approximations of (\ref{S_{W}QP}) or (\ref{Del_{W}QP}), 
an ordinary classical Boltzmann equation is derived.
But in the resonant leptogenesis, quantum coherence between different flavors of $N_i$
plays an essential role and such an approximation is not valid for the RH neutrinos. 
An evaluation of the r.h.s. of (\ref{dndt0}) by scrutinizing the behavior of the off-diagonal components
of the Wightman functions $G_\gtrless^{ij}$, which is formally solved as
(\ref{GwightmanSec2}), in the expanding universe is the main issue
in the following sections.
\section{Resonant oscillation of RH neutrinos }
\label{Sec-RORN}
\setcounter{equation}{0}
In this section, we study  how the RH neutrinos with almost degenerate masses
behave  in the thermal equilibrium.
Deviation from the thermal equilibrium is investigated in the next section \ref{Sec-POEQ}.

We consider two flavors $i=1,2$ whose masses are almost degenerate. 
The third flavor RH neutrino is assumed to have larger mass.
In order to calculate the evolution of the lepton asymmetry in (\ref{dndt0}), 
we need to know the Wightman functions $G_\gtrless$ of the RH neutrinos.
And, since the KB equation of $G_\gtrless^{ij}$ is formally solved 
by the convolution eq.(\ref{GwightmanSec2}), it is necessary to 
investigate the properties of the retarded (advanced) Green functions $G_{R/A}^{ij}$ first.

We first study  both of the flavor diagonal ($i=j$) and off-diagonal
($i \neq j$) components of  $G_R^{ij}$ in the equilibrium. 
Then we will see the behaviors of the Wightman functions $G_\gtrless^{ij}$
in the thermal equilibrium. Throughout the paper, 
 $G^d$ (also $\Pi^d$ for the self-energy) and $G'$ ($\Pi'$)
 denote the flavor diagonal $i=j$ and 
off-diagonal $i \neq j$ components respectively:
\be
 &G^d & \longleftrightarrow \ \ \mbox{flavor diagonal } , \n
& G' &\longleftrightarrow  \ \ \mbox{flavor off-diagonal } .
\ee
\subsection{Retarded/Advanced propagators}
From (\ref{KBeqG_{RA}}) and  (\ref{G_{0}^(-1)}), $G_{R/A}$ satisfies 
\begin{align}
(i\Slash{\nabla}_{x}-M)G^{ij}_{R/A}(x,y)-\int^{\infty}_{t_{int}}dz^0 d^{3}{\bf z} \ a^3(z^0) \ \Pi^{ik} _{R/A}(x,z)G^{kj}_{R/A}(z,y) &=-\delta_{ij}\delta ^{g}(x-y)\ .\label{KBeqG_{R}}  
\end{align}
We first define the spatial Fourier transform of $G_{R/A}$ by
\begin{align}
G^{ij}_{R/A}(x^0 ,y^0;{\bold q})=\int d^3 ({\bold x}-{\bold y})e^{-i{\bold q}\cdot ({\bold x}-{\bold y})}a^{3/2}(x^0)G^{ij}_{R/A}(x^0,y^0 ,{\bold x}-{\bold y})a^{3/2}(y^0) \ .
\end{align}
Similarly, for the self-energy, we define 
\begin{align}
\Pi ^{ij}_{R/A}(x^0 ,y^0;{\bold q})=\int d^3 ({\bold x}-{\bold y})e^{-i{\bold q}\cdot ({\bold x}-{\bold y})}a^{3/2}(x^0)\Pi ^{ij}_{R/A}(x^0,y^0 ,{\bold x}-{\bold y})a^{3/2}(y^0) \ .
\end{align}
Then using (\ref{covderspinor}),  the KB equation (\ref{KBeqG_{R}})  becomes
\begin{align}
 {\Big \{ }i \gamma^0 &\partial_{x^0} -\frac{{\boldsymbol \gamma}\cdot {\bold q}}{a(x^0)}-M{\Big \}}G_{R/A}(x^0 ,y^0;{\bold q}) \notag \\
&-\int^{\infty}_{t_{int}}dz^0 \ \Pi_{R/A}(x^0 ,z^0;{\bold q}) G_{R/A}(z^0,y^0;{\bold q}) =-\delta (x^0-y^0) \ .
\label{KBeqGRsec3}
\end{align}
This is the basic equation for $G_{R/A}$.

We then decompose the propagator and the self-energy into flavor diagonal and
off-diagonal parts:
\begin{align}
G_{R/A}(x^0 ,y^0;{\bold q}) &\equiv G^{d}_{R/A}(x^0,y^0;{\bold q}) + G'_{R/A}(x^0 ,y^0;{\bold q})\ , \n
\Pi_{R/A}(x^0 ,y^0;{\bold q}) &\equiv \Pi^{d}_{R/A}(x^0,y^0;{\bold q}) + \Pi'_{R/A}(x^0 ,y^0;{\bold q})\ . 
\end{align} 
Using this decomposition, we solve the KB equation (\ref{KBeqGRsec3}) iteratively.

First we define the differential-integral operator $D^{d}_{x^0}$ by
\begin{align}
D^{d}_{x^0} f(x^0) \equiv {\Big \{ }i \gamma^0 \partial_{x^0} -\frac{{\boldsymbol \gamma}\cdot {\bold q}}{a(x^0)}-M{\Big \}} f(x^0) - \int^{\infty}_{t_{int}}dz^0 \ \Pi^{d}_{R/A}(x^0 ,z^0;{\bold q}) \ f(z^0) \ . 
\end{align}
In terms of the operator,
 the flavor diagonal component of the KB equation (\ref{KBeqGRsec3})  becomes
\begin{align}
D^{d}_{x^0} G^{d}_{R/A}(x^0 ,y^0;{\bold q})-\int^{\infty}_{t_{int}}dz^0 \ \Pi'_{R/A}(x^0 ,z^0;{\bold q})G'_{R/A}(z^0 ,y^0;{\bold q}) = -\delta (x^0-y^0) \ . 
\label{KBeqG_{RA}^{d}}
\end{align}
Similarly the KB equation of the
 flavor off-diagonal component is written as
\begin{align}
D^{d}_{x^0} G'_{R/A}(x^0 ,y^0;{\bold q})=\int^{\infty}_{t_{int}}dz^0 \ \Pi'_{R/A}(x^0 ,z^0;{\bold q})G^{d}_{R/A}(z^0 ,y^0;{\bold q}) \ .\label{KBeqG_{RA}^{'}} 
\end{align}

We then introduce 
the kernel $G_{R/A}^{d(0)}$ of the operator $D_{x^0}^{d}$:
\begin{align}
D^{d}_{x^0}G^{d(0)}_{R/A}(x^0 ,y^0;{\bold q}) \equiv -\delta(x^0 -y^0) \ . \label{defG_{RA}^{d0}}
\end{align}
with a retarded (advanced) boundary condition.
Using $G_{R/A}^{d(0)}$, we can integrate the equations (\ref{KBeqG_{RA}^{d}}), 
(\ref{KBeqG_{RA}^{'}}) as
\begin{align}
G^{d}_{R/A}(x^0 ,y^0;{\bold q}) &=G^{d(0)}_{R/A}(x^0 ,y^0;{\bold q}) \notag \\
&-\int^{\infty}_{t_{int}}d\tau d\tau ' \ G^{d(0)}_{R/A}(x^0 ,\tau;{\bold q})\Pi^{'}_{R/A}(\tau ,\tau ';{\bold q})G'_{R/A}(\tau , y^0;{\bold q}) \ , \label{intKBeqG_{RA}^{d}}
\end{align}
\begin{align}
G'_{R/A}(x^0 ,y^0;{\bold q})=-\int^{\infty}_{t_{int}}d\tau d\tau ' \ G^{d(0)}_{R/A}(x^0 ,\tau;{\bold q})\Pi^{'}_{R/A}(\tau ,\tau ';{\bold q})G^{d}_{R/A}(\tau , y^0;{\bold q}) \ . \label{intKBeqG_{RA}^{'}}
\end{align}
Then we can iteratively solve the above equations 
by expanding it with respect to the small off-diagonal component of 
the Yukawa coupling $(h^{\dag}h)^{'}$ involved in $\Pi^{'}$:
\begin{align}
G^{d}_{R/A} = &G^{d(0)}_{R/A}+ G^{d(2)}_{R/A} + \cdots, \label{G_{RA}^{0+2}}  \\
& G^{d(2)}_{R/A} \equiv G^{d(0)}_{R/A}* \Pi^{'}_{R/A}* G^{d(0)}_{R/A}* \Pi^{'}_{R/A}* G^{d(0)}_{R/A} \ , 
  \n
G'_{R/A} = & -G^{d(0)}_{R/A}* \Pi^{'}_{R/A}* G^{d(0)}_{R/A}+\cdots \label{G_{RA}^{'}} \ .
\end{align}
Here $*$ denotes a convolution in the time-direction. 
The second term $G^{d(2)}_{R/A}$ in the flavor diagonal propagator (\ref{G_{RA}^{0+2}}) is 
the second order of $ (h^{\dag}h)^{'} $
 and smaller than $G^{d(0)}_{R/A}$
or $G'_{R/A}$. 
Hence we drop it and write $G^{d(0)}$ as $G^{d}$ for notational simplicity in the following.

We note that the above integrals do not have the memory effect.
This is because the convolution is written explicitly as, e.g.,
\be
(G_{R} * \Pi_R * G_R)(x^0,y^0) = \int^{x^0}_{y^0} du \int_{y^0}^u \ dv \ G_R(x^0,u) \Pi_R(u,v) G_R(v,y^0)
\ee
and the integration region is limited between $x^0$ and $y^0$. 
Namely, the retarded (advanced) propagators are "local" functions of time during $x^0$
and $y^0$ and insensitive to the past ($t<x^0, y^0$).
This is different from the convolution contained in the Wightman functions (\ref{GwightmanSec2})
in which the integration range of time is extended to the past.
\subsection{Diagonal  $G_{R/A}^d$ in thermal equilibrium}
\label{Diagonal G_{R/A}^d in thermal equilibrium}
We will first look at the flavor diagonal component
of the propagator $G^d_{R/A}(x^0,y^0;{\bf q})$ in the thermal equilibrium
at temperature $T$.  
The scale factor $a$ is also fixed at $a_0=a(x^0)=a(y^0).$
Because of the translational invariance in the time direction,
$G_{R/A}(x^0,y^0;{\bf q})$ can be further Fourier transformed:
\begin{align}
G^{d(eq)}_{R/A}(q)=\int d(x^0 -y^0) e^{+iq_{0}(x^0 -y^0)} G^{d(eq)}_{R/A}(x^0 ,y^0;{\bold q}) \ . 
\end{align}
Then the KB equation (\ref{intKBeqG_{RA}^{d}}) becomes
\begin{align}
 {\Big \{ } \gamma^0 q_0 -\frac{{\boldsymbol \gamma}\cdot {\bold q}}{a_0}-M-\Pi^{d(eq)}_{R/A}(q_0 ,{\bold q}){\Big \}}G^{d(eq)}_{R/A}(q) =-1 \ 
\end{align}
and can be solved 
\begin{align}
G^{d(eq)}_{R/A}(q)&=-\left( \Slash{q}-M- \Pi^{d(eq)}_{R/A}(q) \right)^{-1} . \label{G_{RA}^{d(eq)}(q)}
\end{align}
The real part of the self-energy
gives the mass and wave-function renormalization.
In the following we assume that they are already taken into account in the bare
Lagrangian and focus only on the imaginary part $\Pi^{d}_{\rho}=\Pi^{d}_{R}-\Pi^{d}_{A}=2i \Im (\Pi^{d}_R)$.
The one-loop diagonal self-energy in the thermal equilibrium is expressed as
$\Pi_\rho^d=\gamma^{\mu} \Pi_{\rho ,\mu}^{d} $. From the imaginary part of the 
pole of the propagator $G_R^{d(eq)}(q)$, we see that the decay width $\Gamma_q$ 
of the RH neutrino is given by
\begin{align}
 q \cdot \Pi^{d(eq)}_{\rho}(q)|_{q_0 =\pm \omega_{\bold q}} \equiv \mp i\omega_{q} \Gamma_{q} \ .\label{omegaGamma}
\end{align}

The $i$-th diagonal component
$G^{d(eq)ii}_{R/A}(q)$ becomes
\begin{align}
G^{d(eq)ii}_{R/A}(q)&\simeq -
\frac{\Slash{q}-\Pi^{d(eq)ii}_{R/A}(q)+M_i}{(q_{0}\pm i \Gamma_{i q} /2 )^2-\omega_{i q}^2 } 
\simeq \left\{ \begin{matrix} \displaystyle
\sum_{\epsilon=\pm} 
\frac{i Z_{\epsilon }^{i}}{q_0- \Omega_{\epsilon i} } \\
\displaystyle \sum_{\epsilon=\pm} 
\frac{i Z_{\epsilon }^{i}}{q_0- \Omega^{*}_{\epsilon i} } \end{matrix} \right. \label{G_{RA}^{d(eq)}(q)2}
\end{align}
where
\be
\Omega_{\epsilon i} \equiv \epsilon \omega_{i q}-i\Gamma_{iq}/2
\ee
and
\begin{align}
Z_{\epsilon }^{i}=\frac{i\epsilon}{2 \omega_{iq}} (\Slash{q}_{\epsilon i}+M_{i}),\ \ \Slash{q}_{\epsilon i} \equiv \epsilon \omega_{i q}\gamma^{0}-{\bold q}\cdot {\boldsymbol \gamma}/a_0 \ .  \label{defZ}
\end{align}
In the real time representation,  it becomes\footnote{
 In the present analysis, we expand various quantities
  with respect to $(h^\dagger h)'$. Hence the propagator
 of $i$-th flavor is almost identified with the propagator of  the $i$-th mass eigenstate
   up to higher order terms of 
 $(h^\dagger h)'$.  
Propagations of a single $N_i$ corresponds to propagations of a single mass eigenstate
 with mass $M_i$ and width $\Gamma_i$.}
\begin{align}
&G_{R}^{d(eq)ii}(x^0,y^0;{\bold q}) =
+\Theta (x^0 -y^0) \sum_{\epsilon=\pm} Z_{\epsilon }^{i}e^{-i\Omega _{\epsilon i}(x^0 -y^0)} \ , \n
&G_{A}^{d(eq)ii}(x^0,y^0;{\bold q})
=
-\Theta (y^0 -x^0) \sum_{\epsilon=\pm} Z_{\epsilon }^{i}e^{-i\Omega^*_{\epsilon i}(x^0 -y^0)} \ . 
\label{G_{RA}^{d(eq)}(q)3}
\end{align}

$\Gamma_q$ is multiplied by the Lorentz boost factor as 
$\Gamma_{ q}\simeq  (M/\omega_{ q}) \times \Gamma$
where $\Gamma \equiv \Gamma_{q=0}$ is the decay width of the RH neutrino.
In the present paper, we consider a situation that two RH neutrinos
are almost degenerate in their masses  
\begin{align}
\Delta M \equiv |M_i -M_j|\simeq \Gamma \ . 
\end{align}
In the following, we sometimes use the averages denoted by quantities without the flavor index $i,j$
\begin{align}
M=\frac{M_i +M_j}{2}\ ,\ \ \omega_{q}=\frac{\omega_{iq} +\omega_{jq}}{2}\ ,\ \ \Omega_{\epsilon}=\frac{\Omega_{\epsilon i} +\Omega_{\epsilon j}}{2} \ ,\ \ etc.
\end{align}
\subsection{Off-diagonal $G_{R/A}^{\prime}$ in thermal equilibrium}
We then study  the behavior of the 
flavor off-diagonal component  $G^{'(eq)}_{R/A}$ of the retarded (advanced)
 propagators in the thermal equilibrium.
From (\ref{G_{RA}^{'}}), it is given by
\begin{align}
G_{R/A}^{'(eq)ij}(x^0,y^0;{\bold q})
=-\int \frac{dq_0}{2\pi} e^{-iq_{0}(x^0 -y^0)}G_{R/A}^{d(eq)ii}(q)\Pi^{'(eq)ij}_{R/A}(q)G_{R/A}^{d(eq)jj}(q) \ . 
\end{align}
The $q_0$ integration can be performed by summing residues of the poles.
Eq.  (\ref{G_{RA}^{d(eq)}(q)2}) shows that the retarded propagator 
$G_{R}^{d(eq)ii}$ has poles at $q_0= \Omega_{\pm, i}$ and 
the advanced propagator  $G_{A}^{d(eq)jj}$  has poles at  $q_0= \Omega_{\pm, j}^*$.
The self-energy $\Pi_{R/A}$ consists of the SM lepton and the Higgs propagator, and hence
it has poles at $q_0=\epsilon_{\ell} \omega_p + \epsilon_\phi \omega_k \mp i \Gamma_{\ell \phi}/2$
with a large imaginary part.
Because of this, the residues of the poles of the self-energy are suppressed by
$\Gamma_i/\Gamma_{\ell\phi} \ll 1$.
Noting the relation
\be
\frac{1}{q_0 -\Omega_{\epsilon i}} \frac{1}{q_0 -\Omega_{\epsilon' j}}
= \frac{1}{\Omega_{\epsilon i} - \Omega_{\epsilon' j}} \left(
\frac{1}{q_0 -\Omega_{\epsilon i}} - \frac{1}{q_0 -\Omega_{\epsilon' j}}
\right) \ ,
\label{partialfraction}
\ee
we can see that 
the contribution $\epsilon = - \epsilon'$ is also suppressed by $\Delta M/M$
compared to the $\epsilon =  \epsilon'$ contribution.
Hence, dropping these suppressed contributions, we have
\begin{align}
G_{R}^{'(eq)ij}(x^0,y^0;{\bold q})
\simeq &+\Theta (x^0 -y^0) \sum_{\epsilon} Z_{\epsilon } \Pi^{'(eq)ij}_{R}(\epsilon \omega_{q}) Z_{\epsilon }\frac{-i}{\Omega_{\epsilon i}-\Omega_{\epsilon j}} \n
&\hspace{40 pt}\times \left( e^{-i\Omega _{\epsilon i}(x^0 -y^0)} -e^{-i\Omega _{\epsilon j}(x^0 -y^0)} \right) \
 \label{G_{R}^{'(eq)}} 
\end{align}
and
\begin{align}
G_{A}^{'(eq)ij}(x^0,y^0;{\bold q})
\simeq &-\Theta (y^0 -x^0) \sum_{\epsilon} Z_{\epsilon } \Pi^{'(eq)ij}_{A}(\epsilon \omega_{q}) Z_{\epsilon }\frac{-i}{\Omega^{*}_{\epsilon i}-\Omega^{*}_{\epsilon j}} \n
&\hspace{40pt}\times \left( e^{-i\Omega^{*} _{\epsilon i}(x^0 -y^0)} -e^{-i\Omega^{*} _{\epsilon j}(x^0 -y^0)} \right) \ .
\label{G_{A}^{'(eq)}} 
\end{align}
We  also used the approximation
$\Pi(\Omega_{\epsilon i})\simeq \Pi(\Omega_{\epsilon j}) \simeq \Pi(\epsilon \omega_{q})$
because $\Gamma_i \ll \Gamma_{\ell \phi}$.

The minus signs in the parentheses come from the relative minus sign of the residue in 
(\ref{partialfraction}).
Because of this, the off-diagonal Green functions vanish at $x^0=y^0$:
\be
G_{R/A}^{\prime} (x,y) {\big |}_{x^0=y^0} =0 \ .
\ee
This should generally hold by the definition of $G_{R/A}$ in (\ref{defG_{RA}}) 
because 
$G_\rho^{ij}(x,y)$ is proportional to $\delta^{ij} \delta^3({\bf x}-{\bf y})$ at equal time $x^0=y^0$:
\be
\gamma^0 G^{ij}_R(x,y){\big |}_{x^0=y^0}
=\Theta(x^0-y^0) \gamma^0 G^{ij}_\rho(x,y) {\big |}_{x^0=y^0}
= \frac{i}{2} \delta({\bf x}-{\bf y}) \delta^{ij} \ .
\ee
Note that the flavor off-diagonal components of the retarded (advanced) propagators
are enhanced by the factor $1/(\Omega_i -\Omega_j)$ (or its complex  conjugate).
Such a large enhancement comes from the large mixing  of the
RH neutrinos with  almost degenerate masses.

For the self-energies $\Pi_{R/A} = \Pi_{h} \pm \Pi_{\rho}/2$,
if we use the vacuum value 
$\Pi'_{\rho}(\epsilon \omega_{ q})=-g_w \Re (h^{\dag} h)'i \epsilon \Slash{q}_{\epsilon}/(16\pi)$
and  $\Pi'_{h}(\epsilon \omega_{ q})=0$ as in Appendix \ref{AppSE},  the following expressions
\cite{Garny:2011hg} are reproduced:
\begin{align}
G_{R}^{'(eq)ij}(x^0,y^0;{\bold q})\simeq &+\Theta (x^0 -y^0) \sum_{\epsilon}\frac{\Slash{q}_{\epsilon}+M}{2\omega_{ q}}\frac{g_w M^2 \Re (h^{\dag} h)' /(16\pi)}{M_{i}^2 -M_{j}^2 -i\epsilon (M_{i}\Gamma_{i} -M_{j}\Gamma_{j})}\notag \\
&\hspace{40pt}\times \left( e^{-i\Omega _{\epsilon i}(x^0 -y^0)} -e^{-i\Omega _{\epsilon j}(x^0 -y^0)} \right) \ ,
\n
G_{A}^{'(eq)ij}(x^0,y^0;{\bold q})\simeq &-\Theta (y^0 -x^0) \sum_{\epsilon}\frac{\Slash{q}_{\epsilon}+M}{2\omega_{q}}\frac{- g_w M^2 \Re (h^{\dag} h)' /(16\pi)}{M_{i}^2 -M_{j}^2 +i\epsilon (M_{i}\Gamma_{i} -M_{j}\Gamma_{j})} \notag \\
&\hspace{40pt}\times \left( e^{-i\Omega^{*} _{\epsilon i}(x^0 -y^0)} -e^{-i\Omega^{*} _{\epsilon j}(x^0 -y^0)} \right) \ . 
\label{thermalGRA2}
\end{align}
Here we have used the relation
\begin{align}
\frac{\epsilon }{2\omega_{ q}}\frac{1}{\Omega_{\epsilon i}-\Omega_{\epsilon j}}&\simeq \frac{1}{\omega_{iq}^2 -\omega_{j q}^2 -i\epsilon (\omega_{i q}\Gamma_{iq} -\omega_{j q}\Gamma_{j q})}\notag \\
&\simeq \frac{1}{M_{i}^2 -M_{j}^2 -i\epsilon (M_{i}\Gamma_{i} -M_{j}\Gamma_{j}) } \ 
\label{Omega-Omega}
\end{align}
which is valid for $\omega_q \simeq \omega_{iq} \simeq \omega_{jq}$ and  $\omega_q \Gamma_q \simeq M \Gamma$.
Hence, the enhancement factor
$1/(\Omega_i -\Omega_j)$ corresponds to taking the regulator $(M_i \Gamma_i -M_j \Gamma_j)$ is obtained.
As shown in  section \ref{Sec-offDWinTE}, the same enhancement factor, that is, the same regulator
appears in the off-diagonal Wightman function in the thermal equilibrium.
For the deviations of the off-diagonal Wightman functions out of equilibrium, however,
we show in  section \ref{Sec-ODWoutofeq} that
the enhancement factor is changed to be $1/(\Omega_i -\Omega_j^*)$. This
corresponds to the regulator $(M_i \Gamma_i + M_j \Gamma_j).$

Finally we note the validity of the expansion with respect to the 
 off-diagonal components of the Yukawa couplings $(h^{\dag}h)'$.
From the expressions (\ref{thermalGRA2}),
the iterative expansions (\ref{G_{RA}^{0+2}}) and (\ref{G_{RA}^{'}}) turn out to be valid 
when the real part of the off-diagonal components of Yukawa coupling $\Re (h^{\dag}h)'$ 
is smaller than the mass difference $|M_i-M_j|/M \simeq \Gamma / M \sim (h^{\dag}h)^{d}_{ii}$. 
Hence the expansion is understood as an expansion of the ratio $(h^\dagger h)'/(h^\dagger h)^d.$\footnote{
In \cite{Garny:2011hg}, numerical analysis has been done beyond this parameter region.}

\subsection{Wightman functions}
The Wightman functions can be solved as (\ref{GformalSol}) or
 (\ref{GwightmanSec2}). 
If we take terms up to the first order of $(h^{\dag}h)^{'}$, 
 the flavor diagonal  component is given by
\begin{align}
G^{d ii}_{\wightman}=&- G^{d ii}_{R}* \Pi^{d ii}_{\wightman} *G^{d ii}_{A }\  \label{intKBeqG_{W}^{d}}.
\end{align}
Similarly the flavor off-diagonal component is given by
\begin{align}
G^{' ij}_{\wightman}=&- G^{' ij}_{R}* \Pi^{djj}_{\wightman} *G^{djj}_{A} 
- G^{dii}_{R}* \Pi^{d ii}_{\wightman } *G^{'ij}_{A }\label{intKBeqG_{W}^{'}1} 
- G^{dii}_{R}* \Pi^{'ij}_{\wightman } *G^{d jj}_{A }\  .
\end{align}
By using (\ref{G_{RA}^{'}}) and (\ref{intKBeqG_{W}^{d}}), 
(\ref{intKBeqG_{W}^{'}1}) can be also rewritten as 
\begin{align}
G^{' ij}_{\wightman }=&- G^{d ii}_{R}* \Pi^{' ij}_{R} *G^{djj}_{\wightman}
- G^{dii}_{\wightman}* \Pi^{'ij}_{A } *G^{djj}_{A}
- G^{dii}_{R}* \Pi^{'ij}_{\wightman} *G^{djj}_{A}  \label{intKBeqG_{W}^{'}2}
\end{align}
which makes it clear that the off-diagonal part of the self-energy causes the flavor mixing of the RH neutrino.\footnote{
This form is convenient for the systematic derivation of the Boltzmann equation from the KB equation in the hierarchical mass spectrum \cite{Frossard:2012pc}, in which the diagonal components of the Wightman propagator are identified as the on-shell external line of the RH neutrinos.
In this paper, we are focusing on the resonant mass spectrum,
and we use this form, without such an assumption, to solve the off-diagonal components of the Wightman propagator.}

\subsection{Diagonal Wightman $G_\gtrless^d$ in thermal equilibrium}
In the thermal equilibrium, 
the Wightman function  can be easily obtained by using the KMS relation.
From (\ref{intKBeqG_{W}^{d}}), the diagonal component
$G^{d(eq)}_{\wightman}$ can be written as 
\begin{align}
G_{\wightman}^{d(eq)}(x^0,y^0;{\bold q})=&-\int \frac{dq_0}{2\pi} e^{-iq_{0}(x^0-y^0)} G_{R}^{d(eq)}(q)\Pi^{d(eq)}_{\wightman }(q)G_{A}^{d(eq)}(q) \ . 
\end{align}
Let $f(q)$ be the thermal distribution function for 
the RH neutrinos. Note that $f(q)$ is a function of $q^0$, which is 
 not equal to the on-shell energy $\omega_{\bf q}.$
The KMS relation for the self-energy function is
\begin{align} 
\Pi^{(eq)}_{\wightman}(q)=-i\left\{ \begin{matrix} 1-f(q_0) \\
-f(q_0) \end{matrix} \right\} \Pi^{(eq)}_{\rho}(q) \ . \label{KMSforSE}
\end{align}
Using the solution of the KB equation for the spectral density 
 $G_{\rho} \equiv G_R-G_A =-G_{R}* \Pi_{\rho}* G_{A}$, we have
\begin{align}
& G_{\wightman}^{d(eq)}(x^0,y^0;{\bold q}) \n
&= -\int \frac{dq_0}{2\pi} e^{-iq_{0}(x^0-y^0)} (-i)\left\{ \begin{matrix} 1-f(q_0) \\
-f(q_0) \end{matrix} \right\} G_{R}^{d(eq)}(q)\Pi^{d(eq)}_{\rho }(q)G_{A}^{d(eq)}(q) \notag \\
& =+\int \frac{dq_0}{2\pi} e^{-iq_{0}(x^0-y^0)} (-i)\left\{ \begin{matrix} 1-f(q_0) \\
-f(q_0) \end{matrix} \right\} \left[ G_{R}^{d(eq)}(q)- G_{A}^{d(eq)}(q) \right] 
\label{G_{W}^{d(eq)}0} \ .
\end{align}
It is nothing but the KMS relation (\ref{KMSforG0}) for the Green function.

Performing the $q_{0}$ integration, it becomes
\begin{align}
G_{\wightman}^{d(eq)ii}(x^0,y^0;{\bold q}) \simeq  \sum_{\epsilon}  \left\{ \begin{matrix} 1-f^{\epsilon}_{i q} \\
-f^{\epsilon}_{i q} \end{matrix}\right\}(-i) Z^{i}_{\epsilon }  
 &{\Big (} \Theta (x^0 -y^0) e^{-i\Omega _{\epsilon i}(x^0 -y^0)} \label{G_{W}^{d(eq)}1} \n
& +\Theta (y^0 -x^0) e^{-i\Omega^{*} _{\epsilon i}(x^0 -y^0)} {\Big ) } \ .
\end{align}
Here we have dropped the contributions from poles of the distribution function $f(q_0)$
since they are suppressed by $\Gamma /T \ll 1$. 
Furthermore we used the distribution function
\begin{align}
f_{i p}^{\epsilon}\equiv f(q_0 =\epsilon \omega_{i q})=\frac{1}{e^{\epsilon \omega_{i q}/T}+1} 
\end{align}
by dropping the imaginary part of  the pole $\Omega_{\epsilon i}$ in $f(q)$ because 
it is suppressed again by the factor $\Gamma \ll T$.
Recall that it satisfies the relation $(1-f_{i p}^{\epsilon})=+f_{i p}^{-\epsilon}$.

\subsection{Off-diagonal Wightman $G_\gtrless^{\prime}$ in thermal equilibrium}
\label{Sec-offDWinTE}
Next we calculate the flavor off-diagonal component 
$G^{'(eq)}_{\wightman}$ in the thermal equilibrium.
The off-diagonal component also satisfies the KMS relation and we have
\begin{align}
 G^{'(eq)ij}_{\wightman} & (x^0,y^0;{\bold q})=+\int \frac{dq_0}{2\pi} e^{-iq_{0}(x^0-y^0)} (-i)\left\{ \begin{matrix} 1-f(q_0) \\
-f(q_0) \end{matrix} \right\} G_{\rho }^{'(eq)ij}(q)  \n
=&+\int \frac{dq_0}{2\pi} e^{-iq_{0}(x^0-y^0)} (-i)\left\{ \begin{matrix} 1-f(q_0) \\
-f(q_0) \end{matrix} \right\} \left[ G_{R}^{'(eq)ij}(q)- G_{A}^{'(eq)ij}(q) \right] \ .
\label{G_{W}^{'(eq)}0} 
\end{align}
Performing $q_0$ integration, it becomes
\begin{align}
G^{'(eq)ij}_{\wightman} & (x^0,y^0;{\bold q})=  \sum_{\epsilon} Z_{\epsilon } \Pi^{'(eq)ij}_{R}(\epsilon \omega_{ q}) Z_{\epsilon }\frac{-i}{\Omega_{\epsilon i}-\Omega_{\epsilon j}} \n
& \times (-i){\bigg [ }\left\{ \begin{matrix} 1-f^{\epsilon}_{iq} \\
-f^{\epsilon}_{iq} \end{matrix} \right\} e^{-i\Omega _{\epsilon i}(x^0 -y^0)}  - \left\{ \begin{matrix} 1-f^{\epsilon}_{jq} \\
-f^{\epsilon}_{jq} \end{matrix} \right\}e^{-i\Omega _{\epsilon j}(x^0 -y^0)} {\bigg ]} 
\label{G_{W}^{'(eq)}1} 
\end{align}
for $x^0 >y^0$. We have used similar approximations by dropping suppressed contributions
by $\Gamma/T$ and $\Gamma/\Gamma_{\ell \phi}$.

The off-diagonal component of the
thermal Wightman functions are enhanced by the same factor $1/(\Omega_{\epsilon i}-\Omega_{\epsilon j})$ as in (\ref{G_{R}^{'(eq)}}). Hence the flavor oscillation of the Wightman function in the thermal equilibrium
is enhanced by a factor with the regulator $M_i \Gamma_i - M_j \Gamma_j.$

At the temperature $T\gg \Delta M$ we have in mind,  
$f_{i}$ and $f_{j}$ can be almost identified. Writing
$f_{i}\simeq f_{j}\simeq f$, we have
\begin{align}
G^{'(eq)ij}_{\wightman}(x^0,y^0;{\bold q})= & \Theta(x^0 -y^0)\sum_{\epsilon} Z_{\epsilon} \Pi^{'(eq)ij}_{R}(\epsilon \omega_{ q}) Z_{\epsilon} \frac{-1}{\Omega_{\epsilon i}-\Omega_{\epsilon j}} 
\notag \\ &\times 
\left\{ \begin{matrix} 1-f^{\epsilon}_{q} \\
-f^{\epsilon}_{q} \end{matrix} \right\} \left(  e^{-i\Omega _{\epsilon i}(x^0 -y^0)}  -e^{-i\Omega _{\epsilon j}(x^0 -y^0)} \right) \notag \\
+& \Theta(y^0 -x^0)\sum_{\epsilon} Z_{\epsilon} \Pi^{'(eq)ij}_{A}(\epsilon \omega_{ q}) Z_{\epsilon}
\frac{-1}{\Omega^{*}_{\epsilon i}-\Omega^{*}_{\epsilon j}}  \n
&\times \left\{ \begin{matrix} 1-f^{\epsilon}_{q} \\
-f^{\epsilon}_{q} \end{matrix} \right\} \left(  e^{-i\Omega^{*} _{\epsilon i}(x^0 -y^0)}  -e^{-i\Omega^{*} _{\epsilon j}(x^0 -y^0)} \right) \ .
 \label{G_{W}^{'(eq)}2} 
\end{align}
The off-diagonal Wightman functions in the thermal equilibrium
 vanishes at the equal time $x^0=y^0$:
\be
\lim_{x_0 \rightarrow y_0} G_\gtrless^{\prime(eq)ij}(x^0,y^0;{\bf q}) 
\propto (\Omega_i -\Omega_j) (x^0-y^0) \sim \Delta M (x^0-y^0) \rightarrow 0 \ .
\label{ODWinTEvanish}
\ee
Later this property becomes very important to evaluate the deviation of 
the off-diagonal component of the Wightman function when the 
system is out of thermal equilibrium.

\subsection{Summary of this section}
In this section, we calculated various propagators of the RH neutrinos in the thermal equilibrium.
We especially focused on the resonant enhancement of the flavor oscillation of $N_i$.
Retarded or advanced propagators are composed of two propagating modes, $i$ and $j$ flavors.
The flavor diagonal components are given by (\ref{G_{RA}^{d(eq)}(q)2}) or (\ref{G_{RA}^{d(eq)}(q)3}).
Since their masses are almost degenerate, the flavor off-diagonal component is largely enhanced
due to their oscillation as in (\ref{G_{R}^{'(eq)}}) or (\ref{G_{A}^{'(eq)}}).
The enhancement factor is proportional to $1/(\Omega_i -\Omega_j)$ (or its complex conjugate)
where $\Omega_i=\omega_i -i\Gamma_i/2$ 
 and gives the regulator $R_{ij}=M_i \Gamma_i - M_j \Gamma_j$
to the enhancement factor. Similarly, the resonant enhancement of Wightman functions is calculated.
In the thermal equilibrium, because of the KMS relation, the behavior of the Wightman functions
is the same as the retarded (advanced) Green functions. The flavor diagonal component $G_\gtrless^d$
is given by  (\ref{G_{W}^{d(eq)}1}) while the off-diagonal component $G_\gtrless^{'}$
is given by (\ref{G_{W}^{'(eq)}1}). A very important property of $G_\gtrless^{'}$ is that it
vanishes at the equal time  as (\ref{ODWinTEvanish}).
\section{Propagators out of equilibrium}
\label{Sec-POEQ}
\setcounter{equation}{0}

Now we study effects of the expanding universe into account.
First we summarize  various time-scales in the system.
An important time
 scale is given by the Hubble expansion rate $H$ of the universe.
Other scales are the decay widths of the SM particles $\Gamma_\phi, \Gamma_\ell$
and of the RH neutrino $\Gamma_i$. 
Another important time scale in the resonant leptogenesis
 is given by the mass difference $\Delta M$ of the RH neutrinos
because it gives the frequency of the flavor oscillation.

In type I sea-saw model studied in the present paper,
the decay width $\Gamma_i$ of the RH neutrino is approximately given 
by  $\Gamma_i \sim (h^{\dag}h)_{ii} M_i/8 \pi$.
The ratio of $\Gamma_i$
to the Hubble parameter (\ref{HubbleParameter})  at temperature $T=M_i$ is 
rewritten 
in terms of the ``effective neutrino mass" $\tilde{m}_i$ as (see e.g. \cite{Fong:2013wr})
\be
K_i = \frac{\Gamma_i}{H(M_i)}
= \frac{\tilde{m}_i}{10^{-3} {\rm eV}}\ , \ \ \ \tilde{m}_i  \equiv \frac{ (h^\dagger h)_{ii} v^2}{M_i} \ .
\ee
where $v$ is the scale of the EWSB. 
Hence if we take the Yukawa coupling
so as to $\tilde{m}_i \sim 0.1$ eV, the ratio becomes $K_i \sim 100$.
This corresponds to the strong washout regime.
The Yukawa coupling itself is very small ($h \sim 10^{-5}$ for $M \sim 10 \ {\rm TeV}$), and we have the 
inequalities
\begin{align}
\Gamma_{\phi} , \Gamma_{\ell}\gg \Gamma_i \gg H \ . \label{scales}
\end{align}

\subsection{Deviation of self-energy from the thermal value}
Under the  condition (\ref{scales}), we can expand the scale factor as
\begin{align}
a_{(X)}= a_{(t)} + a_{(t)} H_{(t)} (X-t) +\cdots \ . \label{aGE} 
\end{align}
The other physical quantities such as temperature are correlated with 
the change of the scale factor, and can be similarly expanded. 

In order to calculate the out-of-equilibrium 
behavior of various Green functions in the expanding
universe, we need to evaluate the change of the self-energies $\Pi(x,y)$.
The self-energy of the RH neutrino is a rapidly decreasing function with the relative time as
$\sim e^{-\Gamma_{\ell \phi} (x^0-y^0)}$
due to the SM gauge interactions. 
So in the leading order approximation, the self-energy $\Pi(x,y)$
can be evaluated by the thermal value with the local temperature at the
center-of-mass time $x^0 \sim y^0 \sim X_{xy}.$
Therefore it is  convenient to write the self-energy
 as
\be
 \Pi(x^0,y^0;{\bf q}) 
 =   \Pi(X_{xy};s_{xy};{\bf q}) \simeq
  \Pi^{(eq)}(X_{xy};s_{xy};{\bf q}) \ , 
 \label{SE-out-of-eq} 
 \ee
 where
 \be
   X_{xy}\equiv \frac{x^0+y^0}{2}\ , \ \ s_{xy} \equiv x^0-y^0 \ .
\ee
The first equation of of (\ref{SE-out-of-eq}) is the definition of $\Pi(X;s;{\bf q})$.
In the second equality, we replaced $\Pi$ by its thermal value $\Pi^{(eq)}$
since the SM leptons and Higgs are in the thermal equilibrium
and the self-energy of the RH neutrinos is well approximated by its thermal value.
$\Pi^{(eq)}(X_{xy};s)$ means  the thermal self-energy in the thermal equilibrium
evaluated at  time $X_{xy}$.

In evaluating   the Wightman function $G_\gtrless$ of the RH neutrinos,  
we need to know a difference of the self-energy $\Pi(u,v)$
from the thermal value at a later time $t$.
For example, in (\ref{GwightmanSec2}), the difference of the
self-energy $\Pi(X_{uv};s)$ at $X_{uv}$ and the thermal value $\Pi^{(eq)}(t;s)$
at $t=X_{xy}$ controls the behavior of $G_\gtrless^{ij}$.
In this case, the time difference between $X_{uv}$ and $t=X_{xy}$ is given by the 
inverse of the decay width $\Gamma_i$ of the RH neutrino $N_i$. Since
\be
\frac{1}{\Gamma_{\ell \phi}} \ll t-X_{uv} \sim \frac{1}{\Gamma_i} \ll \frac{1}{H} \ ,
\label{tXineq}
\ee
the derivative expansion of the self-energy
around the thermal value is a good approximation:
\begin{align}
\Pi(X_{uv};s;{\bold q})  \simeq & \ \Pi^{(eq)}(t;s;{\bold q}) + 
 (X_{uv}-t) \partial_{t} \Pi^{(eq)} (t;s;{\bold q})
 + \Delta_{\mu(X)} \Pi \ . 
\end{align}
The second term is of order ${\mathcal O}(H/\Gamma_i)$ owing to (\ref{tXineq}).
The third term comes from the chemical potential of leptons
 generated by $CP$-violating decay of the RH neutrinos. So it is the genuine deviation of the 
 self-energy from the thermal value at the same time $X_{uv}$.

In this section, we mainly focus on the change of the physical quantities, namely
the second term because the back reaction of the generated lepton asymmetry
to the evolution of the number density of the RH neutrinos is very small.
The effect of the chemical potential becomes important in the generation 
of the lepton asymmetry and is considered in section \ref{Sec5}. 

\subsection{Notice for notations}
As already used in (\ref{SE-out-of-eq}), $\Pi(X;s)$ is the self-energy
at the center-of-mass time $X$ with the relative time $s$.
For the thermal value $\Pi^{(eq)}(X;s)$, $X$ is not necessarily at the center-of-mass time,
but, more generally, denotes the reference time when it is evaluated. $s$ is always the relative time.
For the thermal value, we also use its Fourier transform 
\be
\Pi^{(eq)}(X;q) = \int ds \ \Pi^{(eq)}(X;s) e^{-iqs} \ .
\ee
In order to avoid complications of appearance, we use the same notations $\Pi$ for $\Pi(X;s)$
and its Fourier transform $\Pi(X;q)$. They can be distinguished by their arguments, $s$
or $q$, if necessary. We always use $s$ for the relative time and $q$ for its conjugate frequency.
For the first argument (the reference time), we use  $X$ or $t$.
The same notation is used for the thermal Green functions.
We hope it does not cause any confusion to the readers.

\subsection{Retarded propagator out of equilibrium $\Delta G_R$}
First we study how the retarded (advanced) propagators of the RH
neutrinos deviate from the thermal value in the expanding universe. 
Consider the flavor diagonal component $G_{R/A}^d$ first.
We write the deviation around the thermal value  $G^{d(eq)}$ by $\Delta G^d$:
\begin{align}
G^{d}_{R/A}(X_{xy} ; s_{xy};{\bold q}) = G^{d(eq)}_{ R/A}(t;s_{xy};{\bold q}) + \Delta G^{d}_{ R/A}(X_{xy};s_{xy};{\bold q}) \ .
\end{align}
Note that $\Delta G^{d}_{ R/A}$  depends on the 
reference time $t$ at which the equilibrium value is evaluated.
It is calculated in Appendix \ref{Sec-app-retarded} and  given by
\begin{align}
\Delta G^{d}_{ R} (x^0,y^0;{\bold q})
\simeq   \Theta( s_{xy})\sum_{\epsilon}&{\bigg [} 
\partial_t \left( Z_\epsilon e^{-i \Omega_{\epsilon}s_{xy}}\right)
(X_{xy}-t)
\n
&
-i\frac{H_{(t)} M}{4\omega^2_{q}}\gamma^{0} \frac{{\boldsymbol \gamma}\cdot {\bold q}}{a_{(t)}} \ s_{xy} \  
e^{-i \Omega_{\epsilon q}s_{xy}}
{\bigg ]} \ .
\end{align}
The first term is  the change of the physical parameters such as 
mass or width in  $\Omega_{\epsilon}$ and  $Z_\epsilon$. 
The second term  represents a change of the 
spinor structure due to an expansion of the universe in the propagator during the propagation.
The retarded (advanced) 
propagator does not have the memory effect, and the deviation is essentially
determined by the change of the local temperature.

By taking a variation of (\ref{G_{RA}^{'}}), 
the deviation of the off-diagonal components $G_{R/A}^{\prime}$
can be expressed in terms of the deviation of the diagonal components $G_{R/A}^d$ as
\begin{align}
\Delta G_{R/A}^{'ij} = &
- G_{R/A}^{d(eq)ii}*   \Delta\Pi^{'(eq)ij}_{R/A}*G_{R/A}^{d(eq)jj} 
-\Delta G_{R/A}^{d ii}* \Pi^{'(eq)ij}_{R/A}* G_{R/A}^{d(eq)jj} 
\n
& -  
G_{R/A}^{d(eq)ii}*\Pi^{'(eq)ij}_{R/A}*  \Delta G_{R/A}^{d jj} \ .
\label{DelG_{RA}^{'}} 
\end{align}
The above formula is used to evaluate the deviation of the Wightman functions
of the RH neutrinos in the latter section \ref{Sec-ODWoutofeq}.
Since the above relation (\ref{DelG_{RA}^{'}}) is sufficient for latter calculations
of $\Delta G_\gtrless^{'}$, we do not calculate an explicit form of $\Delta G_R^{'}$ here.
We note that, since
the retarded (advanced)  propagators do not have the memory effect, 
its deviation is essentially
determined by the change of the local temperature.
Also note that the enhancement factor is 
proportional to $1/(\Omega_i -\Omega_j)$ as the Green functions in the 
thermal equilibrium since there is no chance to mix $G_R$ and $G_A$.

\subsection{Diagonal Wightman  out of equilibrium $\Delta G^{d}_{\gtrless}$ }
\label{SecDiagWightOE}
The deviation of the flavor diagonal Wightman function $\Delta G^{d}_{\wightman} (x^0,y^0)$
can be calculated by taking a variation
of  (\ref{intKBeqG_{W}^{d}}):
\begin{align}
\Delta G^{d}_{\wightman} = & 
- \Delta G^{d}_{R} * \Pi^{d (eq)}_{\wightman} *G^{d(eq)}_{A} 
- G^{d(eq) }_{R}* \Pi^{d(eq) }_{\wightman} * \Delta G^{d }_{A} \n
&- G^{d(eq) }_{R}* \Delta \Pi^{d(eq) }_{\wightman} *G^{d(eq) }_{A} \ .
\label{DelG_{W}_0}
\end{align}
There are three terms. 
The first two terms are interpreted as the change of the spectrum 
in the expanding universe contained in $G_{R/A}.$ 
On the other hand, the third term reflects the memory effect.

The third term is explicitly written\footnote{
Since all quantities are already Fourier transformed in the spatial direction with 
momentum ${\bf q}$, we use $u,v$ instead of $u^0,v^0$ to avoid complications.} 
as 
\begin{align}
 -  \int_{-\infty}^{x^0} d u  \int_{-\infty}^{z^0} 
 dv  \ G_{R}^{d(eq)}(x,u) \ 
\Delta \Pi_{\wightman}^{d(eq)} (u,v) \ G^{d(eq)}_{A }(v,z) \ . \label{DelPi(uv)}
\end{align}
This shows that the Wightman function is sensitive to the change of the background
before $x^0$ and $y^0$ unlike the retarded or advanced Green functions.
Writing the self-energy in terms of the center of mass coordinate $X_{uv}=(u+v)/2$
and the relative coordinate $s_{uv}=u-v$, its deviation from the thermal self-energy
at time $t=x^0$ is written as
\begin{align}
\Delta \Pi_{\wightman}^{(eq)} & 
(X_{uv};s_{uv};q)=\int \frac{dq_{0}}{2\pi} e^{-iq_{0}s_{uv}} \partial_{X} \Pi^{(eq)}_{\wightman }(X;q) {\big |}_{X=t} (X_{uv}-t) \n
\simeq &\int \frac{dq_{0}}{2\pi} e^{-iq_{0}s_{uv}} \partial_{X} \left[(-i)\left\{ \begin{matrix} 1-f(q_0) \\
-f(q_0) \end{matrix} \right\} \Pi^{(eq)}_{\rho }(X;q) \right]_{X=t} (X_{uv}-t) \ .
\label{DelPiKMS} 
\end{align}
Note that 
$|s_{uv}| \lesssim 1/\Gamma_{\ell \phi}$ due to the rapid damping of SM leptons and Higgs propagators.
In the second equality the KMS relation for the thermal self-energy (\ref{KMSforSE})
is used. As explained in eq.(\ref{SE-out-of-eq}), the self-energy function out of equilibrium
can be approximated by the equilibrium self-energy $\Pi^{(eq)}$ of (\ref{KMSforSE})
at the local temperature. 
Note that the distribution function 
 $f(q_0)=1/(e^{q_{0}/T}+1)$ is time-dependent through the time-dependence of the temperature
$T=T(X)$.

The calculation of the deviation of the diagonal Wightman function $\Delta G_\gtrless^d$ 
is performed in Appendix \ref{Sec-App-out-of-eq-Wightman-diag}. 
For $x^0>y^0$, it is given by
\begin{align}
\Delta G^{dii}_{\wightman } &( x^0,y^0;{\bold q}) 
 \simeq  (-i) \sum_{\epsilon}  {\Big [}
\left\{ \begin{matrix} 1-f^{\epsilon}_{i q} \\ -f^{\epsilon}_{i q} \end{matrix} \right\} 
 \Delta \hat{G}^{dii}_{R}(x^0,y^0;\epsilon,{\bold q})
\n
&+ d_t  \left\{ \begin{matrix} 1-f^{\epsilon}_{i q} \\
-f^{\epsilon}_{i q} \end{matrix} \right\} 
\left( \frac{-1}{\Gamma _{iq}}+(X_{xy} -t -|s_{xy}|/2) \right) 
  Z_{\epsilon }^{i} e^{-i\Omega_{\epsilon i}(x^0 -y^0)} {\Big ]}
 \label{honbun-deltaGWdiag} 
\end{align}
where
\begin{align}
d_t \equiv \frac{\partial T}{\partial t} \frac{\partial}{\partial T} +\frac{\partial \omega_{q}}{\partial t} \frac{\partial}{\partial \omega_{q}} \ .
\end{align}
Each term of (\ref{honbun-deltaGWdiag})
is classified into three types of terms.

The first term of  $\Delta G_\gtrless^d$ 
in the square bracket reflects the change of the spectrum in the propagators $G_{R}$
and related by the KMS relation (\ref{G_{W}^{d(eq)}0}). It reflects a change of the local
temperature during the period $x^0$ and $y^0$. 

The term proportional to $(X_{xy}-t)$ comes from a difference between the 
distribution function $f_q (t)$ at the reference time $t$
and $f_q (X_{xy})=f_q(t) +(X_{xy}-t)d_{t}f_q$ at time $X_{xy}$. 
The time-dependence of $f_q$ comes from both of the local temperature and 
the physical frequency $\omega_q$ as shown in the definition of 
the derivative operator $d_t$.
The term with $s_{xy}$ is similar. If $x^0 \neq y^0$, the distribution function at $X_{xy}$ is affected by 
the information at the past. 

The most important part is the term proportional to $1/\Gamma_i$, which reflects 
the memory effect of the Wightman function.
Since the Wightman function is written as a convolution
$G^d_{\gtrless}(X_{xy};s_{xy}) = - (G_R * \Pi_\gtrless * G_A)(X_{xy};s_{xy})$,
 they depend on the information in the past at $X_{uv}$ where $X_{xy}-X_{uv} \sim 1/\Gamma_i$ (see (\ref{DelPi(uv)})).
In the expanding universe, the temperature is higher in the past and
the number density of leptons and Higgs are larger than the present density.
Accordingly the number density of the RH neutrinos is also larger by an amount of
\begin{align}
\Delta \left\{ \begin{matrix} 1-f^{\epsilon}_{i q} \\ -f^{\epsilon}_{i q} \end{matrix} \right\} 
&\equiv  d_{t}\left\{ \begin{matrix} 1-f^{\epsilon}_{i q} \\
-f^{\epsilon}_{i q} \end{matrix} \right\} \times \frac{-1}{\Gamma_{iq}} =\frac{d_t f_{iq}^{\epsilon}}{\Gamma_{iq}} \ . \label{defDelf}
\end{align}
Hence the term with $1/\Gamma_i$ is directly related to the memory effect of $G_\gtrless^d.$

In applying $\Delta G_\gtrless$ to the evolution equation of the lepton asymmetry, 
it always appears as a product with the propagators of the SM particles (leptons and Higgs)
as in eq. (\ref{dndt1}). Since these propagators damp quickly with the decay widths $\Gamma_{\ell,\phi}$,
we can drop all the terms in (\ref{honbun-deltaGWdiag}) except the term containing $1/\Gamma_i$.
Furthermore, during the period $1/\Gamma_{\ell \phi}$, 
RH neutrinos are almost stable: $\Gamma_i \ll \Gamma_{\ell \phi}$.
Hence we can replace the frequency $\Omega_i$ by its real part $\omega_i$.

Let us write this simplified form of $\Delta G$  as $\Delta {\mathcal G}$:
\begin{align}
\Delta &{\mathcal G}^{dii}_{\wightman }(x^0,y^0;{\bold q}) \equiv  \sum_{\epsilon}(-i)\Delta \left\{ \begin{matrix} 1-f^{\epsilon}_{i q} \\
-f^{\epsilon}_{i q} \end{matrix} \right\} \times  Z_{\epsilon }^{i} e^{-i \epsilon \omega_{i q }(x^0 -y^0)} \ .\label{defcalG_{W}^{d}}
\end{align}
The definition of $Z^i_\epsilon$ is given in (\ref{defZ}).
$\sum Z^i_\epsilon e^{-i \epsilon \omega_{i q}(x^0-y^0)}$ is nothing but 
$G_\rho^{dii}=G_R^{dii}-G_A^{dii}$ within 
the above  simplification.

As a final remark in this section, we mention that the above simplified form is 
directly obtained from the classical Boltzmann equation as follows.
The Boltzmann equation for the RH neutrino distribution function is given by
\begin{align}
d_t f_{iq} = \frac{2}{2 \omega_{iq}}&\int \frac{d^3 p}{(2\pi)^3}\frac{1}{2\omega_{p}}\int \frac{d^3 k}{(2\pi)^3}\frac{1}{2\omega_{k}} (2\pi)^4 \delta^4 (q-p-k) \notag \\
\times & |{\mathcal M}|_{tree}^2 \left[ (1-f_{iq})f^{(eq)}_{\ell p} f^{(eq)}_{\phi k} -f_{iq}(1-f^{(eq)}_{\ell p})(1-f^{(eq)}_{\phi k}) \right]  \ .
\end{align}
All external momenta are on-shell. Leptons and Higgs are assumed to be in the thermal
equilibrium.
$|{\mathcal M}|_{tree} ^2=g_w (h^{\dag}h)_{ii} (q\cdot p )$ is the 
square of the tree-level decay amplitude of a RH neutrino into a lepton and a Higgs.
The spin in the initial state is averaged and the isospin sum in the final state is performed.
By using the relation 
$(1-f^{(eq)}_{iq})f^{(eq)}_{\ell p} f^{(eq)}_{\phi k} =f^{(eq)}_{iq}(1-f^{(eq)}_{\ell p})(1-f^{(eq)}_{\phi k})$,
it is rewritten as
\begin{align}
d_t f_{iq} &= -\frac{2}{2 \omega_{iq}}\int \frac{d^3 p}{(2\pi)^3}\frac{1}{2\omega_{p}}\int \frac{d^3 k}{(2\pi)^3}\frac{1}{2\omega_{k}} (2\pi)^4 \delta^4 (q-p-k) \notag \\
&\hspace{25pt} \times   \notag |{\mathcal M}|_{tree}^2 \left[ 1-f^{(eq)}_{\ell p}+f^{(eq)}_{\phi k} \right] \left(  f_{iq}-f^{(eq)}_{iq}\right) \notag \\
&=-\Gamma_{iq} \left(  f_{iq}-f^{(eq)}_{iq}\right) \ .  \label{RHBoltzman}
\end{align}
Here, we have used the definition of the decay width (\ref{omegaGamma}) with (\ref{pi_rho}).\footnote{
The factor $[ 1-f^{(eq)}_{\ell p}+f^{(eq)}_{\phi k} ]$ represents the finite density effects, which depend only linearly on the distribution functions  \cite{Garny:2009rv}\cite{Garny:2009qn}\cite{Garny:2010nj}\cite{Beneke:2010wd}\cite{Anisimov:2010dk}\cite{Kiessig:2011fw}.
The RH neutrino interaction rate including all the relevant SM couplings was computed in \cite{Salvio:2011sf}.}
The solution of (\ref{RHBoltzman}) is given by 
\begin{align}
f_{iq}(t) \sim
f^{(eq)}_{iq}(t) -\frac{1}{\Gamma_{i q}} d_t f^{(eq)}_{iq}(t)
\end{align}
and  (\ref{defcalG_{W}^{d}}) is reproduced.

\subsection{Off-diagonal Wightman  out of equilibrium $\Delta G^{\prime}_{\gtrless}$ }
\label{Sec-ODWoutofeq}
We then investigate the deviation of the flavor off-diagonal Wightman function. 
It is most important for generating  the lepton asymmetry.
Since the flavor off-diagonal Wightman function is a sum
of three terms as in (\ref{intKBeqG_{W}^{'}2}), its variation contains 9 terms  (\ref{deltaWightman9}).
Details of the calculations are given in Appendix \ref{Sec-dGoffW}.
6 terms containing  $\Delta G_{R/A}^{d}$ or $\Delta \Pi_{R/A}^{'(eq)}$
reflect the change of the spectrum 
$\Omega _{\epsilon }=\epsilon \omega_{q} \mp i\Gamma_{q}/2$ during the decay of $N_i$.
The change of the distribution functions is contained in the 3 terms with
$\Delta G_{\wightman}^{d}$ and  $\Delta \Pi_{\wightman}^{'(eq)}$. 
In Appendix  \ref{AnotherDerivation}, 
we give a different derivation of $\Delta G_\gtrless^{d}$ and $\Delta G_\gtrless^{'}.$

After lengthy calculations in Appendix \ref{Sec-dGoffW},  we get (\ref{DelG_{W}^{'}1}). For $x^0>y^0$,
$\Delta G_\gtrless^{' ij}$  becomes 
\begin{align} 
 \Delta G^{'ij}_{\wightman} &(x^0,y^0 ;{\bold q}) \n
 \simeq &{\Bigg [}\sum_{\epsilon} Z_{\epsilon} \Pi^{'(eq)ij}_{R}(\epsilon \omega_{ q}) Z_{\epsilon}
 \Delta \left\{ \begin{matrix} 1-f^{\epsilon}_{j q} \\ -f^{\epsilon}_{j q} \end{matrix} \right\}\frac{1}{\Omega_{\epsilon i}-\Omega^{*}_{\epsilon j}} \ e^{-i\Omega_{\epsilon }s_{xy}} \n
&- \sum_{\epsilon} Z_{\epsilon } \Pi^{'(eq)ij}_{A}(\epsilon \omega_{ q}) Z_{\epsilon}\Delta \left\{ \begin{matrix} 1-f^{\epsilon}_{i q} \\ -f^{\epsilon}_{i q} \end{matrix} \right\}\frac{1}{\Omega_{\epsilon i}-\Omega^{*}_{\epsilon j}} \ e^{-i\Omega_{\epsilon }s_{xy}} {\Bigg ]} \ . 
\label{Honbun-DeltaGWightoff}
\end{align}
In this expression, we have assumed that the reference time $t$ is very close to $X_{xy}$, and 
the conditions $|X_{xy} -t|, |s_{xy}| \lesssim 1/\Gamma_{\ell \phi} $ are satisfied.
Such conditions appear when we use the   
Wightman functions in evaluating the evolution equation of the lepton number. 
We also took the leading order terms with respect to $\Gamma/\Gamma_{\ell \phi} \sim \Gamma/T$.  (\ref{Honbun-DeltaGWightoff}) is  of order
 $(H/\Gamma)$.\footnote{
Higher order contributions in the gradient expansion are of order $H/T$ as found in \cite{Garny:2010nz}.
Since  $H/T \ll H/\Gamma$, we do not consider such terms here.}
 
We have also identified $\Omega_i\simeq \Omega_j$ in $e^{-i\Omega_{\epsilon} s_{xy}}$ since the mass difference $\Delta M$ and the widths $\Gamma_i$
are much smaller than the typical scale of $1/|s_{xy}|=\Gamma_{\ell \phi}$.

Here is an important comment.
As discussed in (\ref{Sec-offDWinTE}), the off-diagonal components of the Wightman function
in the thermal equilibrium 
(\ref{G_{W}^{'(eq)}1}) is enhanced by a large factor $1/(\Omega_i -\Omega_j)$
because of the resonant oscillation between flavors.
But  in the limit $x_0 \rightarrow y_0$ it  vanishes as in (\ref{ODWinTEvanish}). 
Both of these properties are related to the behavior of $G_{R/A}^{'}$ through the
KMS relation  and the 
fact that $G_\gtrless^{'}$ is separated into the retarded and advanced propagators
as in (\ref{G_{W}^{'(eq)}0}). 

The deviation $\Delta G_\gtrless^{'ij}$  does not satisfy either properties.
First, the enhancement factor is replaced by $1/(\Omega_i -\Omega_j^*)$.
Second, $\Delta G_\gtrless^{'ij}$ does not vanish in the limit $x^0 \rightarrow y^0$:
\be
\lim_{x_0 \rightarrow y_0} \Delta G_\gtrless^{'ij}(x^0,y^0;{\bf q}) \neq 0 \ .
\ee

The replacement of the enhancement factor by  $1/(\Omega_i -\Omega_j^*)$
 reflects the  mixing between the retarded and advanced propagators. 
Such mixing is naturally generated  because the off-diagonal component of 
the Wightman function is solved as in (\ref{intKBeqG_{W}^{'}2}) to contain
both types of Green functions.
Since the retarded and advanced propagators have poles 
at $q_{0}=\Omega_{\epsilon i}$ and $q_{0}=\Omega^{*}_{\epsilon j}$ respectively, 
the appearance of the term $1/(\Omega_{\epsilon i}-\Omega^{*}_{\epsilon j})$
by $q_0$ integration can be naturally understood.
In the equilibrium case, since the retarded and advanced propagators are
decoupled by the KMS relation,  such mixings of poles at $q_{0}=\Omega_{\epsilon i}$ and
at $q_{0}=\Omega^{*}_{\epsilon j}$ disappear  in the final result of $G_\gtrless^{'ij}$
so that  the enhancement factor becomes
$1/(\Omega_{\epsilon i}-\Omega_{\epsilon j})$
or $1/(\Omega^{*}_{\epsilon i}-\Omega^{*}_{\epsilon j})$.

When we use $\Delta G_\gtrless^{'ij}(x^0,y^0)$ in the evolution equation of the lepton number,
the arguments $x^0, y^0$ are restricted to the 
region $s_{xy}=x^0-y^0 < 1/\Gamma_{\ell \phi} \sim 1/T$ as mentioned above. 
During such short period, the decay of $N_i$ is neglected and we can
safely replace $\Omega_{\epsilon i}$ in $e^{-i \Omega_{\epsilon } s_{xy}}$ by its real part $\omega_{\epsilon }$.
We write the simplified version of  $\Delta G_\gtrless^{'ij}$ as  $\Delta {\mathcal G}_{\wightman}^{'ij}$: 
\begin{align}
\Delta {\mathcal G}^{'ij}_{\wightman}(x^0,y^0;{\bold q})  \simeq & \sum_{\epsilon}
e^{-i\epsilon \omega_{q} (x^0 -y^0)}\  \frac{\omega_{ q}\epsilon }{(M^{2}_{i}-M_j^2)-i\epsilon (M_{i}\Gamma_{i}+M_{j}\Gamma_{j})}  \n 
\times &{\Bigg \{ } 
Z_{\epsilon} \Pi^{'(eq)ij}_{\rho}(\epsilon \omega_{ q}) Z_{\epsilon}\left[ \Delta \left\{ \begin{matrix} 1-f^{\epsilon}_{j q} \\ -f^{\epsilon}_{j q} \end{matrix} \right\}+\Delta \left\{ \begin{matrix} 1-f^{\epsilon}_{i q} \\ -f^{\epsilon}_{i q} \end{matrix} \right\}\right]  \n
&+ 2 Z_{\epsilon} \Pi^{'(eq)ij}_{h}(\epsilon \omega_{ q}) Z_{\epsilon }\left[ \Delta \left\{ \begin{matrix} 1-f^{\epsilon}_{j q} \\ -f^{\epsilon}_{j q} \end{matrix} \right\}-\Delta \left\{ \begin{matrix} 1-f^{\epsilon}_{i q} \\ -f^{\epsilon}_{i q} \end{matrix} \right\}\right]{\Bigg \} } \ . \label{DelG_{W}^{'}2}
\end{align}
The second term in the square bracket with the real part of the self-energy 
can be dropped by imposing $\Pi_h=0$ by the mass renormalisation.
If we include the effect of the temperature dependent mass, $\Pi_h$ is not always zero.
\subsection{Summary of this section}
In this section we studied the deviation of various Green functions from the thermal equilibrium.
The deviation of the retarded Green function $\Delta G_R^d$ is mainly caused by the 
local change of the physical quantities. It is also true for the diagonal component 
of the Wightman function (\ref{defcalG_{W}^{d}}).
It is because the diagonal component in the time-dependent background is determined
by the distribution function at the local temperature.

In contrast, the off-diagonal components behave differently.
The off-diagonal component of the retarded (advanced) Green functions $\Delta G_R^{'ij}$ is largely enhanced by the factor $1/(\Omega_i-\Omega_j)$ due to the flavor oscillation.

The behavior of the off-diagonal components of the Wightman functions $\Delta G_\gtrless^{'}$ in (\ref{Honbun-DeltaGWightoff}) is different.
First it is not written as a change of the local equilibrium Green function $G_\gtrless^{'(eq)}$ in (\ref{G_{W}^{'(eq)}2}):
\be
(\Delta G_\gtrless)^{'ij} \neq \Delta (G_\gtrless^{'ij}) \ .
\label{noncommutative}
\ee
Eq. (\ref{DelG_{W}^{'}1}) and this property are the main results of this section.
The property (\ref{noncommutative}) 
becomes evident when we notice that
$G_\gtrless^{'ij}$ vanishes  in the leading order approximation 
at $x^0=y^0$ as in (\ref{ODWinTEvanish}) while $\Delta G_\gtrless^{'}$ is nonzero
at the equal time, which produces the lepton asymmetry.
Furthermore, unlike (\ref{G_{W}^{'(eq)}2}),  (\ref{Honbun-DeltaGWightoff}) 
 is enhanced by a factor $1/(\Omega_i-\Omega_j^*)$. 
This means that the resonant enhancement of $\Delta G_\gtrless^{'}$
occurs very differently from the resonant oscillation of $G_\gtrless^{'(eq)}$ which is controlled by
 the KMS relation. 
Or in other words, the process of taking a variation $\Delta$ and the flavor oscillation
do not commute as in (\ref{noncommutative}). We come back to this property in section \ref{Sec-Interpretation}.
\section{Boltzmann eq. from Kadanoff-Baym eq.}
\label{Sec5}
\setcounter{equation}{0}
The evolution equation of the lepton asymmetry is given by the KB equation (\ref{dndt1}).
The r.h.s. is written as a functional of the Wightman functions of  RH neutrinos, SM leptons
and SM Higgs. Since the SM leptons and Higgs are almost 
 in the thermal equilibrium,
their distribution functions are approximated by the thermal values at the local temperature.
But the RH neutrinos decay much slower, and furthermore the RH neutrinos with almost
degenerate masses coherently oscillate between different flavors during their decay.
Hence the Wightman functions  $G_\gtrless^{ij}$ of the RH neutrinos must be treated
in a full quantum mechanical way by using the KB equation,
not by the classical Boltzmann equation.
Once $G_\gtrless^{ij}$ are obtained,  they can be inserted into 
the r.h.s. of  (\ref{dndt1}) to obtain the Boltzmann equation for the lepton asymmetry.

Here we summarize the basic ingredients of the evolution equation for the lepton asymmetry.
The evolution equation is given by
\begin{align}
 \frac{d n_{L}}{dt} +3H n_{L} = 2\Re \sum_{i,j}  \int & \frac{d^3 q}{(2\pi)^3}   \int^{t}_{-\infty}d\tau\ 
 (h^{\dag}h)_{ji} \n
\times & {\bigg [}{\rm tr}{\Big \{ }P_{\rm R} G_{<}^{ij}(t,\tau ;
   {\bold q}  
)P_{\rm L}\pi_{>}(\tau,t ;
{\bold q}  
) {\Big \} } \n
&-{\rm tr}{\Big \{ }P_{\rm R} G_{>}^{ij}(t,\tau ;{\bold q})P_{\rm L}\pi_{<}
(\tau,t ; {\bold q}
) {\Big \} }{\bigg ]} 
\label{dndt2}
\end{align}
where $P_{\rm L} \pi_{\wightman} P_{\rm R} =\widetilde{\pi}_{\wightman}$ as defined in (\ref{defpitilde}).
After Fourier transformation of the r.h.s., 
the frequencies $q_0, p_0,k_0$ of the Green functions, $G_\gtrless (q_0)$
and $S_{\wightman} (p_0)$, $\Delta_{\wightman} (k_0)$ in $\widetilde{\pi}_{\wightman}$, satisfy 
the relation $q_0 =p_0+k_0$. 
Furthermore, in the thermal equilibrium,
the Wightman functions are related to the retarded (advanced) propagators
through the KMS relation (\ref{KMSforG0}), (\ref{KMSforSE}) and (\ref{G_{W}^{d(eq)}0}).
Then, by using the relation 
\be
f_{N}(q_0)(1-f_{\ell}(p_0))(1+f_{\phi}(k_0))= (1-f_{N}(q_0))  f_{\ell}(p_0) f_{\phi}(k_0) \ ,
\ee
two terms in the square bracket cancel each other.
Hence there is no generation of lepton asymmetry
in the thermal equilibrium. 

\newpage
\subsection{Lepton asymmetry out of equilibrium}
In the expanding universe, there are three sources for changing the lepton asymmetry,
and the r.h.s. of (\ref{dndt2}) can be classified into the following three terms:
\begin{align}
\frac{d n_{L}}{dt}&+3H n_{L}=\ \sum_{K=d,\prime}\left( {\mathcal C}_{\Delta f}^{K} +{\mathcal C}_{W}^{K} + {\mathcal C}_{\rm BR}^{K} \right) \ .
\label{dndtccc}
\end{align}
Here we rewrite the sum over $i,j$ into 
the flavor diagonal part $K=d$ and the off-diagonal part $K='$. 
Namely $K=d$ corresponds to a summation of $i=j=1$ and $i=j=2$
while $K='$ corresponds to a summation of $i=1, j=2$ and $i=2, j=1$.

The first term ${\mathcal C}^{K}_{\Delta f}$ comes from the deviation of the Wightman functions
of the RH neutrinos (i.e., the distortion of the distribution function $\Delta f$) 
from the thermal value
\be
(\Delta G_\gtrless)^{'} =G_\gtrless^{'} -G_\gtrless^{'(eq)} \neq 0 \ ,
\ee
and is given by
\begin{align}
 {\mathcal C}^{K}_{\Delta f} = 2\Re \int\frac{d^3 q}{(2\pi)^3}  & \sum_{i,j \in K}  (h^{\dag}h)_{ji} \n
 \times \int^{t}_{-\infty}d\tau\ &{\bigg [}{\rm tr}{\Big ( }P_{\rm R} \Delta {\mathcal G}_{<}^{K ij}(t,\tau ;{\bold q})P_{\rm L}\pi^{(eq)}_{>}(\tau,t ;{\bold p}){\Big ) }
\n 
&-{\rm tr}{\Big ( }P_{\rm R} \Delta {\mathcal G}_{>}^{Kij}(t,\tau ;{\bold q})P_{\rm L}\pi^{(eq)}_{<}(\tau,t ;{\bold p}){\Big ) }{\bigg ]} \ .
 \label{defC_{Delf}}
\end{align}
This generates the lepton asymmetry in the expanding universe.

The second term comes from the deviation of $\pi_\gtrless$:
\be
\Delta \pi_\gtrless = \pi_\gtrless - \pi_\gtrless^{(eq)} 
\ee
which is caused by the deviation of the distribution functions of the SM leptons and Higgs. 
${\mathcal C}^{K}_{W}$ is written as
\begin{align}
{\mathcal C}^{K}_{W}= 2\Re \int\frac{d^3 q}{(2\pi)^3} & 
\sum_{i, j \in K} (h^{\dag}h)_{ji} \n
\times \int^{t}_{-\infty}d\tau\ &{\bigg [}{\rm tr}{\Big ( }P_{\rm R} G_{<}^{K(eq)ij}(t,\tau ;{\bold q})P_{\rm L}\Delta \pi_{>}(\tau,t ;{\bold p}){\Big ) }\n
&-{\rm tr}{\Big ( }P_{\rm R} G_{>}^{K(eq)ij}(t,\tau ;{\bold q})P_{\rm L}\Delta \pi_{<}(\tau,t ;{\bold p}){\Big ) }{\bigg ]} \ .
\label{defC_{W}}
\end{align}
This gives washout effect of the lepton asymmetry.

The third term comes from the back reaction of the
generated lepton asymmetry to $G_\gtrless^{'}$, namely to the distribution function of the RH neutrinos. 
It is written as
\begin{align}
{\mathcal C}^{K}_{\rm BR}= 2\Re \int\frac{d^3 q}{(2\pi)^3} &  \sum_{i,j \in K}(h^{\dag}h)_{ji} \n
\times \int^{t}_{-\infty}d\tau\  &{\bigg [}{\rm tr}{\Big \{ }P_{\rm R} 
\Delta_\mu G_{<}^{K ij}(t,\tau ;{\bold q})P_{\rm L}\pi^{(eq)}_{>}(\tau,t ;{\bold p}){\Big \} } \n
&-{\rm tr}{\Big \{ }P_{\rm R} 
\Delta_\mu G_{>}^{K ij}(t,\tau ;{\bold q})P_{\rm L}\pi^{(eq)}_{<}(\tau,t ;{\bold p}){\Big \} }{\bigg ]} \ . \label{defC_{BR}}
\end{align}
Here $\Delta_\mu G$ is defined as 
 the back reaction of the generated chemical potential of the lepton 
and Higgs to the RH Wightman function.

\subsection{Effect of $\Delta G_\gtrless$ on the lepton asymmetry: ${\mathcal C}_{\Delta f}$}
The deviation of the Wightman function from the equilibrium value generates
the lepton asymmetry out of equilibrium.

First let us look at the contribution of the flavor diagonal ($K=d$) part of ${\mathcal C}^{K}_{\Delta f}$.
Inserting (\ref{defcalG_{W}^{d}})\footnote{We note again that  $(X_{xy}-t)$ and $s_{xy}$ of the arguments
of $G_\gtrless(x^0,y^0)$ are smaller than $1/\Gamma_{\ell \phi}$  due to 
 $\pi_{\wightman}(\tau,t) \sim e^{-(t-\tau)\Gamma_{\ell \phi} /2}$.
 Hence the use of  $\Delta {\mathcal G}$ is justified.
}
 and  (\ref{defPiq}) into (\ref{defC_{Delf}}), we have 
\begin{align}
{\mathcal C}^{d}_{\Delta f} & =\sum_{i}\int \frac{d^3 p}{(2\pi)^3 2\omega_{ p}}\frac{d^3 k}{(2\pi)^3 2\omega_{ k}}\frac{d^3 q}{(2\pi)^3 \omega_{ q}}\ (2\pi)^3 \delta ^3(q-p-k)  \notag \\
&\hspace{20pt}\frac{\Gamma_{\ell \phi}}{(\omega_{ q}-\omega_{ p}-\omega_{ k})^2 +\Gamma_{\ell \phi}^{2}/4}\ g_w (h^{\dag}h)_{ii}(q\cdot p)  \notag \\
&\hspace{25pt}\times {\bigg \{ } \Delta  f_{i q} \left( (1-f_{\ell p})(1+f_{\phi k})-(1-f_{\overline{\ell} p})(1+f_{\overline{\phi} k})\right) \notag \\
&\hspace{60pt}- \Delta (1-f_{i q})  \left( f_{\ell p}f_{\phi k}-f_{\overline{\ell} p}f_{\overline{\phi} k} \right){\bigg \} }
\n 
&=0 \ .
\label{dndtDelG^{d}} 
\end{align}
Here we took all the $\epsilon$'s, $\epsilon$ in (\ref{defcalG_{W}^{d}}) and $\epsilon_\ell, \epsilon_\phi$
in (\ref{defPiq}), the same $\epsilon=\epsilon_\ell=\epsilon_\phi$ because the temperature considered
is not so high that a process like $\phi \rightarrow \ell + N$ does not occur.
Hence the flavor diagonal component does not generate the asymmetry.
In the last equality, we used the relation 
$f_{\ell}= f_{\overline{\ell}}=f^{(eq)}_{\ell},\ f_{\phi}= f_{\overline{\phi}}=f^{(eq)}_{\phi}$
for the thermal distribution function.

Next we calculate  the off-diagonal term  ${\mathcal C}_{\Delta f}^{'}$ with $K='$.
Inserting (\ref{DelG_{W}^{'}2}) into (\ref{defC_{Delf}}), we have
\begin{align}
{\mathcal C}_{\Delta f}^{'}&=\sum_{i,j (i \ne j)}\int \frac{d^3 p}{(2\pi)^3 2\omega_{ p}}\frac{d^3 k}{(2\pi)^3 2\omega_{ k}}\frac{d^3 q}{ (2\pi)^3 2\omega_{ q}}\   \frac{(2\pi)^3 \delta ^3(q-p-k) \Gamma_{\ell \phi}}{(\omega_{q}-\omega_{ p}-\omega_{ k})^2 +\Gamma_{\ell \phi}^{2}/4} g_w \Im (h^{\dag}h)^{2}_{ij} \notag \\
&{\Bigg [}
\frac{ (M^{2}_{i}-M^{2}_{j}) /2}{ (M^{2}_{i}-M^{2}_{j})^2+ (M_{i}\Gamma_{i}+M_{j}\Gamma_{j})^2 } 
  \notag \\
&  {\bigg ( }4i(q\cdot \pi^{(eq)}_{\rho}(\omega_{ q}))(q\cdot p) +4i\left( -M^2 (p\cdot \pi^{(eq)}_{\rho}(\omega_{ q}))+(q\cdot \pi^{(eq)}_{\rho}(\omega_{ q}))(q\cdot p) \right) {\bigg ) }
\n
& {\bigg ( } \left[ \Delta f_{i q} +\Delta f_{j q} \right] (1-f^{(eq)}_{\ell p})(1+f^{(eq)}_{\phi k})- \left[\Delta(1-f_{i q})+\Delta (1-f_{j q}) \right]f^{(eq)}_{\ell p}f^{(eq)}_{\phi k} {\bigg ) } \notag  \\
&+  \frac{ M_{i}\Gamma_{i}+M_{j}\Gamma_{j}}{(M^{2}_{i}-M^{2}_{j})^2+ (M_{i}\Gamma_{i}+M_{j}\Gamma_{j})^2 }
 \notag \\
&   {\bigg ( }4(q\cdot \pi^{(eq)}_{h}(\omega_{ q}))(q\cdot p) +4\left( -M^2 (p\cdot \pi^{(eq)}_{h}(\omega_{ q}))+(q\cdot \pi^{(eq)}_{h}(\omega_{ q}))(q\cdot p) \right) {\bigg ) }
\notag \\
& {\bigg ( } \left[ \Delta f_{i q} -\Delta f_{j q} \right] (1-f^{(eq)}_{\ell p})(1+f^{(eq)}_{\phi k})- \left[\Delta(1-f_{i q})-\Delta (1-f_{j q}) \right]f^{(eq)}_{\ell p}f^{(eq)}_{\phi k} {\bigg ) }  {\Bigg ]} \ .
\label{C^{1}Delf0}  
\end{align}
Here, using the definition of $\pi_{\wightman}$ in (\ref{defPiq}),
we have defined $\pi_{\rho}=i(\pi_>-\pi_<)=(\pi_R-\pi_A)$, $\pi_h=(\pi_R+\pi_A)/2$
and their Fourier transform in the time direction,
to separate the self-energies $\Pi^{'(eq)}_{\rho /h}$ in (\ref{DelG_{W}^{'}2}) into the Yukawa coupling $(h^{\dag} h)'$ and the equilibrium values of $\pi_{\rho/h}$ (see (\ref{Pi_{rho}}) and (\ref{Pi_{h}})). 
If we use the vacuum values\footnote{In the flavored leptogenesis, medium effects play an important role \cite{Drewes:2012ma}\cite{Garbrecht:2010sz}.}
for the self-energy calculated in Appendix \ref{AppSE}, i.e.,
$\pi_{\rho}(\epsilon \omega_{ q})=-g_w i \epsilon \Slash{q}_{\epsilon} /(16\pi)$
and $\pi_{h}(\epsilon \omega_{ q})=0$,
the second term in the square bracket is dropped and 
(\ref{C^{1}Delf0}) is simplified as
\begin{align}
{\mathcal C}_{\Delta f}^{'}=&\sum_{i=1,2}\int \frac{d^3 p}{(2\pi)^3 2\omega_{ p}}\frac{d^3 k}{(2\pi)^3 2\omega_{ k}}\frac{d^3 q}{ (2\pi)^3 \omega_{ q}}\  \frac{(2\pi)^3 \delta ^3(q-p-k)  \Gamma_{\ell \phi}}{(\omega_{ q}-\omega_{ p}-\omega_{ k})^2 +\Gamma_{\ell \phi}^{2}/4} \notag \\
& \times \delta |{\mathcal M}|^2 
{\bigg ( } \Delta f_{i q}  (1-f^{(eq)}_{\ell p})(1+f^{(eq)}_{\phi k})- \Delta(1-f_{i q})f^{(eq)}_{\ell p}f^{(eq)}_{\phi k} {\bigg ) } 
  \label{C^{1}Delf1} 
\end{align}
where
\be
\delta |{\mathcal M}|^2  \equiv 
g_w \Im (h^{\dag}h)^{2}_{ij}(q\cdot p)\frac{g_w M^2}{8\pi} \frac{ M^{2}_{i}-M^{2}_{j}}{(M^{2}_{i}-M^{2}_{j})^2+ (M_{i}\Gamma_{i}+M_{j}\Gamma_{j})^2 }  \ .
\label{G-GCP}
\ee
The factor $\delta |{\mathcal M}|^2$ can be interpreted as the $CP$-asymmetric part of the 
 decay amplitudes, which gives the $CP$-asymmetry of the decay rates $\Gamma_{N_i \to \ell \phi}-\Gamma_{N_i \to \overline{\ell} \overline{\phi}}$.

The term (\ref{C^{1}Delf1}) produces the lepton asymmetry through the $CP$-asymmetric
decay of the RH neutrinos that are out of the thermal equilibrium.
The distortion of the distribution function is given in (\ref{defDelf}).
An important point in  (\ref{C^{1}Delf1})  is that the enhancement factor
of the $CP$-asymmetry is given by
$(M^{2}_{i}-M^{2}_{j})/((M^{2}_{i}-M^{2}_{j})^2+ (M_{i}\Gamma_{i}+M_{j}\Gamma_{j})^2 )$,
and the regulator $R_{ij}$ relevant to the $CP$-asymmetric decay of the RH neutrinos
is given, not by $(M_{i}\Gamma_{i}-M_{j}\Gamma_{j})$, but
 by $(M_{i}\Gamma_{i}+M_{j}\Gamma_{j})$.

\subsection{Washout effect  on the lepton asymmetry: ${\mathcal C}_{W}$}
The term ${\mathcal C}^{K}_{W}$ washes out the generated lepton asymmetry.
In order to calculate $\Delta \pi$, we first perform the
Fourier transform of $\pi_\gtrless(\tau,t;{\bf q})$ defined in (\ref{defPiq}):
\begin{align}
\pi_{\wightman}(q)=& -g_w\sum_{\epsilon_{\ell},\epsilon_{\phi}}\int \frac{d^3 p}{(2\pi)^3 2\omega_{p}}\frac{d^3 k}{(2\pi)^3 2\omega_{k}} 
 \frac{(2\pi)^3 \delta ^3(q-p-k) \ \Gamma_{\ell \phi}}{(q_{0}-\epsilon_{\ell}\omega_{ p}-\epsilon_{\phi}\omega_{ k})^2 +\Gamma_{\ell \phi}^{2}/4} \ \Slash{p}_{\epsilon _{\ell}} {\mathcal D}_{\wightman (p,k)}^{\epsilon_{\ell} \epsilon_{\phi}} 
\end{align}
where ${\mathcal D}_{\wightman (p,k)}^{\epsilon_{\ell} \epsilon_{\phi}}$ is defined in (\ref{dndtD}).
Then $\Delta \pi_\gtrless$ is given by
\begin{align}
\Delta \pi_{\wightman}(q) \equiv & \ \pi_{\wightman}(q) - \pi^{(eq)}_{\wightman}(q) \notag \\
=& -g_w\sum_{\epsilon_{\ell},\epsilon_{\phi}}\int \frac{d^3 p}{(2\pi)^3 2\omega_{ p}}\frac{d^3 k}{(2\pi)^3 2\omega_{ k}}\   \frac{(2\pi)^3 \delta ^3(q-p-k) \Gamma_{\ell \phi}}{(q_{0}-\epsilon_{\ell}\omega_{ p}-\epsilon_{\phi}\omega_{ k})^2 +\Gamma_{\ell \phi}^{2}/4} \ \Slash{p}_{\epsilon _{\ell}} \Delta {\mathcal D}_{\wightman (p,k)}^{\epsilon_{\ell} \epsilon_{\phi}}
\end{align}
where $
\Delta {\mathcal D}_{\wightman (p,k)}^{\epsilon_{\ell} \epsilon_{\phi}} \equiv {\mathcal D}_{\wightman (p,k)}^{\epsilon_{\ell} \epsilon_{\phi}} - {\mathcal D}_{\wightman (p,k)}^{\epsilon_{\ell} \epsilon_{\phi}(eq)}.$

First consider the diagonal component $K=d$.
Inserting (\ref{G_{W}^{d(eq)}1}) into (\ref{defC_{W}}), we have
\begin{align}
{\mathcal C}_{W}^{d}=\sum_{i}\int \frac{d^3 p}{(2\pi)^3 2\omega_{ p}} &\frac{d^3 k}{(2\pi)^3 2\omega_{ k}}\frac{d^3 q}{ (2\pi)^3 \omega_{ q}}
\frac{(2\pi)^3 \delta ^3(q-p-k)  \Gamma_{\ell \phi}}{(\omega_{ q}-\omega_{ p}-\omega_{ k})^2 +\Gamma_{\ell \phi}^{2}/4}\  \n
 g_w (h^{\dag}h)_{ii}(q\cdot p)  
&{\bigg \{ }   f^{(eq)}_{i q} \Delta \left\{ (1-f_{\ell p})(1+f_{\phi k})-(1-f_{\overline{\ell} p})(1+f_{\overline{\phi} k})\right\}  \n
& -  (1-f^{(eq)}_{i q}) \Delta \left\{ f_{\ell p}f_{\phi k}-f_{\overline{\ell} p}f_{\overline{\phi} k} \right\}{\bigg \} }
\ .
\label{C^{0}W}
\end{align}
This gives a washout effect on the generated lepton asymmetry
and it is physically interpreted  as the inverse decay of the RH neutrinos.

Next let us see the flavor off-diagonal component, $K='$.
Because of the property (\ref{ODWinTEvanish}), it vanishes in the leading order approximation:
\begin{align}
{\mathcal C}_{W}^{'}= 0 \ .
\end{align} 
Hence only the diagonal component plays a role of washing out the generated lepton asymmetry.
\subsection{Backreaction of the generated lepton asymmetry: ${\mathcal C}_{\rm BR}$}
Finally let us see the back reaction of the generated lepton number asymmetry (i.e.,
the nonzero chemical potential of the SM leptons) to the Wightman functions of the
RH neutrinos.

By using (\ref{Pi_{W}}) and the flavor symmetry $S^{\alpha \beta} = \delta^{\alpha \beta} S$,
the deviation of the self-energy in the presence of the chemical potential is written as
\begin{align}
\Delta_{\mu_{(t)}} \Pi^{ij}_{\wightman} (q) =& \int ds e^{+iq_0 s}
\Delta_{\mu_{(t)}}  \Pi^{ij}_{\wightman}(X=t;s;{\bold q}) \notag \\
=&(h^{\dag}h)_{ij} P_{\rm L}\Delta \pi_{\wightman}(q)+(h^{\dag}h)^{*}_{ij} P_{\rm R}\Delta \overline{\pi}_{\wightman}(q)  \ . 
\label{chempi}
\end{align}
$\Delta \overline{\pi}_{\wightman}$ is the $CP$-conjugate of 
$\Delta \pi_{\wightman}$ and obtained by changing the sign of the chemical potential of the SM leptons
and the Higgs.   $\Delta_\mu G^d_{\wightman}(q)$ is given by replacing
$\Pi_{\wightman}^d$ in (\ref{intKBeqG_{W}^{d}}) by $i=j$ component of (\ref{chempi}),
 and the contribution of the flavor diagonal component is shown to vanishes:
\begin{align}
{\mathcal C}_{\rm BR}^{d}= 0 \ . \label{C^{0}BR}
\end{align}
Similarly 
the off-diagonal contribution becomes 
\begin{align}
{\mathcal C}_{\rm BR}^{'}=&\sum_{i}\int \frac{d^3 p}{(2\pi)^3 2\omega_{ p}}\frac{d^3 k}{(2\pi)^3 2\omega_{ k}}\frac{d^3 q}{(2\pi)^3 \omega_{ q}}\ 
\frac{(2\pi)^3 \delta ^3(q-p-k)   \Gamma_{\ell \phi}}{(\omega_{ q}-\omega_{ p}-\omega_{ k})^2 +\Gamma_{\ell \phi}^{2}/4}\  \n 
& g_w (q\cdot p)  (-1)\frac{g_w M^2}{16\pi}(\Im (h^{\dag}h)_{ij})^2 \frac{(M_{i}\Gamma_{i}+M_{j}\Gamma_{j})}{(M^{2}_{i}-M^{2}_{j})^2 +(M_{i}\Gamma_{i}+M_{j}\Gamma_{j})^{2}} \notag \\
& \times {\bigg [ } f^{(eq)}_{i q} \Delta \left( (1-f_{\ell p})(1+f_{\phi k})-(1-f_{\overline{\ell} p})(1+f_{\overline{\phi} k}) \right) 
\notag \\
&\hspace{20pt} -  (1-f^{(eq)}_{i q})  \Delta \left\{ f_{\ell p}f_{\phi k}-
f_{\overline{\ell} p}f_{\overline{\phi} k} \right\} {\bigg ] } \ .
\label{C^{1}BR}
\end{align}
Details of the calculations are given in Appendix \ref{Sec-App-calwashout}.
In the above calculations, we took the weak coupling limit discussed in Appendix \ref{AppSE}.
This term represents the effect of back reaction of the generated lepton asymmetry on the
Wightman functions of the RH neutrinos. Such a term appears because 
we first solved the propagators of the RH neutrinos in the background of the SM leptons and the Higgs.
The relative sign of the back reaction to the washout effect ${\mathcal C}_W^d$ in (\ref{C^{0}W}) is opposite so that
the back reaction tends to reduce the washout of the generation of lepton asymmetry.
If we solve the KB equations for the lepton asymmetry and the Wightman functions of the RH neutrinos
simultaneously, the generated lepton asymmetry (namely the effect of the chemical potential) 
makes the RH neutrinos further away from the equilibrium. It is the reason why the back reaction 
reduces the washout.

\subsection{$CP$-violating parameter}
The $CP$-violating parameter can be read off from (\ref{C^{1}Delf1}). 
$\delta |{\mathcal M}|^2$ of (\ref{G-GCP}) gives the $CP$-asymmetry of the decay rates 
$\Gamma_{N_i \to \ell \phi}-\Gamma_{N_i \to \overline{\ell} \overline{\phi}}$.
Since the tree decay amplitude is given by 
$|{\mathcal M}|^2_{tree} = g_w (h^{\dag}h)_{ii}(q\cdot p)$,
the $CP$-violating parameter $\varepsilon_i$ is given by 
\begin{align}
\varepsilon_{i}&\equiv \frac{\Gamma_{N_i \to \ell \phi}-\Gamma_{N_i \to \overline{\ell} \overline{\phi}}}{\Gamma_{N_i \to \ell \phi}+\Gamma_{N_i \to \overline{\ell} \overline{\phi}}} \notag \\
&=\frac{\sum_{j(\ne i)}g_w \Im (h^{\dag}h)^{2}_{ij}(q\cdot p)\frac{g_w M^2}{8\pi}\frac{ M^{2}_{i}-M^{2}_{j}}{(M^{2}_{i}-M^{2}_{j})^2+ (M_{i}\Gamma_{i}+M_{j}\Gamma_{j})^2 }}{2\times g_w (h^{\dag}h)_{ii}(q\cdot p)} \notag \\
&=\sum_{j(\ne i)}\frac{\Im (h^{\dag}h)^{2}_{ij}}{(h^{\dag}h)_{ii}} \frac{g_w M^2}{16\pi}\frac{ M^{2}_{i}-M^{2}_{j}}{((M^{2}_{i}-M^{2}_{j}))^2+ (M_{i}\Gamma_{i}+M_{j}\Gamma_{j})^2 } \notag \\
&= \sum_{j(\ne i)}\frac{\Im (h^{\dag}h)^{2}_{ij}}{(h^{\dag}h)_{ii}(h^{\dag}h)_{jj}} \frac{ (M^{2}_{i}-M^{2}_{j})M_{i}\Gamma_{j}}{(M^{2}_{i}-M^{2}_{j})^2+ (M_{i}\Gamma_{i}+M_{j}\Gamma_{j})^2 } \times (1+{\mathcal O}(\Delta M /M)) \ .
\label{CPVparameterfinal}
\end{align}
Hence the  regulator discussed in the introduction is given by
\be
R_{ij}=M_i \Gamma_{i}+M_{j}\Gamma_{j} \ .
\ee
The result is consistent with the result obtained in \cite{Garny:2011hg}.
In the paper \cite{Garny:2011hg}, the $CP$-violating parameter is obtained indirectly from the 
generated lepton asymmetry in a static background with an out-of-equilibrium initial condition.
In our calculation, we directly obtained the same result
in the expanding universe. It shows that the result obtained by Garny {\it et al.} is 
universal and can be applied to the thermal resonant leptogenesis.
\subsection{Summary of this section}
By using $\Delta G_\gtrless^{'ij}$ calculated in the previous section \ref{Sec-POEQ}
in the r.h.s. of (\ref{dndt2}), we obtained the evolution equation (\ref{dndtccc}) with three
terms. ${\mathcal C}_{\Delta f}^{'}$ generates the lepton asymmetry and corresponds
to  the $CP$-asymmetric decay of the RH neutrinos. ${\mathcal C}_{W}$ gives the washout effects on 
the generated lepton numbers. ${\mathcal C}_{\rm BR}$ is the effect of the back reactions of the generated 
lepton asymmetry on the distribution functions of the RH neutrinos. 
From  ${\mathcal C}_{\Delta f}^{'}$, we extracted the $CP$-asymmetric parameter $\varepsilon_i$
given in (\ref{CPVparameterfinal}). The enhancement factor due to the degenerate masses
is regularized with an regulator $R_{ij}=M_i \Gamma_i + M_j \Gamma_j$, which  reflects
 the enhancement factor of $\Delta G_\gtrless^{'}$.

\section{Physical interpretation of the regulators}
\label{Sec-Interpretation}
\setcounter{equation}{0}
In this section, we give a physical interpretation of the appearance of the regulator 
$R_{ij}= |M_{i}\Gamma_{i}+M_{j}\Gamma_{j}|$ instead of  $|M_{i}\Gamma_{i}-M_{j}\Gamma_{j}|$
in the flavor off-diagonal component of the Wightman function $\Delta G_\gtrless^{'}$.

The Wightman Green functions of the RH neutrinos are
 solved as in (\ref{intKBeqG_{W}^{d}}) and (\ref{intKBeqG_{W}^{'}1})
 in terms of the retarded, advanced propagators and the 
self-enegies.
\begin{figure}
\begin{center}
\includegraphics[width=0.9\textwidth ]{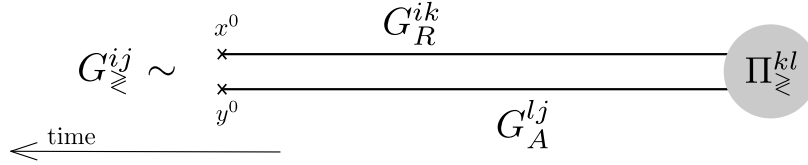}
\caption{
The information of the Wightman functions of the RH neutrinos are encoded in the 
self-energies $\Pi\gtrless$ in the past and transferred from the past to $t=x^0,y^0$
by the retarded and advanced Green functions.
}
\label{FigWightman}
\end{center}
\end{figure}
These equations mean that the information of the distribution function of the RH neutrinos
 in the Wightman functions $G_\gtrless^{ij}$ are encoded in the self-energies $\Pi_\gtrless^{kl}$
in the past, and transferred from the past to the present $t=x^0,y^0.$
The self-energies $\Pi_\gtrless^{kl}$ encode the information of the distributions functions of the
SM leptons and the Higgs in the past  (see Fig.\ref{FigWightman}).
In the flavor diagonal case of (\ref{intKBeqG_{W}^{d}}), all flavors of the RH neutrinos are the same
in the leading order approximation.
On the other hand, in the flavor off-diagonal case of (\ref{intKBeqG_{W}^{'}1}),
the flavor oscillation plays an important role. 

Here we note that, as shown in (\ref{G_{R}^{'(eq)}}) and (\ref{G_{A}^{'(eq)}}),
 $G_{R/A}^{' ij}$ is a coherent sum of two terms, each of which corresponds to a propagation 
 of the $i$-th (or $j$-th) flavor RH neutrino.
We divide it as follows:
\be
G_R^{'ij} = {\big [} G_R^{'ij} {\big]}_{i} +{\big [} G_R^{'ij} {\big]}_{j} \ .
\ee
A precise definition is given in (\ref{offshellG_{R}}).

 \subsection{On-shell and off-shell separation of $G_\gtrless^{'(eq)}$}
Now let's investigate $G_{\wightman}^{'ij}$.
By looking at the  first term of (\ref{intKBeqG_{W}^{'}1}),  it contains $G_A^{djj}$ which describes
the propagation of the $j$-th RH neutrino. The propagator $G_R^{'ij}$ in the first term contains
both of the propagations of $i$-th and $j$-th flavor neutrinos. 
If the $j$-th neutrino propagates in $G_R^{'ij}$,
only a single ($j$-th) neutrino propagates from the past,  when the decay/inverse-decay 
represented by $\Pi_\gtrless$ takes place, to the present  at $t=x^0, y^0.$
We call this type of contributions the ``on-shell" contributions.\footnote{
See the footnote of the section \ref{Diagonal G_{R/A}^d in thermal equilibrium}.
Propagations of a single $N_i$ corresponds to propagations of a single mass eigenstate
 with mass $M_i$ and width $\Gamma_i$.
 It is why we call this contrition as ``on-shell".}
These contributions  are all taken into account in the classical Boltzmann equation.

On the contrary, if the $i$-th neutrino propagates in $G_R^{'ij}$, 
two different flavors propagate from the past to the present. This type of 
contributions are essentially ``off-shell". In the classical Boltzmann equation, 
we first calculate the S-matrix elements of various processes and the external lines
are taken to be on-shell. Hence this type of ``off-shell" 
contributions  are not taken into account by ordinary methods.\footnote{In the evolution equation of the lepton number, ``off-shell'' contributions can be interpreted as the interference terms in the (inverse)decay process of the superposition of different mass eigenstates.}
Separation of various Green functions, especially $\Delta G_\gtrless^{'ij}$,
are calculated in Appendix \ref{SecAppOnOffShell}.

For $G_\gtrless^{'}$ in eq.(\ref{intKBeqG_{W}^{'}1}), 
on-shell contributions come from $j$-th propagation 
${\big [} G^{'ij}_{R} {\big ]}_{j}$ of
$G_R^{'ij}$ in the first term and the $i$-th propagation 
${\big [}  G^{'ij}_{A} {\big ]}_{i} $
of $G_A^{'ij}$ in the second term.
All the other terms, $i$-th propagation ${\big [} G^{'ij}_{R} {\big ]}_{i}$ of
$G_R^{'ij}$ in the first term, the $j$-th propagation 
${\big [}  G^{'ij}_{A} {\big ]}_{j} $
of $G_A^{'ij}$ in the second term give the off-shell contributions.
The third term is off-shell since different mass eigenstates propagate in $G_R^{dii}$ and $G_A^{djj}$. 
$ G_R^{'(eq)}$ is separated into 
\be
 G_R^{'(eq)} = 
{\big [} G_R^{'(eq)} {\big]}_{\mbox{on-shell}} +{\big [} G_R^{'(eq)} {\big]}_{\mbox{off-shell}} \ .
\ee
If we neglect the off-shell terms and take only the on-shell terms, 
${\big [} G_R^{'(eq)} {\big]}_{\mbox{on-shell}} $ becomes ($x^0 > y^0$)
\begin{align}
& {\big [} G^{'(eq)ij}_{\wightman}(x^0,y^0;{\bold q}) {\big ]}_{\mbox{on-shell}} \n
& = \sum_{\epsilon} Z_{\epsilon} \Pi^{'(eq)ij}_{R}(\epsilon \omega_{ q}) Z_{\epsilon}\frac{+i}{\Omega_{\epsilon i}-\Omega_{\epsilon j}} (-i)\left\{ \begin{matrix} 1-f^{\epsilon}_{jq} \\
-f^{\epsilon}_{jq} \end{matrix} \right\}e^{-i\Omega _{\epsilon j}(x^0 -y^0)} \notag \\
&\hspace{10pt}+ \sum_{\epsilon} Z_{\epsilon} \Pi^{'(eq)ij}_{A}(\epsilon \omega_{ q}) Z_{\epsilon}\frac{-i}{\Omega^{*}_{\epsilon i}-\Omega^{*}_{\epsilon j}} (-i)\left\{ \begin{matrix} 1-f^{\epsilon }_{iq} \\
-f^{\epsilon}_{iq} \end{matrix} \right\} e^{-i\Omega _{\epsilon i}(x^0 -y^0)} \ .
\label{onshell(eq)} 
\end{align}
Note that the sum of the on-shell contributions do not vanish even at $x^0 =y^0$ and $f_i \simeq f_j$:
\be
\lim_{x^0 \rightarrow y^0} {\big [} G^{'(eq)ij}_{\wightman}(x^0,y^0;{\bold q}) {\big ]}_{\mbox{on-shell}} \neq 0 \ .
\label{Gnotvanish}
\ee
It is different from the property 
of the full contributions given in (\ref{G_{W}^{'(eq)}1}).
\subsection{On-shell and off-shell separation of  $\Delta G_\gtrless^{'}$}
We next investigate $\Delta G_\gtrless^{'}$.
We  show that neglecting the off-shell contribution in $\Delta G_\gtrless^{'}$,
we get an enhancement factor for the $CP$-violating parameter with a regulator
$|M_{i}\Gamma_{i}-M_{j}\Gamma_{j}|$.

In Appendix \ref{App-onshellDeltaG}, we separate $\Delta G_\gtrless^{'}$
into on-shell and off-shell contributions:\footnote{
This is analogous to the separation  in \cite{Garny:2011hg},
in which the authors emphasized an importance of the first principle calculation to keep the quantum coherence between the different flavor RH neutrinos. 
Calculating the evolution of the generated lepton number under a non-equilibrium initial condition in the flat space-time, they found  two different behaviors  of the generated lepton number.
One is the ordinary term common in the conventional Boltzmann equation.
The other term is specific to the quantum treatment by the quantum KB approach.
The latter oscillates in time and  reduces the eventual  lepton number.
``Off-shell'' contribution here corresponds to the latter effect. 
However, note that in the present case the $CP$-violating parameter, and hence the resulting lepton number does not oscillate.
the oscillatory behavior is averaged out because the deviation from the equilibrium is caused by the expansion of the universe,
and its expansion rate $H$ is much smaller than the oscillation scale $\Delta M \simeq \Gamma$.
This averaging also occurs in the analysis by \cite{De Simone:2007pa} in the strong washout regime.}
\be
\Delta G_\gtrless^{'} = {\big [} \Delta G^{'ij}_{\wightman} {\big ]}_{\mbox{on-shell}}
+ {\big [} \Delta G^{'ij}_{\wightman} {\big ]}_{\mbox{off-shell}} \ .
\ee
The full $\Delta G_\gtrless^{'} $ is given in (\ref{DelG_{W}^{'}1}).
The on-shell contribution is given by  (for $x^0 >y^0$)
\begin{align}
{\big [} \Delta G^{'ij}_{\wightman}&(x^0,y^0;{\bold q}) {\big ]}_{\mbox{on-shell}}
\n
= &\sum_{\epsilon} Z_{\epsilon} \Pi^{'(eq)ij}_{R}(\epsilon \omega_{ q}) Z_{\epsilon}(-i)\Delta \left\{ \begin{matrix} 1-f^{\epsilon}_{j q} \\ -f^{\epsilon}_{j q} \end{matrix} \right\}\frac{i}{\Omega_{\epsilon i}-\Omega_{\epsilon j}} \ e^{-i\Omega_{\epsilon }s_{xy}}\notag \\
&+\sum_{\epsilon} Z_{\epsilon } \Pi^{'(eq)ij}_{A}(\epsilon \omega_{ q}) Z_{\epsilon }(-i)\Delta \left\{ \begin{matrix} 1-f^{\epsilon}_{iq} \\ -f^{\epsilon}_{iq} \end{matrix} \right\}\frac{-i}{\Omega^{*}_{\epsilon i}-\Omega^{*}_{\epsilon j}} \ e^{-i\Omega_{\epsilon }s_{xy}} \ .
 \label{Delonshell22} 
\end{align}
This on-shell contribution has two important properties.
First, it satisfies
\be
{\big [} \Delta G^{'ij}_{\wightman} {\big ]}_{\mbox{on-shell}} =
\Delta {\big [}  G^{'ij}_{\wightman} {\big ]}_{\mbox{on-shell}}
\label{commutative}
\ee
where ${\big [}  G^{'ij}_{\wightman} {\big ]}_{\mbox{on-shell}}$
is given in (\ref{onshell(eq)}).
The on-shell contribution of (\ref{Delonshell}) is simply obtained 
by replacing $f^{(eq)}$ by its variation $\Delta  f$ in (\ref{onshell(eq)}).
This replacing means that the process of the flavor oscillations
and the process of taking a variation from the thermal values are commutative
if we neglect the off-shell contributions.
For full quantum calculations, (\ref{G_{W}^{'(eq)}2}) cannot be obtained by such a replacement
from (\ref{G_{W}^{'(eq)}1}). This is because the flavor oscillations and the deviation from the
thermal values are coherently mixed and these processes are not commutable.
Namely, dropping the off-shell contributions corresponds to neglecting the 
interference between the flavor oscillations and the deviation of the distribution functions
from the thermal equilibrium.

Second, compared with the full result (\ref{DelG_{W}^{'}1}), the enhancement factor
$1/(\Omega_i-\Omega_j^*)$ is replaced by $1/(\Omega_i-\Omega_j)$.
It is related to the above non-commutativity of taking $\Delta$ and flavor oscillation effects.

By inserting the on-shell formula (\ref{onshell(eq)}) and (\ref{Delonshell})
into (\ref{defC_{Delf}}), and supposing 
 $\pi_{\rho}(\epsilon \omega_{ q})=-g_w i \epsilon \Slash{q}_{\epsilon} /(16\pi)$, \ $\pi_{h}(\epsilon \omega_{q})=0$, 
 we have an on-shell approximation  ${\big [} {\mathcal C}_{\Delta f}^{'} {\big ]}_{\mbox{on-shell}} $ of ${\mathcal C}_{\Delta f}^{'}$:
\begin{align}
{\big [} {\mathcal C}_{\Delta f}^{'} {\big ]}_{\mbox{on-shell}} \simeq
 \sum_{i=1,2} &\int \frac{d^3 p}{(2\pi)^3 2\omega_{ p}}\frac{d^3 k}{(2\pi)^3 2\omega_{ k}}\frac{d^3 q}{(2\pi)^3 \omega_{ q}}\   \frac{(2\pi)^3 \delta ^3(q-p-k) 
 \Gamma_{\ell \phi}}{(\omega_{ q}-\omega_{ p}-\omega_{ k})^2 +\Gamma_{\ell \phi}^{2}/4}  \notag \\
&  g_w \Im (h^{\dag}h)^{2}_{ij}(q\cdot p)
 \frac{g_w M^2}{8\pi} \frac{ M^{2}_{i}-M^{2}_{j}}{(M^{2}_{i}-M^{2}_{j})^2+ (M_{i}\Gamma_{i}-M_{j}\Gamma_{j})^2 }   \notag \\
&\times {\bigg \{ } 
\Delta f_{i q}  (1-f^{(eq)}_{\ell p})(1+f^{(eq)}_{\phi k})- \Delta(1-f_{i q}) f^{(eq)}_{\ell p}f^{(eq)}_{\phi k}{\bigg \} } 
\ . 
\end{align}
Hence the regulator $|M_{i}\Gamma_{i}+M_{j}\Gamma_{j}|$ is replace by 
$|M_{i}\Gamma_{i}-M_{j}\Gamma_{j}|$ if we take only the on-shell terms.
It is not valid in general, especially in the resonant leptogenesis.
If the masses are hierarchical, it becomes identical with the correct value in (\ref{C^{1}Delf1}).

\subsection{Summary of this section}
As  emphasized above, if we neglect the off-shell contributions that are not included 
in the ordinary Boltzmann type analysis, we get a  result (\ref{Delonshell22}) 
which is different from the correct one given in  (\ref{Honbun-DeltaGWightoff}).
The only difference is the enhancement factor, and if the mass difference is much larger 
than the width they coincide. But the difference is enlarged when the masses are almost
degenerate. This reflects the fact that the flavor oscillation becomes important only for
degenerate masses.  Another important point is that the property of the
noncommutativity (\ref{noncommutative})  in the full result disappears
if we take only the on-shell contributions as  in (\ref{commutative}).
The noncommutativity is related to the vanishing of $G_\gtrless^{'}$
at the equal time (\ref{ODWinTEvanish}). For the on-shell contributions, 
${\big [} G_\gtrless^{'} {\big ]}_{\mbox{on-shell}}$ does not vanish as shown in (\ref{Gnotvanish}).

Based on this observation, we give another derivation of the properties of $\Delta G_\gtrless^{'}$
in Appendix \ref{AnotherDerivation} by directly solving the KB equations.
If we assume the vanishing condition (\ref{DeldotAB^{'}}) 
 of $ G_\gtrless^{'}$ which is equivalent to (\ref{ODWinTEvanish}), we show that
 the enhancement factor with a regulator $M_i \Gamma_i + M_j \Gamma_j$ appears
 as in (\ref{DelAB^{'}}). On the other hand, if we erroneously assume that it does not vanish,
 it leads to a much larger enhancement factor. 
\section{Summary}
\setcounter{equation}{0}
We investigate the Kadanoff-Baym equations of the thermal resonant leptogenesis
in the expanding universe. 
The lepton asymmetry is generated in the $CP$-asymmetric decay of the RH neutrinos which are 
deviated from the thermal equilibrium.  
If the RH neutrinos have almost degenerate masses, they coherently oscillate during their decay
into the SM leptons and the Higgs.
In such a situation, the classical Boltzmann equation is not valid because 
the decays and the inverse-decays of the RH neutrinos cannot be separated into different
processes, and the full quantum mechanical approach is necessary.
A systematic approach is given by solving the KB equations.

Kadanoff-Baym approach to the resonant leptogenesis was performed in the previous
analysis \cite{Garny:2011hg}. In the paper, the authors studied the coherent oscillation
of the RH neutrinos in a time-independent background with a non-equilibrium initial condition.
In the process of approaching the equilibrium, the RH neutrinos coherently oscillate and decay 
into the SM particles. From the generated amount of the lepton number, they extracted the 
$CP$-violating parameter $\varepsilon_i$ of the $i$-th RH neutrino (\ref{CPVparameter})
and obtained the regulator $R_{ij}=|M_i \Gamma_i + M_j \Gamma_j|$.
Since the resonant enhancement is the key to the resonant leptogenesis, it is 
essential to obtain the correct form of the regulator.

In the present paper, by extending the analysis in the static background \cite{Garny:2011hg} to the thermal
resonant leptogenesis in the expanding universe, we obtain an analytical expression 
of the evolution equation of the lepton number asymmetry. 
The $CP$-violating parameter is obtained as in (\ref{CPVparameterfinal}).
The regulator we obtained is consistent with the result \cite{Garny:2011hg}.

The difference between the regulator $R_{ij}=|M_i \Gamma_i + M_j \Gamma_j|$
and $R_{ij}=|M_i \Gamma_i - M_j \Gamma_j|$ obtained by \cite{Buchmuller:2000nd}
comes from different forms of the enhancement factors of flavor off-diagonal components
of the RH neutrino propagators. 
Since the RH neutrinos have almost degenerate masses, they coherently oscillate very much.
We show that the resonant oscillations between different flavors have two different
types, one  proportional to $1/(\Omega_i-\Omega_j^*)$ and the other proportional to 
$1/(\Omega_i-\Omega_j)$.
Here $\Omega_i=\omega_{iq} - i \Gamma_{iq}/2$ is a position of the 
pole of the $i$-th RH neutrino. 
In the thermal equilibrium, the resonant oscillations in the flavor off-diagonal Green functions
have the type $1/(\Omega_i-\Omega_j)$ ( or its complex conjugate)
as shown in (\ref{G_{R}^{'(eq)}}) or in (\ref{G_{W}^{'(eq)}2}).
Since $1/(\Omega_i-\Omega_j)$ is rewritten by (\ref{Omega-Omega}), the enhancement
of flavor oscillation corresponds to the regulator 
 $R_{ij}=|M_i \Gamma_i - M_j \Gamma_j|$.
However, the deviation of the off-diagonal components of out of equilibrium has a
different enhancement factor $1/(\Omega_i-\Omega_j^*)$ as shown in (\ref{Honbun-DeltaGWightoff}),
which corresponds to the regulator  $R_{ij}=|M_i \Gamma_i + M_j \Gamma_j|$.
Physical interpretation of the change of the regulator is given in Section \ref{Sec-Interpretation}.
The off-shell contributions to the off-diagonal component to the Wightman functions are
essential which can be incorporated only by the KB equations.
The property (\ref{ODWinTEvanish})
 is also important  for this change of the regulator. 
As we show in Appendix \ref{AnotherDerivation}, if we erroneously assume that the 
off-diagonal component of the Wightman function is non-vanishing and its deviation
is given by the change of the local temperature, it leads to much more enhanced 
oscillation similar to the regulator  $R_{ij}=|M_i \Gamma_i - M_j \Gamma_j|$. 

In the present paper, we focus on the formalism of the thermal resonant leptogenesis
and derivations of the $CP$-violating parameter in the expanding universe.
Phenomenological investigations of the amount of the generated lepton asymmetry and the lower bound
of the leptogenesis scale consistent with the neutrino oscillation data are left
for future investigations. 

\section*{Acknowledgments}
The research of SI is supported by Grant-in-Aid for Scientific Research (C) No. 23540329 and (A) No. 23244057.
This work was also partially supported by
``The Center for the Promotion of Integrated Sciences (CPIS)'' of 
Sokendai.
The research of MY is supported by the Grant-in-Aid for Scientific research from the Ministry of Education, Science, Sports, and Culture, Japan, No. 23740208, and No. 25003345. 

\appendix

\section{CTP formalism \label{AppCTP} }
\setcounter{equation}{0}
In this appendix, we briefly review  the closed time path (CTP) formalism.

In equilibrium field theories, we implicitly assume that the initial and final states
 asymptotically approach the ground state of the free Hamiltonian.
But this does not hold in general, especially in time-dependent backgrounds such
as the evolving universe. The final state is generally different from 
the initial state.
The CTP formalism, or the Schwinger-Keldish formalism, is the general formalism
to calculate physical quantities for time-dependent wave functions. 

Suppose that a system is described by a Hamiltonian $\hat{H_0}+\hat{H_1}$, where $\hat{H_0}$ 
 and $\hat{H_1}$ are free and  interaction Hamiltonians, and that the system is
in the initial state $|\psi_i \rangle$ at  time $t=t_i$.
In the interaction picture, 
the expectation value of an observable $\hat{\mathcal O}$ at time $t$ is given by
 \be
{\mathcal O} (t) = \langle \psi_i ^I (t) |  \hat{{\mathcal O}}^I(t) |\psi_i ^I (t) \rangle=
 \langle \psi_i | U^{ I}(t_i,t) \hat{{\mathcal O}}^I(t) U^{ I}(t,t_i)|\psi_i \rangle \ .
\label{OpEvo}
\ee
Here the operator in the interaction picture $\hat{{\mathcal O}}^I(t)$ is related to 
the operator in the Heisenberg picture as
\be
\hat{{\mathcal O}}^H (t)&=&U^I(t,t_i) \hat{{\mathcal O}}^I (t) U^{I \dagger }(t,t_i) \ , \n
U^I(t,t') &=& T \exp \left( -i\int_{t'}^t dt'' \hat{H_1}^I(t'') \right) \ .
\label{IntRepTEO}
\ee

In equilibrium cases,  the final state at time $t=t_f$ is assumed to be
proportional to the initial state
$
U^{I}(t_f,t_i) |\psi_i \rangle =  e^{i \theta} |\psi_i \rangle
$ where $\theta(t_f,t_i)$ is a c-number phase. Then
we can factorize  ${\mathcal O} (t) $ of (\ref{OpEvo}) as
\be
{\mathcal O} (t) &=& \langle \psi_i | U^{I}(t_i,t_f) |\psi_i \rangle
\langle \psi_i | U^{I}(t_f,t)  \hat{{\mathcal O}}^I(t) U^{ I}(t,t_i)|\psi_i \rangle \n
&=& e^{-i \theta} \langle \psi_i | U^{I}(t_f,t)  \hat{{\mathcal O}}^I(t) U^{ I}(t,t_i)|\psi_i \rangle \ .
\ee
Similarly, an expectation value of the time-ordering product of two operators
$ \hat{{\mathcal O}}_1^I (t_1) $ and $ \hat{{\mathcal O}}_2^I(t_2) $ is given by
\be
{\mathcal O}(t_1,t_2) = e^{-i \theta} \langle \psi_i | T \left( \hat{{\mathcal O}}_1^I(t_1)  \hat{{\mathcal O}}_2^I(t_2) U^{I}(t_f,t_i) 
\right) |\psi_i  \rangle \ .
\label{VacVacTr2}
\ee
This formula gives an ordinary perturbative expansion of
correlation functions in equilibrium field theories.
Namely, if we take $t_i \rightarrow -\infty$ and $t_f \rightarrow \infty$,
the interaction vertices $\hat{H}^I(t) $ are inserted in $-\infty<t<\infty$.

\begin{figure}
\begin{center}
\includegraphics[width=0.7\textwidth ]{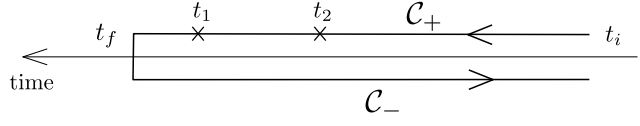}
\caption{
A closed time path ${\mathcal C}$ from $t_i$ to $t_f$ and then back to $t_i$.
(i.e., ${\mathcal C}={\mathcal C}_+ + {\mathcal C}_-$.)
Operators are inserted at $t=t_1$ and $t_2$ for the time ordered product ${\mathcal O}(t_1,t_2) $.
Interaction vertices are inserted everywhere on the CTP.
}
\label{FigCTP}
\end{center}
\end{figure}

In non-equilibrium cases where the final state is no longer proportional to the
initial state, the factorization property  does not hold and
we have
\be
{\mathcal O}(t_1,t_2) = \langle \psi_i  | U^{I}(t_i,t_f) 
T \left( \hat{{\mathcal O}}_1^I(t_1)  \hat{{\mathcal O}}_2^I(t_2) U^{I}(t_f,t_i) \right) | \psi_i
\rangle \ .
\label{NoneqOpEvo}
\ee
In perturbative expansions, 
the interaction vertices are inserted not only on the path ${\mathcal C}_+$
from $t_i$ to $t_f$, but also on the backward path ${\mathcal C}_-$ from $t_f$ to $t_i$.
Figure \ref{FigCTP} shows the closed time path (CTP), 
${\mathcal C}={\mathcal C}_+ + {\mathcal C}_-.$
In this formalism, the final state is not specified at all and we can calculate time-dependence
of various quantities as in (\ref{NoneqOpEvo}). 
The time-ordering $T_C$ is  defined on the CTP as a path-ordering 
along ${\mathcal C}={\mathcal C}_+ + {\mathcal C}_-.$
\section{Evolution equations of various propagators}  \label{AppKB}
\setcounter{equation}{0}
We define various propagators and give a brief derivation of their evolution equations.
In the following we consider a real (Majorana) fermion field
$\hat{\psi}$ and write its conjugate by 
$\overline{\hat{\psi}}=\hat{\psi}^{t}C$ where
$C=i\gamma^2 \gamma^0$. 
In the CTP formalism,
a generating function of time-ordered products of operators is given by
\begin{align}
Z[\overline{J}] =& e^{i W[\overline{J}]} = \left\langle {\rm T}_{\mathcal C}\ e^{ i\int _{\mathcal C}d^{4}x\sqrt{-g}\ \overline{J}(x) \hat{\psi} (x) } \right\rangle  \notag \\
=&\int d\Psi_{+} d\Psi_{-} \langle \Psi_{+} |\hat{\rho}_{(t_i)}| \Psi_{-} \rangle  \int {\mathcal D}'\psi \  e^{i S[\psi] +i\int _{\mathcal C}d^{4}x\sqrt{-g}\  \overline{J}(x) \psi (x) }
\end{align}
where $\hat\rho(t_i)=|\psi_i \rangle \langle \psi_i |$.
The path integral $\int {\mathcal D}'\psi$ denotes an integration of the Grassmann variables 
$\psi_\pm$ on ${\mathcal C}_\pm$
with the fixed boundary conditions $\psi_\pm(t_i)=\Psi_\pm$.
The integrations of $\Psi_\pm$ represent an weighting by the initial wave function $|\psi_i \rangle$.
The source  $\overline{J}(x)$ is defined on ${\mathcal C}={\mathcal C}_+ + {\mathcal C}_-$.
The 1PI effective action is obtained from the generating function $W[\overline{J}]$ 
of the connected Green function by the Legendre transformation.
Defining the classical field by the left-derivative of $W[\overline{J}]$ with respect to the Grassmannian source $\overline{J}$:
\be
\Psi (x) =+\frac{\delta W[\overline{J}]}{\delta  \overline{J}(x)} \ ,
\ee
we have
\begin{align}
\Gamma[\Psi] = W[\overline{J}]-\int_{\mathcal C}d^4 x_g \overline{J}(x)\Psi(x) \ . 
\end{align}
For notational simplicity, we use the following abbreviation
\be
d^4 x_g \equiv d^4 x \sqrt{-g(x)}
\ee
unless the explicit dependence of the measure on $x$ is necessary.
The stationary condition  $(\delta / \delta \Psi(x) ) \Gamma [\Psi ]=0$ at $\overline{J}=0$
gives the equation of motion of $\Psi$.

By taking the second derivative of the effective action $\Gamma$ with respect to 
$\Psi$, we obtain 
the Schwinger-Dyson (SD) equation
\begin{align}
i G^{-1}(x,y) 
=iG^{-1}_{0}(x,y) -i \Pi(x,y) \ .
\label{SD00}
\end{align}
$\Pi$ is the self-energy and only 1PI diagrams contribute to it.
The  connected Green function $G$ on ${\mathcal C}$ is defined by
\begin{align}
G(x,y) = & \ \langle {\rm T}_{\mathcal C}\hat{\psi}(x) \overline{\hat{\psi}}(y) \rangle
 \n
=& \ {\rm \Theta}_{\mathcal C}(x^0 -y^0)G_{>}(x,y)+{\rm \Theta}_{\mathcal C}(y^0 -x^0)G_{<}(x,y) \ ,
\label{CTP-G}
\end{align}
where $\Theta_C(x^0-y^0)$ is the step function on ${\mathcal C}$ and
\be
G_{<}(x,y) \equiv  -\langle \overline{\hat{\psi}}(y) \hat{\psi}(x) \rangle \ , \ \ 
 G_{>}(x,y) \equiv \langle \hat{\psi}(x) \overline{\hat{\psi}}(y) \rangle
 \ee
are the Wightman Green functions.
$iG^{-1}_{0}(x,y) 
= iG_{0(x)} ^{-1} \delta^{g}_{\mathcal C}(x-y)$
is an inverse of the free propagator 
and $\Pi (x,y)$  is the self-energy of the fermion field $\psi$.

The statistical propagator $G_F(x,y)$ and the spectral function $G_{\rho}(x,y)$ are
defined  by
\begin{align}
G_{F}(x,y)=& \ \frac{1}{2}\left( G_{>}(x,y)+G_{<}(x,y) \right)  =\frac{1}{2}\langle [ \hat{\psi}(x) , \overline{\hat{\psi}}(y) ] \rangle \ ,  \\
G_{\rho}(x,y)=&  \ i\left( G_{>}(x,y)-G_{<}(x,y) \right)  =i\langle \{ \hat{\psi}(x) , \overline{\hat{\psi}}(y) \} \rangle \ .
\label{Grho=Gwightman}
 \end{align}
$G_F$ contains information of the distribution function of the specified state 
while $G_\rho$ depends only on the spectrum of the system.
In this sense, $G_F$ is dynamical  while $G_\rho$ is kinematical.
Especially, $G_\rho (x,y)$ becomes proportional to the spatial delta-function 
$\delta^{(3)} ({\bf x}-{\bf y})$
in the equal-time limit.
We further define the retarded and advanced Green functions by
\begin{align}
G_{R/A}(x,y)=\pm \Theta (\pm (x^0 -y^0)) G_{\rho} (x,y) .
\label{GRA=Grho}
\end{align}
They are related to $G_\rho$ as
\be
G_R(x,y) -G_A(x,y) &=&  G_\rho(x,y) \ , \n
 G_R(x,y) + G_A(x,y) &=&  { \rm sign} (x^0 -y^0) G_\rho(x,y) \ .
 \label{GRARHO}
\ee
In terms of $G_F$ and $G_\rho$, the Green function (\ref{CTP-G})  can be written as
\begin{align}
G(x,y)  
=&G_{F}(x,y)-\frac{i}{2}{\rm sign}_{\mathcal C}(x^0 -y^0) G_{\rho}(x,y) \ ,
\label{pathorderedG} 
\end{align}
where the sign-function on ${\mathcal C}$ is defined by
\be
{\rm sign}_{\mathcal C}(x^0 -y^0)=\Theta_{\mathcal C}(x^0 -y^0)-\Theta_{\mathcal C}(y^0 -x^0) \ .
\ee

By convoluting (\ref{SD00}) with the full propagator $G$, we have
\begin{align}
iG_{0(x)} ^{-1}G(x,y)-i\int_{\mathcal C}d^4 z_g  \ \Pi(x,z) G(z,y)=i\delta^{g}_{\mathcal C}(x-y) \ . \label{SDeq}
\end{align}
Here $\delta ^{g}_{\mathcal C}(x-y)$ is the delta-function
on $\mathcal C$ with the space-time metric $g$, and satisfies 
$\int_C d^4 z_g  \delta ^{g}_{\mathcal C}(x-y) =1.$
By denoting $x$ on ${\mathcal C}_\pm$ as $x_\pm$ respectively, the delta-function on ${\mathcal C}$
can be expressed by a $2 \times 2$ matrix:
\be
\delta ^{g}_{\mathcal C}(x_a -y_b)=c_{ab} \delta^{g}(x-y) \ , \ \ \ c_{ab}={\rm diag}(1,-1)
\ee
where $a,b $ takes $+$ or $-$.
The minus sign on ${\mathcal C}_-$ comes from the backward integral of the time variable
and corresponds to  the anti-time-ordering of the Green function
$G$ in (\ref{CTP-G}). 
$\delta^{g}(x-y)$ is an ordinary delta-function for $(x-y)$.

The 2-point function $G(x,y)$ of (\ref{CTP-G}) with $x,y \in {\mathcal C}$ can be similarly decomposed 
(depending on whether $x,y$ are on ${\mathcal C}_+$ or ${\mathcal C}_-$) into a $2 \times 2$
matrix form as
\begin{align}
G_{ab}(x,y) =
\left( \begin{matrix}G_{++} (x,y) &G_{+-}(x,y) \\ G_{-+}(x,y) &G_{--}(x,y) \end{matrix} \right)
=  \left( \begin{matrix}G_{\rm T}(x,y)&G_{<}(x,y)\\ G_{>}(x,y)&G_{\widetilde{\rm T}}(x,y) 
\end{matrix} \right)
\end{align}
where  
${\rm T}, \widetilde{{\rm T}}$ denote time and anti-time orderings respectively, and 
\be
G_{ \rm T}(x,y) &=\ \Theta (x^{0} -y^{0}) G_{>}(x,y) +\Theta (y^{0} -x^{0}) G_{<}(x,y) \ , \n
G_{\widetilde{\rm T}}(x,y) &=\ \Theta (x^{0} -y^{0}) G_{<}(x,y) +\Theta (y^{0} -x^{0}) G_{>}(x,y) \ .
\ee
$\Theta(x^0-y^0)$ is the ordinary step-function.
By using (\ref{pathorderedG}) and (\ref{GRARHO}), we have
\be
G_{ab}(x,y) &=& \left( \begin{matrix} G_{F} -\frac{i}{2} {\rm sign}(x^0-y^0)G_\rho&
G_{F} +\frac{i}{2}  G_\rho
\\ G_{F} -\frac{i}{2} G_\rho
&G_{F} + \frac{i}{2} {\rm sign}(x^0-y^0)G_\rho
\end{matrix} \right) \n
&=& \left( \begin{matrix} G_{F} -\frac{i}{2} (G_R+G_A)&
G_{F} +\frac{i}{2}  (G_R-G_A)
\\ G_{F} -\frac{i}{2} (G_R-G_A)
&G_{F} + \frac{i}{2} (G_R+G_A)
\end{matrix} \right) \n
&=& U^{ t}\left( \begin{matrix}0&G_A\\ G_R &G_F\end{matrix} \right)U 
\label{gabmatrix}
\ee
where 
\be
U\equiv\left( \begin{matrix}-i/2&i/2\\ 1&1\end{matrix} \right) \ .
\ee

We also decompose the self-energy
$\Pi (x,y)$ as
\begin{align}
\Pi (x,y)=&\ {\rm \Theta}_{\mathcal C}(x^0 -y^0)\Pi _{>}(x,y)+{\rm \Theta}_{\mathcal C}(y^0 -x^0)\Pi _{<}(x,y)\notag \\
=&\ \Pi _{F}(x,y)-\frac{i}{2}{\rm sign}_{\mathcal C}(x^0 -y^0) \Pi _{\rho}(x,y) \ . \label{pathorderedPi}
\end{align}
Defining $\Pi_{R/A}$ by
\begin{align}
\Pi_{R/A}(x,y)=\pm \Theta (\pm (x^0 -y^0)) \Pi_{\rho} (x,y) \ , 
\label{PiRAApp}
\end{align}
the matrix form of the self-energy is obtained as
\begin{align}
\Pi_{ab} = \left( \begin{matrix}\Pi_{\rm T}&\Pi_{<}\\ \Pi_{>}&\Pi_{\widetilde{\rm T}}\end{matrix} \right)=U^{t}\left( \begin{matrix}0&\Pi_A\\ \Pi_R &\Pi_F\end{matrix} \right)U \ . 
\label{piabmatrix}
\end{align}

Using these matrix forms of $G_{ab}$ and $\Pi_{ab}$,
the equation (\ref{SDeq}) becomes
\begin{align}
iG_{0(x)} ^{-1}G_{ab}(x,y)-i\int_{t_i}^{t_f}d^4 z_g  \ \Pi_{ac}(x,z) c_{cd} G_{db}(z,y)=i\delta^{g}(x-y) c_{ab} \ .
\end{align}
The matrix $c_{cd}$ between $\Pi$ and $G$ comes from the backward integration of the time 
variable in the original integral in (\ref{SDeq}). 
By multiplying $(U^{t})^{-1}$ on the left and  $U^{-1}$ on the right,
using (\ref{gabmatrix}) and (\ref{piabmatrix}) and noting
\be
U c\ U^t =  \left( \begin{matrix}  0 & -i \\ -i & 0 \end{matrix}   \right) \ ,
\ee
we obtain the following set of the evolution equations:
\begin{align}
& iG_{0(x)} ^{-1}G_{F}(x,y) -\int^{\infty}_{t_{int}}d^{4}z_g  \ \Pi _{R}(x,z)G_{F}(z,y) 
-\int^{\infty}_{t_{int}}d^{4}z_g  \ \Pi _{F}(x,z)G_{A}(z,y) =0 \ , \label{KBeqG_{F}App}  
\end{align}
\begin{align}
& iG_{0(x)} ^{-1}G_{R/A}(x,y)-\int^{\infty}_{t_{int}}d^{4}z_g 
\ \Pi _{R/A}(x,z)G_{R/A}(z,y) &=-\delta ^{g}(x-y) \ . \label{KBeqG_{RA}App}
\end{align}
From the equations (\ref{KBeqG_{RA}App}), we obtain the evolution equation for the spectral density $G_{\rho}=G_{R}-G_{A}$:
\begin{align}
iG_{0(x)} ^{-1}G_{\rho}(x,y) &-\int^{\infty}_{t_{int}}d^{4}z_g \ \Pi _{R}(x,z)G_{\rho}(z,y) -\int^{\infty}_{t_{int}}d^{4}z_g \ \Pi _{\rho}(x,z)G_{A}(z,y) =0 \ . \label{KBeqG_{rho}App} 
\end{align}
The Wightman Green function $G_{\wightman}=G_F \mp (i/2)G_{\rho}$ satisfies
\begin{align}
iG_{0(x)} ^{-1}G_{\wightman}(x,y) &-\int^{\infty}_{t_{int}}d^{4}z_g \ \Pi _{R}(x,z)G_{\wightman}(z,y) -\int^{\infty}_{t_{int}}d^{4}z_g \ \Pi _{\wightman}(x,z)G_{A}(z,y) =0 \label{KBeqG_{W}App} \ .
\end{align}

\section{2PI formalism}
\label{App2PI}
\setcounter{equation}{0}
In this appendix, we give a brief review of a systematic approach to evaluate the 
self-energy based on the 2PI formalism (see, e.g., \cite{Berges:2004yj} for more details).
The generating functional $Z[\overline{J},R]$  in the presence of sources $J(x)$ and 
$R(x,y)$\footnote{
Note that the Majorana condition $R(x,y)=CR^{t}(y,x)C^{-1}$
is not imposed on the source field $R(x,y)$.
}
is given by
\begin{align}
Z[\overline{J},R] =& e^{i W[\overline{J},R]} \notag \\
=&\left\langle {\rm T}_{\mathcal C}\ e^{ i\int _{\mathcal C}d^{4}x_g\  \overline{J}(x) \hat{\psi} (x) +\frac{i}{2}\int _{\mathcal C}d^{4}x_g d^{4}y_g\ \overline{\hat{\psi}} (x) R(x,y) \hat{\psi} (y)  } \right\rangle  \ .
\end{align}
By taking a variation with respect to the source fields $J(x)$ and $R(y,x)$, we have
\begin{align}
\frac{\delta  W[\overline{J},R]}{\delta \overline{J}_{\zeta}(x)} &= + \Psi_\zeta (x) \ , \n
\frac{\delta  W[\overline{J},R]}{\delta R_{\eta \zeta}(y,x)} &=
 - \frac{1}{2}\left(\Psi_{\zeta}(x) \overline{\Psi}_{\eta}(y)  + G_{\zeta\eta}(x,y) \right) \ .
\end{align}
Here $\zeta ,\eta$ represent Spinor indices.

$\overline{\Psi}$ is defined as $\overline{\Psi}(x)\equiv \Psi^{t}(x)C$
and the connected Green function $G$ is given by
\be
G_{\zeta \eta}(x,y) = i\frac{\delta ^2 W[\overline{J},R]}{ \delta \overline{J}_{\zeta}(x) \delta J_{\eta}(y) } \ .
\ee
By taking the Legendre transform of 
$W[\overline{J},R]$ with respect to the sources
$J, R$, we obtain the effective action in the presence of source fields
\begin{align}
\Gamma[\Psi,G] &\equiv 
W[\overline{J},R] - \int_{\mathcal C}d^4 x_g \ \overline{J}_{\zeta}(x)  \frac{\delta  W[\overline{J},R]}{\delta \overline{J}_{\zeta}(x)} -\int_{\mathcal C}d^4 x_g d^4 y_g R_{\zeta \eta}(x,y) \frac{\delta  W[\overline{J},R]}{\delta R_{\zeta \eta}(x,y)} \n
&= W[\overline{J},R]-\int_{\mathcal C}d^4 x_g  \overline{J}\Psi -\frac{1}{2}\int_{\mathcal C}d^4 x_g d^4 y_g \overline{\Psi}(x) R(x,y) \Psi(y) + \frac{1}{2}{\rm Tr}GR \  .
\end{align}
${\rm Tr}$ in the last term represents a trace in the Spinor indices and an integration
over the closed time path ${\mathcal C}$.

Now  we decompose the effective action into 
\begin{align}
\Gamma[\Psi ,G] =S(\Psi ) &-\frac{i}{2}{\rm Tr}\ln G^{-1}-\frac{i}{2}{\rm Tr} G_{0}^{-1}G +\Gamma_{2}[\Psi,G] \ . 
\end{align}
The first term is the classical action. The second and the third term are `1-loop' type contributions to the effective action. 
The meaning of the decomposition can be understood by
 taking a functional derivative\footnote{In taking the functional derivative with respect to $G$, 
the Majorana condition $G(x,y)=CG^{t}(y,x)C^{-1}$ should be used after setting the source field $R$  zero.
}
 with respect to $G$:
\begin{align}
\left. \frac{\delta \Gamma[\Psi ,G]}{\delta G_{\eta \zeta}(y,x)} \right|_{R=0}
=\frac{i}{2}G_{\zeta \eta}^{-1}(x,y)-\frac{i}{2}
G_{0, \zeta \eta}^{-1}(x,y)+ \frac{\delta \Gamma_{2}(\Psi ,G)}{\delta G_{\eta \zeta}(y,x)} 
= 0 \  . 
\label{SD2PI}
\end{align}
Compared with the SD equation (\ref{SD00}), the last term can be identified as
the self-energy $\Pi$:
\begin{align}
\Pi_{\zeta\eta}(x,y;\Psi,G)&=-2i\frac{\delta \Gamma_{2}(\Psi,G)}{\delta G_{\eta \zeta }(y,x)} \ .\label{PiGamma_{2}}
\end{align}
In this way,
the proper self-energy $\Pi$ is obtained by differentiating $\Gamma_2$
with respect to the full propagator $G$. 
Since the proper self-energy $\Pi$ is calculated as a sum of  contributions from 1PI diagrams,
$\Gamma_2$ becomes a sum of contributions from 2PI diagrams with respect to the full propagator. 
In another word, the proper self-energy can be systematically obtained by taking
a functional derivative of 2PI diagrams (in which all internal lines are full propagators) 
with respect to the full propagator.

The SD equation (\ref{SD2PI}) can be interpreted as a self-consistent
equation for the full propagators.
By rewriting this equation on the forward time line ${\mathcal C}_+$, we obtain the set of 
KB equations (\ref{KBeqG_{F}App}), (\ref{KBeqG_{RA}App}), (\ref{KBeqG_{rho}App}), (\ref{KBeqG_{W}App}) which can be interpreted as equations for the full propagators.

For Dirac or Weyl fermions, we introduce additional source terms $$+i\int _{\mathcal C}d^{4}x_g\  \overline{\hat{\psi}}(x) J (x)+i\int _{\mathcal C}d^{4}x d^{4}y\ \overline{\hat{\psi}} (x) R(x,y) \hat{\psi} (y) \ . $$
The self-energy is similarly obtained as a functional of the full propagators:
\begin{align}
\Pi_{\zeta\eta}(x,y;\Psi,\overline{\Psi} ,G)&=-i\frac{\delta \Gamma_{2}(\Psi,\overline{\Psi},G)}{\delta G_{\eta \zeta }(y,x)} \ . \label{SigmaGamma_{2}} 
\end{align}

\section{Self-energies $\Sigma, \Pi$}
\setcounter{equation}{0}
\label{AppSE}
In this appendix, using the 2PI formalism, 
we give an expression of the self-energy function for the RH neutrino 
$\Pi(x,y)$ in terms of the lepton and Higgs propagators. 
The simplest and the most important contribution to  the 2PI effective action
$\Gamma _2$ in the model (\ref{Lint}) 
is given by the 2-loop diagram of Figure \ref{Fig2PI-1}. 
\begin{figure}
\begin{center}
\includegraphics[width=0.25\textwidth ]{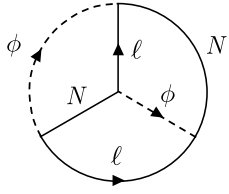}
\caption{
Another example of 2PI diagrams.
}
\label{Fig2PI-2}
\end{center}
\end{figure}
The second simplest 2PI diagram is given by Figure \ref{Fig2PI-2}.
Note that each internal line represents a {\it full}
 propagator of the SM lepton, Higgs and the RH  neutrino. 
If the RH neutrinos have almost degenerate mass, we need to use the resummed 
propagators for the RH neutrinos. Once resummed, we can use an ordinary perturbative
expansion with respect to the Yukawa coupling $h_{i \alpha}$. 
Hence it will not be a bad approximation to use the simplest 2PI diagram to evaluate 
the self-energy. 

In terms of the full propagators, the contribution from the diagram
Figure \ref{Fig2PI-1} becomes
\begin{align}
\Gamma_{2}^{({\rm 2loop})}[G,S,\Delta] =&\ \frac{i}{2}h^{\dag}_{i\alpha}h_{\beta j}\int_{\mathcal C}d^4w_g d^4z_g \ \epsilon_{a'a}\epsilon_{bb'} \Delta_{a'b'}(w,z) 
\n
&\times {\rm tr}\left[ P_{\rm R} \left( 
G^{ji}(z,w)+ CG^{t, ij}(w,z)C^{-1}
\right) P_{\rm L}S^{\alpha \beta}_{ab}(w,z) \right]  \ .
\end{align}
Here $G,S,\Delta$ are full propagators of the RH neutrino, the SM lepton doublet
and the Higgs doublet respectively.
$(i,j), (\alpha, \beta), (a,b,a',b')$ represent the flavor indices of the RH neutrino,
the flavor indices of the leptons and the $SU(2)_L$ indices of the SM doublets respectively.

By using the formula (\ref{SigmaGamma_{2}}), 
the self-energy of the SM lepton doublet is given by taking a functional derivative of $\Gamma_2$
with respect to the lepton propagator $S$:
\begin{align}
\Sigma^{\alpha \beta}_{ab}(x,y)=&\ h_{\alpha i}h^{\dag}_{j\beta}P_{\rm R} G^{ij}(x,y)P_{\rm L} \epsilon_{aa'}\epsilon_{b'b}\Delta_{b'a'}(y,x) \notag \\
=&-\delta_{ab} h_{\alpha i}h^{\dag}_{j\beta}P_{\rm R} G^{ij}(x,y)P_{\rm L}\Delta (y,x)\ . 
\end{align}
Here we have used the Majorana property $G^{ij}(x,y)=CG^{t ji}(y,x)C^{-1}$
of the RH neutrinos.
In the second equality, we have used the fact that the lepton and the Higgs propagators
are $SU(2)_L$ symmetric and proportional to $\delta_{ab}$, $S_{ab}=S\delta_{ab}, \Delta_{ab}=\Delta \delta_{ab}$,
in the early universe where the $SU(2)_L$ symmetry  is restored. 
This is indeed the case in the era of the lepton asymmetry generation through the decay
of the RH neutrino.
Similarly the self-energy of the RH neutrino is obtained by taking a functional 
derivative of $\Gamma_2$ with respect to $G$:
\begin{align}
\Pi^{ij}(x,y)=
&{\large \{ }h^{\dag}_{i\alpha}h_{\beta j}P_{\rm L}S^{\alpha \beta}_{ab}(x,y)P_{\rm R} \Delta_{a'b'}(x,y) \n
&\ + h^{\dag}_{j\alpha}h_{\beta i}P_{\rm R}P \overline{S}^{\beta \alpha}_{ba}(\overline{x},\overline{y})PP_{\rm L} \overline{\Delta}_{b'a'}(\overline{x},\overline{y}) {\large \} } \epsilon_{a'a} \epsilon_{bb'} \notag \\
=&-g_{w} h^{\dag}_{i\alpha}h_{\beta j}P_{\rm L}S^{\alpha \beta}(x,y)P_{\rm R} \Delta(x,y) \notag \\
&\ -g_{w} (h^{\dag}_{i\alpha}h_{\beta j})^{*}P_{\rm R}P \overline{S}^{\alpha \beta}(\overline{x},\overline{y})PP_{\rm L} \overline{\Delta}(\overline{x},\overline{y}) \ 
\end{align}
where $P=\gamma^0$. In the first equality, we have used
\be
\overline{S}^{\beta \alpha}_{ba}(\overline{x},\overline{y})=CPS^{t \alpha \beta}_{\ ab}(y,x)(CP)^{-1}\ , \ \ 
\overline{\Delta}_{ba}(\overline{x},\overline{y})=\Delta_{ ab}(y,x) \ .
\ee
In the second equality,  $SU(2)_L$ symmetry of $S$ and $\Delta$ is used. 
Decomposing these self-energies into the Wightman functions as in 
(\ref{pathorderedPi}), we have
\begin{align}
\Sigma^{\alpha \beta}_{ab\ \wightman}(x,y)=&-\delta_{ab} h_{\alpha i}h^{\dag}_{j\beta}P_{\rm R} G_{\wightman}^{ij}(x,y)P_{\rm L}\Delta_{\wightmaninverse} (y,x)  
\equiv \ \delta_{ab} \Sigma^{\alpha \beta}_{\wightman}(x,y) \ ,\label{Sigma_{W}} \\
\Pi^{ij}_{\wightman}(x,y)
=&-g_{w} h^{\dag}_{i\alpha}h_{\beta j}P_{\rm L}S_{\wightman}^{\alpha \beta}(x,y)P_{\rm R} \Delta_{\wightman}(x,y)  \n
&\ -g_{w} (h^{\dag}_{i\alpha}h_{\beta j})^{*}P_{\rm R}P \overline{S}_{\wightman}^{\alpha \beta}(\overline{x},\overline{y})PP_{\rm L} \overline{\Delta}_{\wightman}(\overline{x},\overline{y}) \ .
\label{Pi_{W}}
\end{align}

In the following, we derive the self-energy of the RH neutrino
$\Pi^{(eq)}$ under an assumption that the lepton and the Higgs are in the thermal equilibrium.
The approximation is justified in the leading order calculation 
since the SM leptons and the Higgs particles interact faster
than the Hubble expansion rate in the era of the leptogenesis. 
See (\ref{GammaSMggH}).
Hence the deviation from the equilibrium can be neglected
in the calculation of $\Pi$.
In the equilibrium,
 the lepton and the Higgs propagators  become $CP$-symmetric and satisfy
\be
\overline{S}(x,y)=S(x,y)\ ,\ \ \ \overline{\Delta}(x,y)=\Delta (x,y) \ .
\ee
By using the quasi-particle approximation for the propagators 
(\ref{S_{W}QP}) and (\ref{Del_{W}QP}), the Fourier transform of the  
self-energy $\Pi_\rho =i(\Pi_{>}-\Pi_{<})=\Pi_R-\Pi_A$ of the RH neutrino becomes
\begin{align}
\Pi^{(eq)ij}_{\rho}(q)=& \left( \Re (h^{\dag}h)_{ij}-i \Im (h^{\dag}h)_{ij} \gamma_5 \right) \pi^{(eq)}_{\rho }(q)  \ , \label{Pi_{rho}} 
\end{align}
where
\begin{align}
\pi^{(eq)}_{\rho}(q) =&(- g_w)\sum_{\epsilon_{\ell},\epsilon_{\phi}}\int 
\frac{d^3 p}{(2\pi)^3 2\omega_{ p}}\frac{d^3 k}{2\omega_{ k}}\  
\delta ^3(q-p-k)\notag \\
& \times \frac{i \Gamma_{\ell \phi}}{(q_{0}-\epsilon_{\ell}\omega_{ p}-\epsilon_{\phi}\omega_{ k})^2 +\Gamma_{\ell \phi}^{2}/4} \ \Slash{p}_{\epsilon _{\ell}} {\mathcal D}_{\rho (p,k)}^{\epsilon_{\ell} .\epsilon_{\phi}(eq)} \ .
\label{pi_rho} 
\end{align}
${\mathcal D}_{\rho (p,k)}$ is given by
\begin{align}
{\mathcal D}_{\rho (p,k)}^{\epsilon_{\ell} \epsilon_{\phi}} \equiv 
{\mathcal D}_{> (p,k)}^{\epsilon_{\ell} \epsilon_{\phi}}-{\mathcal D}_{< (p,k)}^{\epsilon_{\ell} \epsilon_{\phi}} 
= (-1)^{\epsilon_{\ell}} (-1)^{\epsilon_{\phi}}
\left( 1 -f_{\ell p}^{\epsilon_{\ell}}+f_{\phi k}^{\epsilon_{\phi}} \right) 
\end{align}
with the definition of ${\mathcal D}_{\gtrless (p,k)}$ in (\ref{dndtD}),
and 
satisfy the relation ${\mathcal D}_{\rho (p,k)}^{-\epsilon_{\ell} -\epsilon_{\phi}(eq)} =-{\mathcal D}_{\rho (p,k)}^{+\epsilon_{\ell} +\epsilon_{\phi}(eq)}$ and
is followed by the relation of 
\be
\pi^{(eq)}_{\rho}(-q_0 ,{\bold q}) = +\gamma^0 \pi^{(eq)}_{\rho}(+q_0 ,{\bold q}) \gamma^0 \  . \label{sign-dependence of pi_{rho}}
\ee
In the calculation we have used the integral
(in the limit $t_{int} \rightarrow -\infty$) 
\be
2 \int_{-\infty}^t d\tau 
e^{i(- q^0 +\epsilon_{\ell}\omega _{p} +\epsilon_{\phi}\omega _{k} + i \Gamma_{\ell \phi}   /2) (t-\tau)}
= \frac{\Gamma_{\ell \phi} -2i(q^0 -\epsilon_{\ell}\omega _{p} -\epsilon_{\phi}\omega _{k}) }
{(q^0 -\epsilon_{\ell}\omega _{p} -\epsilon_{\phi}\omega _{k})^2+ \Gamma_{\ell \phi}^2 /4} \ . 
\n
\ee
The contribution from the boundary at $\tau= - \infty$  vanishes because of the damping factor $\sim e^{-\Gamma_{\ell \phi} (t-\tau) /2}.$ 

In the weak coupling limit of the SM gauge couplings, 
$\Gamma_{\ell \phi}$ becomes much less than the typical energy transfer $(q_0 -\epsilon_{\ell}\omega_{p}-\epsilon_{\phi}\omega_{k})\sim T$ where $T$ is the temperature at which the leptogenesis occurs.
In such a limit, the above integral  becomes proportional to 
$\delta(q_0 -\epsilon_{\ell}\omega_{p}-\epsilon_{\phi}\omega_{k})$, and the exact energy conservation
is satisfied instead of the Lorentz type in (\ref{pi_rho}).
Furthermore, in order to simplify the form of the self-energy (\ref{pi_rho}), we neglect 
the medium effects (e.g., the Pauli exclusion of the SM lepton and the induced emission of the Higgs) 
encoded in  ${\mathcal D}_{\rho}$ in (\ref{pi_rho}) and drop the distribution function $f$.

Adopting these two simplifications of the weak coupling limit and neglecting the medium effects,  
the self-energy  (\ref{pi_rho}) reduces to the vacuum one:
\be
\pi_{\rho}(q)\to \frac{-i g_w }{16\pi}\Theta (q^2){\rm sign}(q_0) \Slash{q} \ .
\label{SEsimplified}
\ee
Since the main purpose of the present paper is to obtain the effect of quantum oscillations
of almost degenerate RH neutrinos, we  use this simplified form of the self-energy.
The full treatment is investigated by using the integral form (\ref{pi_rho}) of the self-energy
instead of (\ref{SEsimplified}).

Similarly, for $2\Pi_{h}= \Pi_R +\Pi_A$, we have
\begin{align}
\Pi^{(eq)ij}_{h}(q)=& \left( \Re (h^{\dag}h)_{ij}-i \Im (h^{\dag}h)_{ij} \gamma_5 \right) \pi^{(eq)}_{h }(q)  \ , \label{Pi_{h}} 
\end{align}
where
\begin{align}
\pi^{(eq)}_{h}(q) =&  (-g_w) \sum_{\epsilon_{\ell},\epsilon_{\phi}}\int \frac{d^3 p}{(2\pi)^3 2\omega_{ p}}\frac{d^3 k}{ 2\omega_{ k}}\  \delta ^3(q-p-k)\notag \\
&\times \frac{ -(q_{0}-\epsilon_{\ell}\omega_{ p}-\epsilon_{\phi}\omega_{ k})}{(q_{0}-\epsilon_{\ell}\omega_{ p}-\epsilon_{\phi}\omega_{ k})^2 +\Gamma_{\ell \phi}^{2}/4} \ \Slash{p}_{\epsilon _{\ell}} 
{\mathcal D}_{\rho (p,k)}^{\epsilon_{\ell} \epsilon_{\phi}(eq)} \ .
\end{align}
It satisfies the relation
\be
\pi^{(eq)}_{h}(-q_0 ,{\bold q}) = -\gamma^0 \pi^{(eq)}_{h}(+q_0 ,{\bold q}) \gamma^0 \ . \label{sign-dependence of pi_{h}}
\ee
Note that $\pi_{\rho}^{(eq)}(q)$ is pure imaginary while 
$\pi_{h}^{(eq)}(q)$ is  real.
The real part $\pi_{h}(q)$ contains a diverging integral which  is subtracted by
the mass renormalisation. 
In the body of the paper, we have implicitly assumed that the self-energy $\pi_h^{(eq)}(q)$
is already regularized.
The imaginary part $\pi_\rho^{(eq)}$ gives a decay width of the RH neutrino.

\section{ Kramers-Moyal product }
\setcounter{equation}{0}
\label{AppKMproduct}
The convolution  $*$ is defined on bi-local functions $f(x,y)$ and $g(x,y)$:
\be
(f*g)(x,y) \equiv \int dz \ f(x,z) g(z,y) \ .
\ee
The Wigner representation of bi-local functions is also defined as the Fourier transform
of the relative coordinate as 
\be
\widetilde{f}(X;p) \equiv \int dx \ e^{ipx} f(X+\frac{x}{2}, X-\frac{x}{2}) \ .
\ee
Then it is straightforward to show that
\begin{align}
&(\widetilde{f*g})(X;p) \n
& =\int \frac{dp_1}{2 \pi} \frac{dp_2}{2\pi} dX_1 dX_2
\widetilde{f}(X+\frac{X_2}{2};p+p_1) \widetilde{g}(X-\frac{X_1}{2};p+p_2) e^{-i(p_1 X_1+p_2 X_2)} \n
&= \int \frac{dp_1}{2 \pi} \frac{dp_2}{2\pi} dX_1 dX_2
e^{p_1 \partial^f_p+X_2 \partial^f_X/2} \widetilde{f}(X;p) e^{p_2 \partial^g_p-X_1 \partial^g_X/2} 
\widetilde{g}(X;p) 
e^{-i(p_1 X_1+p_2 X_2)} \n
&=
e^{\frac{i}{2}(\partial^f_p \partial^g_X - \partial^f_X \partial^g_p  )} \widetilde{f}(X;p)  \widetilde{g}(X;p)
\label{KMproductApp}
\end{align}
where $\partial^f_X$ is a $X$derivative on the function $f$.
The non-commutative product is called the Kramers-Moyal product.
In the leading order approximation of the derivative expansion, the commutator
of the $*$-product is reduced to the Poisson bracket:
\be
[\widetilde{f},\widetilde{g}]_* (X;p) \sim i \left( \partial_p \widetilde{f} \partial_X \widetilde{g}
 -\partial_X \widetilde{f} \partial_p \widetilde{g}     \right) \ .
\label{PoissonBracket}
\ee
\section{Useful identities}
\label{App-Useful}
\setcounter{equation}{0}
The Green functions are written in terms of  convulsions, and by taking a variation, 
various functions are inserted in the integral.  
Thus we ofter encounter the following types of integrals:
\be
& \int d\tau \ G_1(x-\tau) G_2(\tau-x) \ f(\tau) \ .  \label{Useful01} \\
& \int du dv \ G_1(x-u) \Pi(u-v)   G_2(v-y) \ f_1(X_{uv})  f_2(X_{xu}) f_3(X_{vy}) \ .
\label{Useful02}
\ee
Here  $G_i$ or $\Pi$ are assumed to be functions of the relative coordinate only.
$X_{uv}=(u+v)/2$ etc. are the center of mass coordinates.
In order to evaluate these integrals, we consider the following identities.
For the integral (\ref{Useful01}), 
\begin{align}
& \int \frac{dq_1}{2\pi}\frac{dq_2}{2\pi} \widetilde{G}(q_1)
\widetilde{G}(q_2)\int d\tau e^{-iq_1(x-\tau) -iQ\tau -iq_2 (\tau-y)} \n
&= \int \frac{dq}{2\pi} \widetilde{G}_1(q+Q/2) \widetilde{G}_2 
(q-Q/2) e^{-iq (x-y) -iQ X_{xy} } \ . \label{intexp1} 
\end{align}
By acting $(i\partial _Q-t)$ on both sides and setting $Q=0$, we obtain
\be
&& \int d\tau \ G_1(x-\tau) G_2(\tau-y) \times (\tau -t) \n
&& = \int \frac{dq}{2\pi}
\left( \frac{i}{2} (\partial_q^{(1)}-\partial_q^{(2)}) +X_{xy} -t \right)
\widetilde{G}_1(q) \widetilde{G}_2(q)  e^{-iq (x-y)}   \ .
\label{Useful11}
\ee
$\partial^{(i)}$ are derivatives acting on $G_i.$
Taking higher derivatives with respect $Q$, we can obtain other  relations.
For the next type integral (\ref{Useful02}), we start from the following identity:
\begin{align}
& \int \frac{dq}{2\pi} \frac{dq_1}{2\pi}\frac{dq_2}{2\pi} F(q_1,q,q_2)
\int du dv \ e^{-iq_1 (x-u)  -iq(u-v)  -iq_2 (v-y) }
e^{-iQ_1 X_{xu} -i X_{uv} Q - iQ_2 X_{vy}}
\notag\\
&= \int \frac{dq}{2\pi}\frac{dq_1}{2\pi}\frac{dq_2}{2\pi} F(q_1 ,q, q_2) (2\pi) \delta (q-(q_1 +q_2)/2 +(Q_1 -Q_2)/4) \notag \\
&\times \int dX_{uv} e^{-iq_1 (x-X) -iQ_1 (x+X_{uv})/2 -iX_{uv }Q -iq_2 (X_{uv}-y) -iQ_2 (X_{uv}+y)/2} \notag \\
&= \int \frac{dq}{2\pi} F(q+\frac{Q_1}{2} +\frac{Q}{2} ,q,q-\frac{Q_2}{2} -
\frac{Q}{2}) e^{-i(q+\frac{Q_1}{2} -\frac{Q_2}{2})s_{xy} -i(Q_1 +Q_2 +Q)X_{xy}} \ . \label{Usefulint}
\end{align}
To  relate with the integral (\ref{Useful02}), we set
$F(q_1,q,q_2)=\widetilde{G}(q_1) \widetilde{\Pi} (q)\widetilde{G}(q_2)$.
By acting $(i\partial _Q-t)$ and setting $Q=Q_1=Q_2=0$, we have
\be
&& \int du dv \ G_1(x-u) \Pi(u-v) G_2(v-y) \ (X_{uv}-t) \n
&& =  \int \frac{dq}{2\pi}
\left( \frac{i}{2} (\partial_q^{(1)}-\partial_q^{(2)}) +X_{xy} -t \right)
\widetilde{G}_1(q) \widetilde{\Pi}(q) \widetilde{G}_2(q)  e^{-i q (x-y)}   \ . \n
\label{Useful21}
\ee
If $\Pi(u-v)=\delta(u-v)$, the identity is reduced to (\ref{Useful11}).
By acting $(i\partial _{Q_1}-t)$ and setting $Q=Q_1=Q_2=0$, we have
\be
&& \int du dv \ G_1(x-u) \Pi(u-v) G_2(v-y) \ (X_{xu}-t) \n
&& =   \int \frac{dq}{2\pi}
\left( \frac{i}{2} \partial^{(1)} +x -t \right)
\widetilde{G}_1(q) \widetilde{\Pi}(q) \widetilde{G}_2(q)  e^{-i q (x-y)} 
 \ .
\label{Useful22}
\ee
Here note that $x=X_{xy}+s_{xy}/2$ 
where $s_{xy}=x-y.$
By acting $(i\partial _{Q_2}-t)$ and setting $Q=Q_1=Q_2=0$, we have
\be
&& \int du dv \ G_1(x-u) \Pi(u-v) G_2(v-y) \ (X_{vy}-t) \n
&& =   \int \frac{dq}{2\pi}
\left(- \frac{i}{2} \partial^{(2)} +y -t \right)
\widetilde{G}_1(q) \widetilde{\Pi}(q) \widetilde{G}_2(q)  e^{-i q (x-y)}
\ .
\label{Useful23}
\ee
Note that $y=X_{xy}-s_{xy}/2$.
\section{Calculation of $\Delta G_{R/A}^{d}$}
\setcounter{equation}{0}
 \label{Sec-app-retarded}
The deviation of the retarded (advanced) propagators out of equilibrium can be 
calculated by taking a variation of  (\ref{KBeqG_{RA}}).
In this section, we consider the diagonal component $G_{R/A}^d$.
By dropping the higher order term $\Pi^\prime G^\prime$, 
the diagonal component of the  equation (\ref{KBeqG_{RA}}) is written symbolically as
\be
D^d_{R/A} G^d_{R/A} = - \delta_{xy} \ .
\label{DdGd}
\ee
where 
\begin{align}
D^{d}_{R/A}G^{d}_{R/A} =& \left( i \gamma^0 \partial_{x^0} - \frac{{\boldsymbol \gamma}\cdot {\bold q}}{a(x^0)}-M\right) G^{d}_{R/A}(x^0 ,y^0 ;{\bold q}) \n
&-\int dz^0 \ \Pi^d_{R/A}(x^0,z^0;{\bf q})G^{d}_{R/A}(z^0 ,y^0 ;{\bold q}) \  .
\label{operatorD}
\end{align}

Then expand $D$ and $G$ around the thermal equilibrium values (at the reference time $t$) as
$D=D_t^{(eq)}+\Delta D$ and $G=G_t^{(eq)}+\Delta G$. 
Inserting
them into above we have
\be
D_{R/A}^{d(eq)} \Delta G^d_{R/A} + \Delta D^d_{R/A} \ G^d_{R/A}  = 0 
\ee
where we used $D_{R/A}^{d(eq)} G_{R/A}^{d(eq)} =- \delta_{xy}$ and dropped the 
higher order term $\Delta D  \Delta G$ of deviations from the equilibrium.
It can be solved as
\be
\Delta G^d_{R/A} = G_{R/A}^{d(eq)} \ \Delta D^d_{R/A} \ G_{R/A}^{d(eq)} \ .
\ee
The deviation  $\Delta D_{R/A}$ comes from the change of the 
physical momentum through $a(t)$ and the change of the self-energy
$\Pi_{R/A}$. Thermal corrections to the mass $M$ are included in $\Pi$.

Writing the integral explicitly, $\Delta G^d$ becomes
\begin{align}
 \Delta G^d_{R/A}(x^0,y^0;{\bf q})  =\int du dv \ G_{R/A}^{d(eq)}(t;x^0- u) \ \Delta D^d_{R/A}(u,v)\
G_{R/A}^{d(eq)}(t;v,-y^0;{\bf q})\ . \label{DeltaGRformal}
\end{align}
For notational simplicity, we did not write the reference time $t$ explicitly in $\Delta D_{R/A}(u,v)$.

Note that the equilibrium self-energy $\Pi_{R/A}^{d(eq)}(t;u-v)$
is a function of the relative coordinate $u-v$ with the temperature at time $t$.
Since the self-energy  $\Pi_{R/A}^{d(eq)}(t;u-v)$ is obtained by loop integrals
of the SM particles, it decreases rapidly  as
 $e^{-\Gamma_{\ell \phi} (u-v)/2}$. 
Hence we can adopt derivative expansions of 
 the self-energy $\Pi(u,v)$ around  the thermal value $\Pi_{R/A}^{(eq)}(t;u-v)$. 

From (\ref{operatorD}), we have
\begin{align}
 \Delta & D^d_{R/A}(u,v) =  D^d_{R/A}(u,v)  -D_{R/A}^{d(eq)}(t;u-v)  \n
& \sim \left[ \frac{{\boldsymbol \gamma}\cdot {\bold q}}{a_{(t)}}H_{(t)} \delta(u-v)
-\partial_{t} \Pi^{d(eq)}_{R/A}(t;u-v)\right](X_{uv} -t) \ . \label{DeltaDRexplicit}
\end{align}
The first term is the change of the physical momentum. The second term 
is the change of the background SM plasma, i.e. the change of the distribution functions and the 
spectrum of the SM leptons and the Higgs. Let us write the Fourier transform 
of the coefficient in the square bracket of (\ref{DeltaDRexplicit}) with respect to $(u-v)$ as 
\be
\zeta _{R/A}(t;q) \equiv \frac{{\boldsymbol \gamma}\cdot {\bold q}}{a_{(t)}}H_{(t)}-\partial_t 
\widetilde{\Pi}^{d(eq)}_{R/A} (t;q) \ . \label{zeta}
\ee

The retarded (advanced ) Green functions in the (local) equilibrium are given by
\begin{align}
G^{d(eq)}_{R/A}(X;q) = -\left( \gamma^0 q_0 
- \frac{{\boldsymbol \gamma}\cdot {\bold q}}{a_{(X)}}-\Pi^{d(eq)}_{R/A}(X; q)-M \right)^{-1} 
=\frac{-1}{\Slash{Q}_{R/A} -M}
\label{GRAexplicit}
\end{align}
where
$\Slash{Q}_{R/A}\equiv \Slash{q}-\Pi_{R/A}(q)$.
They satisfy 
\be
& (\Slash{Q}_R +M)(\Slash{Q}_R -M) = (q_0-\Omega_{+})(q_0 -\Omega_{-}) \ , \n
& (\Slash{Q}_A +M)(\Slash{Q}_A -M) =  (q_0-\Omega_{+}^*)(q_0 -\Omega_{-}^*) \ .
\ee
We also define 
$G_{RS}^{d(eq)}={\rm tr}\{ G_{R}^{d(eq)} \}/4 = -M/(Q_R^2-M^2)$.

Physical quantities are  determined by the local temperature at time $X$. 
Then by taking derivatives with respect to 
$X$ and $q_0$, we have
\begin{align}
\partial_t G^{d(eq)}_{R/A} (t;q)&\equiv \left. \partial _{X} G^{d(eq)}_{R/A}(X;q)\right|_{X =t} 
=G^{d(eq)}_{R/A} (t;q) \zeta _{R/A} (t;q) G^{d(eq)}_{R/A} (t;q) \ ,
\label{GzetaG} \\
\partial_{q_0} G^{d(eq)}_{R/A} (t;q)
&\equiv \left. \partial _{q_0} G^{d(eq)}_{R/A}(X;q)\right|_{X =t} 
=G^{d(eq)}_{R/A} (t;q) \xi _{R/A} (t;q) G^{d(eq)}_{R/A} (t;q) \label{GxiG}
\end{align}
where 
\be
\xi _{R/A}(t;q) \equiv \gamma^{0}-\partial_{q_0} \widetilde{\Pi}^{d(eq)}_{R/A} (t;q) \ . \label{xi}
\ee
By using (\ref{GRAexplicit}), 
 $\partial_{q_0} G^{d(eq)}_{R} (t;q)$ is written as
\begin{align}
\partial_{q_0} G^{d(eq)}_{R} (t;q)=
\frac{G_{R S}^{d(eq)}}{M} \left( \{ \xi_R , \Slash{Q}_R \}(\Slash{Q}_R +M)\frac{G_{R S}^{d(eq)}}{M} + \xi_R \right) \ .\label{partial_q G}
\end{align}

Let us now calculate $\Delta G_{R/A}^d(x^0,y^0))$ for $x^0> y^0.$
Inserting (\ref{DeltaDRexplicit}) into (\ref{DeltaGRformal}) and
using the identities (\ref{Useful11}) and (\ref{Useful21}), we get 
\begin{align} 
& \Delta G^d_{R/A}(x^0,y^0;{\bf q})
 \equiv  \int \frac{dq_0}{2\pi} \Delta \widetilde{G}^{d}_{R/A} (X_{xy};q) e^{-iq_0 s_{xy}} \equiv U_1 +U_2 \ , \n
&U_1 = \int \frac{dq_0}{2\pi} \frac{i}{2} e^{-iq_0 s_{xy}} {\Big (}
\partial_{q_0} G^{d(eq)}_{R/A} (t;q)
\zeta_{R/A}  G_{R/A}^{d(eq)} (t;q)  \n
& \hspace{100pt}
-G_{R/A}^{d(eq)} (t;q)\zeta_{R/A} \partial_{q_0} G^{d(eq)}_{R/A} (t;q) {\Big )} \  , \n
&U_2 =\int \frac{dq_0}{2\pi} \partial_t G^{d(eq)}_{R/A} (X_{xy} -t) e^{-iq_0 s_{xy}} \ .
\label{defDeltildeG_{RA}^{d}App}
\end{align}
By using (\ref{partial_q G}), $U_1$ becomes
\begin{align}
U_1&= \int \frac{dq_0}{2\pi} \frac{i}{2} \left( \frac{G_{R S}^{d(eq)}}{M} \right)^2 \left(
\xi_{R} \zeta_{R}  (\Slash{Q}_{R}+M)
-(\Slash{Q}_{R}+M) \zeta_{R} \xi_{R} \right) e^{-iq_0 s_{xy}} \n
&\simeq  \int \frac{dq_0}{2\pi} \frac{i}{2} \frac{2HM}{(q_0-\Omega_{+})^2(q_0 -\Omega_{-})^2} \gamma^0 \frac{{\bold \gamma}\cdot{\bold q}}{a} e^{-iq_0 s_{xy}} \n
&\simeq \Theta (s_{xy})\sum_{\epsilon}(-i)\frac{HM}{4\omega_q^2} \gamma^0 \frac{{\bold \gamma}\cdot{\bold q}}{a} s_{xy} e^{-i\Omega_{\epsilon} s_{xy}}\ . \label{U_1}
\end{align}
In the first equality, the triple pole contributions are canceled out each other.
In the second line, only the lowest order terms in Yukawa couplings are taken.
In the third line, 
we have performed the $q_0$ integration 
and neglected  higher order terms in $\Gamma/M \simeq h^{\dag}h$.\footnote{
In the third line, we picked up only a term containing $\partial_{q_0} e^{-iq_0 s_{xy}}$
because $s_{xy} \sim 1/\Gamma \gg 1/M$ and it 
becomes much larger than the other terms.}
$U_2$ can be calculated much easier because 
the time derivative in $U_2$ is commutable with the $q_{0}$ integral. Then we get
\be
U_2= \partial_{X} \left. G_{R}^{d(eq)}(X;s_{xy})\right|_{X=t} (X_{xy}-t) \ . \label{U_2}
\ee 

Summing up (\ref{U_1}) and (\ref{U_2}),
we get the time representation of the deviation of the retarded(advanced) propagator:
\begin{align}
\Delta & G^{d}_{ R/A} (x^0,y^0;{\bold q})
\equiv \Theta(\pm s_{xy}) \sum_{\epsilon}\Delta \hat{G}^{d}_{R/A}(x^0,y^0;\epsilon,{\bold q})
\n
&\simeq \pm \Theta(\pm s_{xy})\sum_{\epsilon}{\bigg [} 
\partial_t \left( Z_\epsilon e^{-i (\epsilon \omega_{\epsilon q}\mp i\Gamma_{q}/2) s_{xy}}\right)
(X_{xy}-t)
\n
&\hspace{75pt}
-i\frac{H_{(t)} M}{4\omega^2_{q}}\gamma^{0} \frac{{\boldsymbol \gamma}\cdot {\bold q}}{a_{(t)}} \ (x^0-y^0) \  
e^{-i (\epsilon \omega_{\epsilon q}\mp i\Gamma_{q}/2) s_{xy}}
{\bigg ]} \ .
\label{App-deltaGRA}
\end{align}
The first term comes from $U_2$,
It is written as the change of the mass and width in  $\Omega_{\epsilon}=\epsilon \omega_{\epsilon q} - i\Gamma_{q}/2$
and the physical quantities in $Z_\epsilon$. 
The second term comes from $U_1$.
It represents a change of the 
spinor structure in the propagator during the propagation in the expanding universe.

\section{Calculation of $\Delta G^d_\rho$}
\setcounter{equation}{0}
The deviation of the spectral density from the equilibrium value is 
obtained by taking a variation of the relation
$G_{\rho}=-G_{R}*\Pi_{\rho}*G_{A}$:
\begin{align}
\Delta G^{d}_{\rho}=&- \Delta G^{d}_{R}* \Pi^{d (eq)}_{\rho} *G^{d(eq)}_{A} 
- G^{d(eq) }_{R}* \Pi^{d(eq) }_{\rho} *\Delta G^{d }_{A} \n
&- G^{d(eq) }_{R}* \Delta \Pi^{d(eq) }_{\rho} *G^{d(eq) }_{A} \ .
\label{DelG_{rho}^{d}0} 
\end{align}
In this appendix we explicitly evaluate these terms since it is similar to and instructive for more
complicated calculations of $\Delta G_\gtrless^{d}$ in Appendix \ref{Sec-App-out-of-eq-Wightman-diag}.

By using (\ref{defDeltildeG_{RA}^{d}App}) and  (\ref{Useful22}), the first term in the r.h.s. of $\Delta G^{d}_{\rho}$ is given by
\be
&& - (\Delta G^{d}_{R}* \Pi^{d (eq)}_{\rho} *G^{d(eq)}_{A} )(x^0,y^0) \n
&& = - \int du dv \ \Delta G^{d}_{R} (x^0,u) \Pi^{d (eq)}_{\rho} (u,v) G^{d(eq)}_{A} (v,y^0) 
 \equiv T_{11}+T_{12}+T_{13}+T_{14} \ , \n
&&T_{11}=-\int \frac{dq_0}{2\pi}\frac{i}{2}\left( \partial_{q_0}G_{R}^{d(eq)}\zeta_{R}G_{R}^{d(eq)} - G_{R}^{d(eq)}\zeta_{R} \partial_{q_0}G_{R}^{d(eq)} \right)\Pi_{\rho}^{d(eq)}G_{A}^{d(eq)}e^{-iq_0 s_{xy}} \ , \n
&&T_{12}=- \int \frac{dq_0}{2\pi}\frac{i}{2} \partial_{q_0}\partial_{t}G_{R}^{d(eq)}\Pi_{\rho}^{d(eq)}G_{A}^{d(eq)}e^{-iq_0 s_{xy}} \ , \n
&&T_{13}=-  
\int \frac{dq_0}{2\pi} \partial_{t}G_{R}^{d(eq)}\Pi_{\rho}^{d(eq)}G_{A}^{d(eq)} (X_{xy}-t)
 e^{-iq_0 s_{xy}} \ , \n
&&T_{14}=-
\int \frac{dq_0}{2\pi} \partial_{t}G_{R}^{d(eq)}\Pi_{\rho}^{d(eq)}G_{A}^{d(eq)} 
\frac{s_{xy}}{2}  e^{-iq_0 s_{xy}} \ .
\label{E4-1}
\ee
$(T_{12}+T_{13}+T_{14})$ came from $U_2$ while $T_{11}$ came from $U_1$.
By using (\ref{defDeltildeG_{RA}^{d}App}) and (\ref{Useful23}), the second term is
\begin{align}
& - G^{d(eq) }_{R}* \Pi^{d(eq) }_{\rho} *\Delta G^{d }_{A} 
\equiv T_{21}+T_{22}+T_{23}+T_{24} \ ,
 \n
&T_{21}=-\int \frac{dq_0}{2\pi}\frac{i}{2}G_{R}^{d(eq)}\Pi_{\rho}^{d(eq)}\left( \partial_{q_0}G_{A}^{d(eq)}\zeta_{A}G_{A}^{d(eq)} - G_{A}^{d(eq)}\zeta_{A} \partial_{q_0}G_{A}^{d(eq)} \right)e^{-iq_0 s_{xy}} \ , \n
&T_{22}=-\int \frac{dq_0}{2\pi}\frac{-i}{2} G_{R}^{d(eq)}\Pi_{\rho}^{d(eq)}\partial_{q_0}\partial_{t}G_{A}^{d(eq)}e^{-iq_0 s_{xy}} \ , \n
&T_{23}=-\int \frac{dq_0}{2\pi} G_{R}^{d(eq)}\Pi_{\rho}^{d(eq)}\partial_{t}G_{A}^{d(eq)} \left((X_{xy}-t)  \right)e^{-iq_0 s_{xy}} \ , \n
&T_{24}=\int \frac{dq_0}{2\pi} G_{R}^{d(eq)}\Pi_{\rho}^{d(eq)}\partial_{t}G_{A}^{d(eq)}  
\frac{s_{xy}}{2} e^{-iq_0 s_{xu}} \ . \label{E4-2}
\end{align}
Finally, by using (\ref{DelPiKMS}) and (\ref{Useful21}), the third term becomes
\begin{align}
&- G^{d(eq)}_{R {\bold q}} * \Delta \Pi_{\rho {\bold q}}^{d(eq)} * G_{A {\bold q}}^{d(eq)} \equiv T_{31}+T_{32} \ ,\n
&T_{31}=-\int \frac{dq_0}{2\pi}\frac{i}{2}\left( \partial_{q_0}G_{R}^{d(eq)}\partial_{t} \Pi_{\rho}^{d(eq)}G_{A}^{d(eq)} - G_{R}^{d(eq)}\partial_{t} \Pi_{\rho}^{d(eq)} \partial_{q_0}G_{A}^{d(eq)} \right)e^{-iq_0 s_{xy}}  \ , \n
&T_{32}=-\int \frac{dq_0}{2\pi} G_{R}^{d(eq)}\partial_{t} \Pi_{\rho}^{d(eq)}G_{A}^{d(eq)} \left(X_{xy}-t \right)e^{-iq_0 s_{xy}} \ .\label{E4-3}
\end{align}
Note that $\Pi_\rho=\Pi_R-\Pi_A$ and hence $\partial_t \Pi_\rho=-(\zeta_R-\zeta_A).$

Let us look at the above terms separately.
First, the terms proportional to $(X_{xy}-t)$ become $T_a$:
\begin{align}
T_a &\equiv T_{13}+T_{23}+T_{32} =
-\int \frac{dq_0}{2\pi} \partial_{t} \left( G_{R}^{d(eq)} \Pi_{\rho}^{d(eq)}G_{A}^{d(eq)}\right) \left(X_{xy}-t \right)e^{-iq_0 s_{xy}}\notag \\
&=\int \frac{dq_0}{2\pi} \partial_{t} \left( G_{R}^{d(eq)} -G_{A}^{d(eq)} \right) \left(X_{xy}-t \right) e^{-iq_0 s_{xy}} \ .\label{t}
\end{align}
Next, the terms proportional to $s_{xy}$ can be rewritten as $T_b + T_{44}$:
\begin{align}
&T_{14}+T_{24} \equiv T_{b}+T_{44} \ , \n
&T_{44}=-\int \frac{dq_0}{2\pi}\frac{-i}{2} \partial_{q_0}\left(\partial_{t}G_{R}^{d(eq)}\Pi_{\rho}^{d(eq)}G_{A}^{d(eq)}-G_{R}^{d(eq)}\Pi_{\rho}^{d(eq)}\partial_{t}G_{A}^{d(eq)} \right)  e^{-iq_0 s_{xy}} 
 \ , \n
&T_b = -\int \frac{dq}{2\pi} \partial_{q_0}\left[ \frac{+i}{2} \left( \partial_{t}G_{R}^{d(eq)}\Pi_{\rho}^{d(eq)}G_{A}^{d(eq)}-G_{R}^{d(eq)}\Pi_{\rho}^{d(eq)}\partial_{t}G_{A}^{d(eq)} \right) e^{-iq_0 s_{xy}}\right] \ . \label{s2}
\end{align}
$T_b$ is a total derivative and vanishes, but we keep it for later convenience in calculating $\Delta G_\gtrless^d$
in the next section. 

The other terms and $T_{44}$ are combined to become
\begin{align}
 T_c \equiv & T_{44} + T_{11}+T_{12}+T_{21}+T_{22}+T_{31}  \n
=&-\int \frac{dq_0}{2\pi}\frac{i}{2}\left( \partial_{q_0}G_{R}^{d(eq)}\zeta_{R}G_{R}^{d(eq)} - G_{R}^{d(eq)}\zeta_{R} \partial_{q_0}G_{R}^{d(eq)} \right)e^{-iq_0 s_{xy}} \n
&-\int \frac{dq_0}{2\pi}\frac{-i}{2}\left( \partial_{q_0}G_{A}^{d(eq)}\zeta_{A}G_{A}^{d(eq)} - G_{A}^{d(eq)}\zeta_{A} \partial_{q_0}G_{A}^{d(eq)} \right)e^{-iq_0 s_{xy}} \ . \label{T_c} 
\end{align}
where we used the relation $G_\rho=G_R-G_A=-G_R \Pi_\rho G_A$ and $\partial_t \Pi=\zeta_A-\zeta_R.$

As a result,  $\Delta G_\rho^d$ becomes 
\begin{align}
& \Delta G^{d}_{\rho}(x^0,y^0 ;{\bold q})= T_a +T_b+ T_c  =T_a+T_c\n
& =\int \frac{dq_0}{2\pi} \left[ \Delta G^{d(eq)}_{R} (X_{xy};q) - \Delta G^{d(eq)}_{A} (X_{xy};q) \right] e^{-iq_0 s_{xy}} \n
& =\Delta G^{d}_{R}(x^0,y^0 ;{\bold q}) -\Delta G^{d}_{A}(x^0,y^0 ;{\bold q}) \ ,
\label{DelG_{rho}^{d}} 
\end{align}
as expected.
Here we used  (\ref{defDeltildeG_{RA}^{d}App}).
In the second equality, $T_b$ is dropped since it is a total derivative. 
In the calculation of $\Delta G^d_{\gtrless}$, $T_b$ is modified to contain
a function $f(q_0)$ outside  the derivative $\partial_{q_0}$, and contributes to the final result. 

\section{Calculation of $\Delta G^d_{\gtrless}$} 
\label{Sec-App-out-of-eq-Wightman-diag}
\setcounter{equation}{0}
Diagonal components of the Wightman functions are obtained in the leading order approximation as
$G^d_{\gtrless} = - G_R^d * \Pi^d_{\gtrless} * G^d_A$.
As explained in (\ref{SE-out-of-eq}), the self-energy function can be safely
replaced by its equilibrium value at the local temperature:
\be
\Pi_{\gtrless} \longrightarrow \Pi^{(eq)}_{\gtrless}= 
(-i)\left\{ \begin{matrix} 1-f(q_0) \\ -f(q_0) \end{matrix} \right\}  \Pi^{(eq)}_\rho \ .
\label{PiWightmanEq}
\ee
The distribution function $f(q_0)=1/(e^{q_0/T(t)}+1)$ is time-dependent through the  temperature $T(X)$.

The calculation of the deviation $\Delta G_\gtrless^d$ is parallel to the calculation 
of $\Delta G_\rho^d$ in the previous section. The only difference is that 
$\Pi^{d(eq)}_\rho$ is replaced by $\Pi^{d(eq)}_\gtrless$ in (\ref{DelG_{rho}^{d}0}).
Especially in (\ref{E4-3}), $\partial_t \Pi_\rho$ is replaced by
\be
\partial_t \Pi_\gtrless = 
(-i) \partial_t \left\{ \begin{matrix} 1-f \\ -f \end{matrix} \right\} \Pi_\rho +
(-i) \left\{ \begin{matrix} 1-f \\ -f \end{matrix} \right\} \partial_t \Pi_\rho
\ee
and we have 
\begin{align}
&\Delta G^{d}_{\wightman}(x^0,y^0 ;{\bold q})=V_1 +V_2 +V_3 +V_4 \ , \n
&V_1 = \int \frac{dq_0}{2\pi}(-i)\left\{ \begin{matrix} 1-f \\
-f \end{matrix} \right\} \left[ \Delta G^{d(eq)}_{R} (X_{xy};q) - \Delta G^{d(eq)}_{A} (X_{xy};q) \right] e^{-iq_0 s_{xy}} \ ,\n
&V_2 = \int \frac{d q_0}{2\pi}(-i)\partial_{q_0}\left\{ \begin{matrix} 1-f \\
-f \end{matrix} \right\} \frac{+i}{2}\left( \partial _t G_{R}^{d(eq)} \Pi_{\rho}^{d(eq)} G_{A}^{d(eq)} -G_{R}^{d(eq)}\Pi_{\rho}^{d(eq)} \partial _t G_{A}^{d(eq)}  \right)e^{-iq_0 s_{xy}} \ ,\n 
&V_3 =\int \frac{d q_0}{2\pi}(-i)\partial_t \left\{ \begin{matrix} 1-f \\
-f \end{matrix} \right\} \frac{-i}{2}\left( \partial _{q_0} G_{R}^{d(eq)} \Pi_{\rho}^{d(eq)} G_{A}^{d(eq)} -G_{R}^{d(eq)}\Pi_{\rho}^{d(eq)} \partial _{q_0} G_{A}^{d(eq)}  \right)e^{-iq_0 s_{xy}} \ ,\n 
&V_4 = \int \frac{d q_0}{2\pi}(-i)\partial_t \left\{ \begin{matrix} 1-f \\
-f \end{matrix} \right\} \left( G_{R}^{d(eq)} -G_{A}^{d(eq)}  \right) (X_{xy}-t)e^{-iq_0 s_{xy}} \label{DelG_{W}^{d}App} \ .
\end{align}
$V_1$ is obtained by inserting the distribution function in the integrand of (\ref{DelG_{rho}^{d}}) for $\Delta G_\rho.$
$V_2$ comes from $T_{b}$-type term of (\ref{s2}) in the previous appendix.  
In the calculation of $\Delta G_\rho$, 
this term vanishes because it is a total divergence.
In the present case, since the distribution function depends on $q_0$, the derivative can act on it
and the term remains. 
$V_3$ and $V_4$ come from time-dependence of the distribution function $f(q_0)$. 

Let us consider the region $x^0 > y^0$, and perform the $q_0$ integration.
First, $V_4$ and $T_{a}$-type  contribution (\ref{t}) in $V_1$ turn out to represent the change of local temperature as
\be
\partial_{X} \left. G_{\wightman}^{d(eq)}(X;s_{xy})\right|_{X=t} (X_{xy}-t) \ . \label{diagonal-local-temperature-dependence}
\ee
$T_{c}$-type contribution (\ref{T_c}) in $V_1$ is of higher order with respect to $\Gamma_i/M$ and 
can be neglected.\footnote{It
is nothing but the one obtained by inserting the distribution function into $U_2$ in the retarded propagator.
In (\ref{U_1}), we  neglected higher order contribution in $\Gamma /M$.
In the present case, 
although $q_0$ derivative, coming from the double pole integral, 
can act on the distribution function $f(q)$,
the contribution becomes a higher order contribution because of $\partial_{q_0} f(q_0) \sim f(q_0)/T \sim f(q_0)/M$,
and can be neglected again.}
Therfore, we get
\begin{align}
V_1 +V_4 \simeq & \sum_{\epsilon} (-i) d_{t} \left\{ \begin{matrix} 1-f^{\epsilon}_{q} \\
-f^{\epsilon}_{q} \end{matrix} \right\} (X_{xy}-t) Z_{\epsilon} e^{-i\Omega_{\epsilon} s_{xy}} \n
&+\sum_{\epsilon} (-i) \left\{ \begin{matrix} 1-f^{\epsilon}_{q} \\
-f^{\epsilon}_{q} \end{matrix} \right\} \Delta \hat{G}_{R}^{d}(x^0 ,y^0 ;\epsilon ,{\bold q}) \ . \label{V_1+4}
\end{align}

Let's move on to $V_3$.
By using the explicit expression (\ref{partial_q G}) of $\partial_{q_0}G_{R/A}^{d(eq)}$, we get
\begin{align}
V_3 =& \int  \frac{d q_0}{2\pi}  \frac{-1}{2} \partial_t \left\{ \begin{matrix} 1-f \\
-f \end{matrix} \right\} \frac{G_{R S}^{d(eq)}}{M}  \frac{G_{A S}^{d(eq)}}{M} e^{-iq_0 s_{xy}} \n
&\times {\Bigg [}  \left( \{ \xi _R , \Slash{Q}_{R}\} (\Slash{Q}_{R} +M)\frac{G_{R S}^{d(eq)}}{M} +\xi_{R} \right) \Pi_{\rho}^{d(eq)} (\Slash{Q}_{A} +M) \n
&\hspace{20pt}-(\Slash{Q}_{R} +M) \Pi_{\rho}^{d(eq)} \left( \{ \xi _A , \Slash{Q}_{A}\} (\Slash{Q}_{A} +M)\frac{G_{A S}^{d(eq)}}{M} +\xi_{A} \right) {\Bigg ]}\n
\equiv & V_{31} +V_{32} +V_{33} \ , \notag
\end{align}
where
\begin{align}
V_{31} = \int \frac{d q_0}{2\pi} & \frac{-1}{2} \partial_t \left\{ \begin{matrix} 1-f \\
-f \end{matrix} \right\} \left( \frac{G_{R S}^{d(eq)}}{M}\right)^2 \left( \frac{G_{A S}^{d(eq)}}{M} \right)^1 e^{-iq_0 s_{xy}} \n
&\times \{ \xi _R , \Slash{Q}_{R}\} (\Slash{Q}_{R} +M) \Pi_{\rho}^{d(eq)} (\Slash{Q}_{A} +M) \ , \n
V_{32} = \int \frac{d q_0}{2\pi} & \frac{-1}{2} \partial_t \left\{ \begin{matrix} 1-f \\
-f \end{matrix} \right\} \left( \frac{G_{R S}^{d(eq)}}{M}\right)^1 \left( \frac{G_{A S}^{d(eq)}}{M} \right)^2 e^{-iq_0 s_{xy}} \n
& \times (-1) \{ \xi _A , \Slash{Q}_{A}\} (\Slash{Q}_{R} +M) \Pi_{\rho}^{d(eq)} (\Slash{Q}_{A} +M) \ , \n
V_{33} = \int \frac{d q_0}{2\pi} & \frac{-1}{2} \partial_t \left\{ \begin{matrix} 1-f \\
-f \end{matrix} \right\} \left( \frac{G_{R S}^{d(eq)}}{M}\right)^1 \left( \frac{G_{A S}^{d(eq)}}{M} \right)^1 e^{-iq_0 s_{xy}} \n
& \times  \left(\xi_{R} \Pi_{\rho}^{d(eq)} (\Slash{Q}_{A} +M) -(\Slash{Q}_{R} +M) \Pi_{\rho}^{d(eq)} \xi_{A} \right) \ . \label{V_3}
\end{align}
Note that the number of the factor $G_{S}/M$ in $V_{33}$ is less 
than those in the other two terms. 
Hence $V_{33}$ turns out to be a higher order with respect to
 $\Gamma /M$ and can be neglected.
The $q_0$ integration of $V_{31}$ and $V_{32}$ can be similarly performed as in  $U_{1}$,
and we get
\begin{align}
V_3 \simeq& \sum_{\epsilon} \frac{\partial T}{\partial t}\frac{\partial}{\partial T} \left\{ \begin{matrix} 1-f^{\epsilon}_{q} \\
-f^{\epsilon}_{q} \end{matrix} \right\} \frac{\{ \Slash{q}_{\epsilon} , \Pi_{\rho}^{d} \}}{4 \epsilon \omega_{q}} \left( \frac{-2}{\Gamma_{q}^2} -\frac{s_{xy}}{\Gamma_{q}} \right) Z_{\epsilon} e^{-i\Omega_{\epsilon} s_{xy}} \n
=&\sum_{\epsilon} (-i)\frac{\partial T}{\partial t}\frac{\partial}{\partial T} \left\{ \begin{matrix} 1-f^{\epsilon}_{q} \\
-f^{\epsilon}_{q} \end{matrix} \right\} \left( \frac{-1}{\Gamma_{q}} -\frac{s_{xy}}{2} \right) Z_{\epsilon} e^{-i\Omega_{\epsilon} s_{xy}} \ . \label{V_3 timerepr}
\end{align}
In the first line, half of the term containing $2/\Gamma_{q}^2$ comes from $V_{32}$, and the other half of it 
arises from the $q_0$ derivative of $G_{AS}/M$.
The term $s_{xy}/\Gamma_{q}$ comes from $\partial_{q_0} e^{-iq_{0} s_{xy}}$ in $V_{31}$. 
Note that time-derivatives in (\ref{V_3}) act on the distribution function only through 
the time dependence of temperature. 
In the second line, we have used the definition of the width.

Noticing that $V_2$ is obtained from $V_3$ by changing its sign
 and exchanging $\partial_{q_0}$ and $\partial _t$.
Then we can immediately get the time representation for $V_2$.
Similarly to (\ref{partial_q G}), we get
\begin{align}
\partial_{t} G^{d(eq)}_{R} (t;q)=&\frac{G_{R S}^{d(eq)}}{M} \left( \{ \zeta_R , \Slash{Q}_R \}(\Slash{Q}_R +M)\frac{G_{R S}^{d(eq)}}{M} + \zeta_R \right) \ . \label{partial_t G}
\end{align} 
An exchange of $\partial_{q_0}$ and $\partial _t$ leads to an exchange of $\xi$ and $\zeta$.
From the definitions of $\xi$ in (\ref{xi}) and $\zeta$ in (\ref{zeta}), 
the exchange results in 
\be
\frac{\partial T}{\partial t}\frac{\partial}{\partial T} \to (-1)\times \frac{H |{\bold q}|^2 }{\epsilon \omega_{q} a^2} \frac{\partial}{\partial (\epsilon \omega_{q})} = \frac{\partial \omega_q}{\partial t}\frac{\partial}{\partial \omega_{q}}
\ee 
in (\ref{V_3 timerepr}) in the leading order. Hence we get the expression of $V_2$ as
\begin{align}
V_2 \simeq&\sum_{\epsilon} (-i)\frac{\partial \omega_q}{\partial t}\frac{\partial}{\partial \omega_{q}} \left\{ \begin{matrix} 1-f^{\epsilon}_{q} \\
-f^{\epsilon}_{q} \end{matrix} \right\} \left( \frac{-1}{\Gamma_{q}} -\frac{s_{xy}}{2} \right) Z_{\epsilon} e^{-i\Omega_{\epsilon} s_{xy}} \ . \label{V_2 timerepr}
\end{align}

Putting  (\ref{V_3 timerepr}) and (\ref{V_2 timerepr}) together,
\be
V_3 + V_2 \simeq \sum_{\epsilon} (-i)d_t \left\{ \begin{matrix} 1-f^{\epsilon}_{q} \\
-f^{\epsilon}_{q} \end{matrix} \right\} \left( \frac{-1}{\Gamma_{q}} -\frac{s_{xy}}{2} \right) Z_{\epsilon} e^{-i\Omega_{\epsilon} s_{xy}} \label{V_3+2}
\ee
is obtained.

Finally, from (\ref{V_1+4})$+$(\ref{V_3+2}), we get the final result:
\begin{align}
\Delta &G^{dii}_{\wightman } ( x^0,y^0;{\bold q}) 
=  \sum_{\epsilon} \left( \Theta(s_{xy}) \Delta G_\gtrless^{(1)} + \Theta(-s_{xy}) \Delta G_\gtrless^{(2)}
\right) \ , \label{DelG_{W}^{d}a}
\end{align}
\begin{align}
\Delta G_\gtrless^{(1)} \simeq & -i d_t 
\left\{ \begin{matrix} 1-f^{\epsilon}_{i q} \\
-f^{\epsilon}_{i q} \end{matrix} \right\} 
\left( \frac{-1}{\Gamma _{iq}}+(X_{xy} -t -|s_{xy}|/2) \right) Z_{\epsilon }^{i} e^{-i\Omega_{\epsilon i}(x^0 -y^0)} \n
&  -i 
\left\{ \begin{matrix} 1-f^{\epsilon}_{i q} \\ -f^{\epsilon}_{i q} \end{matrix} \right\} 
 \Delta \hat{G}^{dii}_{R}(x^0,y^0;\epsilon,{\bold q}) \ , \n
 \Delta G_\gtrless^{(2)} \simeq &
 -i d_t  \left\{ \begin{matrix} 1-f^{\epsilon}_{i q} \\
-f^{\epsilon}_{i q} \end{matrix} \right\}\left( \frac{-1}{\Gamma _{iq}}+(X_{xy} -t -|s_{xy}|/2) \right) Z_{\epsilon }^{i} e^{-i\Omega^{*}_{\epsilon i}(x^0 -y^0)} \n
& +i \left\{ \begin{matrix} 1-f^{\epsilon}_{i q} \\
-f^{\epsilon}_{i q} \end{matrix} \right\} \times \Delta \hat{G}^{dii}_{A}(x^0,y^0;\epsilon,{\bold q}) \ , \notag
\end{align}
where
\begin{align}
d_t \equiv \frac{\partial T}{\partial t} \frac{\partial}{\partial T} +\frac{\partial \omega_{q}}{\partial t} \frac{\partial}{\partial \omega_{q}} \ .
\end{align}

\section{Calculation of $\Delta G_\gtrless^{\prime}$ }
\setcounter{equation}{0}
\label{Sec-dGoffW}
In the leading order approximation, off-diagonal components of the Wightman function are given by 
eq. (\ref{intKBeqG_{W}^{'}2}). Its deviation from the thermal value is obtained by
taking a variation:
\begin{align}
\Delta G^{' ij}_{\wightman } 
=& - \Delta G^{d ii}_{R}* \Pi^{' ij}_{R} *G^{djj}_{\wightman}
- G^{d ii}_{R}* \Delta \Pi^{' ij}_{R} *G^{djj}_{\wightman}
- G^{d ii}_{R}* \Pi^{' ij}_{R} *\Delta G^{djj}_{\wightman} \n
&- \Delta G^{dii}_{\wightman}* \Pi^{'ij}_{A } *G^{djj}_{A}
- G^{dii}_{\wightman}* \Delta \Pi^{'ij}_{A } *G^{djj}_{A}
- G^{dii}_{\wightman}* \Pi^{'ij}_{A } *\Delta G^{djj}_{A} \n
& - \Delta G^{dii}_{R}* \Pi^{'ij}_{\wightman} *G^{djj}_{A}
- G^{dii}_{R}* \Delta \Pi^{'ij}_{\wightman} *G^{djj}_{A}
- G^{dii}_{R}* \Pi^{'ij}_{\wightman} * \Delta G^{djj}_{A} \ . \label{deltaWightman9}
\end{align}
We will calculate these 9 terms in the following sections.

Among all these 9 terms, it will turn out that 
leading order contributions to  $\Delta G_\gtrless^{\prime}$
that eventually remain in the limit 
\be
H \ll \Gamma \ll \Gamma_{\ell \phi} \sim T \sim M  \label{Scales-Appendix}
\ee
come from three terms containing variations of the Wightman functions $\Delta G_\gtrless^d$ or
$\Delta \Pi_\gtrless^{\prime}$, namely the 3rd term, the 4th term and the 8th term in  (\ref{deltaWightman9}).
These terms represent  variation of the distribution function $\Delta f$, and are shown to be
of order ${\mathcal O}(H/\Gamma)$.
After summation of various terms, the other terms become 
${\mathcal O}(H/\Gamma) \times {\mathcal O}(\Gamma/\Gamma_{\ell \phi})$ etc. and 
negligible in the above limit compared to the leading terms.

We first give order estimations of the following quantities.
First, since the distribution function $f(q_0)$ contains a factor $e^{-q_0/T(t)}$, 
the derivatives are
\be
\partial_{t} f(q_0) \sim  \frac{\dot{T}q_0}{T^2} f(q_0) \sim H f(q_0) \ , \ \ \ 
\partial_{q_0} f(q_0) \sim \frac{f(q_0)}{T} \  .
\ee
We have used $\dot{T} \propto HT$ and that the typical frequency in the analysis is $q_0 \sim T$.
On the other hand, the derivatives of Green functions (\ref{GzetaG}) and (\ref{GxiG}) are
\be
\partial_{t} G(t;q_0) \sim \frac{HT}{\Gamma} G(t;q_0) \ , \ \ \ 
\partial_{q_0} G(t;q_0) \sim \frac{G(t;q_0)}{\Gamma} \ .
\label{orderestimationG}
\ee
Green functions contain a factor  $1/(q_0 -\Omega_j)$ or $1/(q_0 -\Omega_j^*)$
where $\Omega =\omega_i-i \Gamma_i/2$. In  $q_0$ integrations, $q_0$
is replaced by positions of poles coming from other similar factors,
and these factors become $1/(\Omega_i-\Omega_j)$ for $i \neq j$ 
or $1/(\Omega_i-\Omega_j^*)$.
Both of them are of order $1/\Delta M \sim 1/\Gamma$.
Acting $\partial_t$, we have
\be
\partial_t \frac{1}{q_0 -\Omega_j} = \frac{\dot{\Omega}_j}{(q_0-\Omega_i)^2} \sim 
\frac{H \Omega_j}{(q_0-\Omega_j)^2} \sim \frac{H T}{(q_0-\Omega_j)^2} \  .
\ee
Then replacing $q_0$ by $\Omega_i$, the first relation of (\ref{orderestimationG})
is obtained. 
The second relation comes from the relation
\be
\partial_{q_0} \frac{1}{q_0 -\Omega_j} =  \frac{-1}{(q_0 -\Omega_j)^2}
\ee
and $1(q_0-\Omega_j) \rightarrow {\mathcal O}(1/\Gamma).$
Hence, for both of the derivatives, $(\partial f)/f$ is smaller than $(\partial G)/G$ by a factor $\Gamma/T$.

$(\partial \Pi )/\Pi $ is also smaller than $(\partial G)/G$. 
It is because of the following reason.
The self-energy $\Pi(q_0)$ contains a factor $1/(q_0 - (\omega_{p}+\omega_{k}\mp i\Gamma_{\ell \phi}/2))$ where $\omega_p$ and $\omega_k$
are energies of the SM lepton and the Higgs propagating in the self-energy
diagram. After performing $q_0$ integration,
$q_0$ is replaced by the position of poles $\Omega_{i,j}$ in $G(q_0)$.
Due to  (\ref{Scales-Appendix}),
$1/(q_0 - (\omega_{p}+\omega_{k}\mp i\Gamma_{\ell \phi}/2))$ becomes of order $1/T\sim 1/ \Gamma_{\ell \phi}$ 
unlike $1/\Gamma$ for $G$.
Therfore, acting derivatives, $(\partial \Pi )/\Pi $ becomes smaller than $(\partial G)/G$ by a factor $\Gamma/T$.

\subsection{Leading contributions}
\label{Sec-relevant}
We first calculate three terms that are obtained by taking variations of the
Wightman-type functions. The other terms, which turn out to be subleading, are evaluated in 
 Appendix \ref{Sec-irrelevant}. 

Since $\Delta G_\gtrless^d$ is written as a sum of four terms
of $V_{1,2,3,4}$ in (\ref{DelG_{W}^{d}App}),
the 3rd term in (\ref{deltaWightman9}) can be also written as a sum of the following
four terms  $V'_{1,2,3,4}$ respectively.
By using   (\ref{DelG_{W}^{d}App}),   (\ref{defDeltildeG_{RA}^{d}App}) and (\ref{Useful23}), 
it becomes
\begin{align}
-\int du dv  G^{d (eq)}_{R}(s_{xu};{\bold q}) \Pi^{'(eq)}_{R}(s_{uv};{\bold q}) \Delta G^{d}_{\wightman}(X_{vy},s_{vy};{\bold q}) = V'_{1}+V'_{2}+V'_{3}+V'_{4}  \  .
\end{align}
The first term $V'_1$ is written as
\begin{align}
V'_{1}= V'_{11}+V'_{12}+V'_{13}+V'_{14} \notag
\end{align}
where
\begin{align}
V'_{11} =&\int \frac{d q_0}{2\pi}(-i) \left\{ \begin{matrix} 1-f \\
-f \end{matrix} \right\}\frac{-i}{2} G_{R}^{d(eq)}\Pi_{R}^{'(eq)}{\bigg \{ }\left[ \partial_{q_0} G^{d(eq)}_{R} \zeta_{R} G_{R}^{d(eq)}-G_{R}^{d(eq)} \zeta_{R} \partial_{q_0} G^{d(eq)}_{R} \right] \n
&-\left[ \partial_{q_0} G^{d(eq)}_{A} \zeta_{A} G_{A}^{d(eq)}-G_{A}^{d(eq)} \zeta_{A} \partial_{q_0} G^{d(eq)}_{A} \right] {\bigg \} }e^{-iq_0 s_{xy}} \ , 
\label{V'11} \\
V'_{12}= &\int \frac{d q_0}{2\pi}(-i) \left\{ \begin{matrix} 1-f \\
-f \end{matrix} \right\}\frac{+i}{2} G^{d(eq)}_{R} \Pi_{R}^{'(eq)} \partial _{q_0} \partial _t G_{\rho}^{d(eq)} e^{-iq_0 s_{xy}} \ ,  
\label{V'12} \\
V'_{13}= &\int \frac{d q_0}{2\pi}(-i) \left\{ \begin{matrix} 1-f \\
-f \end{matrix} \right\}(-1)  G^{d(eq)}_{R}\Pi_{R}^{'(eq)} \partial _t G_{\rho}^{d(eq)}   \left( X_{xy}-t -\frac{s_{xy}}{2} \right) e^{-iq_0 s_{xy}} \ ,
\label{V'13}  \\
V'_{14}= &\int \frac{d q_0}{2\pi}(-i) \partial _{q_0} \left\{ \begin{matrix} 1-f \\
-f \end{matrix} \right\}\frac{+i}{2}G^{d(eq)}_{R} \Pi_{R}^{'(eq)}  \partial _t G_{\rho}^{d(eq)} e^{-iq_0 s_{xy}} \  .\label{V'14}
\end{align}
Only the last term  $V'_{14}$ (\ref{V'14}) becomes ${\mathcal O}(H/\Gamma)$ and remains in the above limit.
The other terms turn out to be negligible.
Using the relation (\ref{GzetaG}), 
the first term $V'_{11}$ looks of order  
${\mathcal O}(HT/\Gamma^2) \sim {\mathcal O}(H/\Gamma) \times {\mathcal O}(T/\Gamma) $.
But the factor ${\mathcal O}(T/\Gamma)$ is cancelled between two terms in the square brackets.
Furthermore, combined with other terms in  Appendix \ref{Sec-irrelevant}, they
become eq.(\ref{GzetaGvanish}) and negligible compared to the leading
terms of ${\mathcal O}(H/\Gamma).$
$V'_{12}$ looks of order ${\mathcal O}(HT/\Gamma^2)$, but again
combined with other terms in Appendix \ref{Sec-irrelevant}, they become
(\ref{sowhat}) and negligible.
$V'_{13}$, which is ${\mathcal O}(H/\Gamma)$, is similarly combined to be (\ref{GXxy}) and (\ref{Gsxy}),
and can be neglected.

$V'_{2}$ and $V'_{3}$ are given by
\begin{align}
V'_{2}=\int \frac{d q_0}{2\pi}(-i) & \partial_{q_0} \left\{ \begin{matrix} 1-f \\
-f \end{matrix} \right\}\frac{-i}{2} G^{d(eq)}_{R} \Pi_{R}^{'(eq)} {\Big[ } \partial_{t} G^{d(eq)}_{R} \Pi^{d(eq)}_{\rho} G_{A}^{d(eq)} \n
&-G_{R}^{d(eq)} \Pi^{d(eq)}_{\rho} \partial_{t} G^{d(eq)}_{A} {\Big ]} e^{-iq_0 s_{xy}} \ ,
\end{align}
and
\begin{align}
V'_{3}=\int \frac{d q_0}{2\pi}(-i)&  \partial_{t} \left\{ \begin{matrix} 1-f \\
-f \end{matrix} \right\}\frac{+i}{2} G^{d(eq)}_{R} \Pi_{R}^{'(eq)}{\Big[ }
 \partial_{q_0} G^{d(eq)}_{R} \Pi^{d(eq)}_{\rho} G_{A}^{d(eq)} \n
&-G_{R}^{d(eq)} \Pi^{d(eq)}_{\rho} \partial_{q_0} G^{d(eq)}_{A} {\Big ]} e^{-iq_0 s_{xy}} \ .
\end{align}
Both of them remain in the above limit.

Finally the last term $V'_{4}$ 
is written as
\begin{align}
V'_{4}=V'_{41}+V'_{42} +V'_{43} \notag
\end{align}
where
\begin{align}
V'_{41}=&\int \frac{d q_0}{2\pi}(-i) \partial _t 
\left\{ \begin{matrix} 1-f \\ -f \end{matrix} \right\}
\frac{+i}{2}G^{d(eq)}_{R} \Pi_{R}^{'(eq)} \partial _{q_0} G_{\rho}^{d(eq)} e^{-iq_0 s_{xy}}  \ ,
\label{V'41} \\
V'_{42}=&\int \frac{d q_0}{2\pi}(-i)\partial _t \left\{ \begin{matrix} 1-f \\
-f \end{matrix} \right\}(-1)G^{d(eq)}_{R} \Pi_{R}^{'(eq)} G_{\rho}^{d(eq)}  
\left( X_{xy}-t -\frac{s_{xy}}{2} \right) e^{-iq_0 s_{xy}} \ ,
\label{V'42} \\
V'_{43}=&\int \frac{d q_0}{2\pi}(-i) \partial _{q_0} \partial _t \left\{ \begin{matrix} 1-f \\
-f \end{matrix} \right\}\frac{+i}{2} G^{d(eq)}_{R} \Pi_{R}^{'(eq)} G^{d(eq)}_{\rho} e^{-iq_0 s_{xy}} \  .
\label{V'43}
\end{align}
Only the first term $V'_{41}$ is ${\mathcal O}(H/\Gamma)$ and remains. The other two terms become negligible
because $V'_{42} = {\mathcal O}(H/\Gamma_{\ell \phi})={\mathcal O}(H/T) \ll {\mathcal O}(H/\Gamma)$ and
$V'_{43} = {\mathcal O}(H/T) \ll {\mathcal O}(H/\Gamma)$.

Similarly, by using  (\ref{DelG_{W}^{d}App}),   (\ref{defDeltildeG_{RA}^{d}App}) and (\ref{Useful22}), 
the 4th term in  (\ref{deltaWightman9}) becomes
\begin{align}
-\int du dv  \Delta G^{d }_{\gtrless}(X_{xu};s_{xu};{\bold q}) \Pi^{'(eq)}_{A}(s_{uv};{\bold q})  G^{d(eq)}_{A}(s_{vy};{\bold q}) = V''_{1}+V''_{2}+V''_{3}+V''_{4} 
\end{align}
where
\begin{align}
V''_{1}=&\int \frac{d q_0}{2\pi}(-i) \left\{ \begin{matrix} 1-f \\
-f \end{matrix} \right\}\frac{-i}{2} {\bigg \{ }\left[ \partial_{q_0} G^{d(eq)}_{R} \zeta_{R} G_{R}^{d(eq)}-G_{R}^{d(eq)} \zeta_{R} \partial_{q_0} G^{d(eq)}_{R} \right] \n
&-\left[ \partial_{q_0} G^{d(eq)}_{A} \zeta_{A} G_{A}^{d(eq)}-G_{A}^{d(eq)} \zeta_{A} \partial_{q_0} G^{d(eq)}_{A} \right] {\bigg \} }\Pi_{A}^{'(eq)}G_{A}^{d(eq)}e^{-iq_0 s_{xy}} \n
+&\int \frac{d q_0}{2\pi}(-i) \left\{ \begin{matrix} 1-f \\
-f \end{matrix} \right\}\frac{-i}{2}  \partial _{q_0} \partial _t G_{\rho}^{d(eq)}\Pi_{A}^{'(eq)} G^{d(eq)}_{A} e^{-iq_0 s_{xy}} \n
+&\int \frac{d q_0}{2\pi}(-i) \left\{ \begin{matrix} 1-f \\
-f \end{matrix} \right\}(-1)  \partial _t G_{\rho}^{d(eq)} \Pi_{A}^{'(eq)} G^{d(eq)}_{A}  \left( X_{xy}-t +\frac{s_{xy}}{2} \right) e^{-iq_0 s_{xy}} \n
+&\int \frac{d q_0}{2\pi}(-i) \partial _{q_0} \left\{ \begin{matrix} 1-f \\
-f \end{matrix} \right\}\frac{-i}{2}\partial _t G_{\rho}^{d(eq)} \Pi_{A}^{'(eq)}  G^{d(eq)}_{A} e^{-iq_0 s_{xy}} \  , \label{V''1}
\end{align}
\begin{align}
V''_{2}=\int \frac{d q_0}{2\pi}(-i) & \partial_{q_0} \left\{ \begin{matrix} 1-f \\
-f \end{matrix} \right\}\frac{-i}{2} {\Big[ } \partial_{t} G^{d(eq)}_{R} \Pi^{d(eq)}_{\rho} G_{A}^{d(eq)} \n
&-G_{R}^{d(eq)} \Pi^{d(eq)}_{\rho} \partial_{t} G^{d(eq)}_{A} {\Big ]}\Pi_{A}^{'(eq)}G_{A}^{d(eq)} e^{-iq_0 s_{xy}} \  , \label{V''2}
\end{align}
\begin{align}
V''_{3}=\int \frac{d q_0}{2\pi}(-i) & \partial_{t} \left\{ \begin{matrix} 1-f \\
-f \end{matrix} \right\}\frac{+i}{2} {\Big[ } \partial_{q_0} G^{d(eq)}_{R} \Pi^{d(eq)}_{\rho} G_{A}^{d(eq)} \n
&-G_{R}^{d(eq)} \Pi^{d(eq)}_{\rho} \partial_{q_0} G^{d(eq)}_{A} {\Big ]}
\Pi_{A}^{'(eq)} G_{A}^{d(eq)} e^{-iq_0 s_{xy}} \ , \label{V''3}
\end{align}
\begin{align}
V''_{4}=&\int \frac{d q_0}{2\pi}(-i) \partial _t \left\{ \begin{matrix} 1-f \\
-f \end{matrix} \right\}\frac{-i}{2} \partial _{q_0} G_{\rho}^{d(eq)}\Pi_{A}^{'(eq)} G^{d(eq)}_{A}  e^{-iq_0 s_{xy}} \n
+&\int \frac{d q_0}{2\pi}(-i)\partial _t \left\{ \begin{matrix} 1-f \\
-f \end{matrix} \right\}(-1) G_{\rho}^{d(eq)}  \Pi_{A}^{'(eq)}G^{d(eq)}_{A} \left( X_{xy}-t +\frac{s_{xy}}{2} \right) e^{-iq_0 s_{xy}} \n
+&\int \frac{d q_0}{2\pi}(-i) \partial _{q_0} \partial _t \left\{ \begin{matrix} 1-f \\
-f \end{matrix} \right\}\frac{-i}{2} G^{d(eq)}_{\rho} \Pi_{A}^{'(eq)} G_{A}^{d(eq)} e^{-iq_0 s_{xy}} \ . \label{V''4}
\end{align}
The arguments are parallel to $V'$ and we do not repeat the discussions.


The last term containing a variation of the Wightman function  comes
from the 8th term in  (\ref{deltaWightman9}). 
It contains $\Delta \Pi_\gtrless^{\prime}$, and becomes
\begin{align}
-\int & du dv \  G^{d (eq)}_{R}(s_{xu};{\bold q}) \Delta \Pi^{'(eq)}_{\wightman}(X_{uv};s_{uv};{\bold q}) G^{d(eq)}_{A}(s_{vy};{\bold q}) =V_1'''+V_2''' \label{V'''}
\end{align}
where
\begin{align}
V_1'''=-&\int \frac{d q_0}{2\pi}(-i) \left\{ \begin{matrix} 1-f \\ -f \end{matrix} \right\}
{\Big [}  G_{R}^{d(eq)} 
\partial _t\Pi_{\rho}^{'(eq)} G^{d(eq)}_{A}
 (X_{xy}-t) \n
&  +
 \frac{i}{2}\left( \partial_{q_0} G^{d(eq)}_{R} \partial _t \Pi_{\rho}^{'(eq)}G_{A}^{d(eq)}- G^{d(eq)}_{R} \partial _t \Pi_{\rho}^{'(eq)} \partial_{q_0}G_{A}^{d(eq)} \right)
 {\Big ]}
 e^{-iq_0 s_{xy}} \ ,  \label{V'''1}
\end{align}
\begin{align}
V_2'''=-&\int \frac{d q_0}{2\pi}(-i) \partial _t \left\{ \begin{matrix} 1-f \\ -f \end{matrix} \right\}
{\Big [}
G^{d(eq)}_{R}  \Pi_{\rho}^{'(eq)}G_{A}^{d(eq)} (X_{xy} -t) \n 
& +
\frac{i}{2}\left( \partial_{q_0} G^{d(eq)}_{R}  \Pi_{\rho}^{'(eq)}G_{A}^{d(eq)}- G^{d(eq)}_{R}  \Pi_{\rho}^{'(eq)} \partial_{q_0}G_{A}^{d(eq)} \right)
{\Big ]}
e^{-iq_0 s_{xy}} \label{V'''2} \ .
\end{align}
Only the second term of $V_2'''$ becomes ${\mathcal O}(H/\Gamma)$ and 
remains in the limit. The first terms of $V_1'''$ and $V_2'''$
are combined to be (\ref{GXxy}), and
 the second term of $V_1'''$ is combined to be 
(\ref{GzetaGvanish}); these are negligibly small.
\subsection{ $\Delta G_\gtrless^\prime$ in the time-representation}
\label{Sec-FTrelevant}
We now calculate  the time-representation of $\Delta G_\gtrless^\prime$ 
by Fourier transforming the leading contributions in the previous subsection \ref{Sec-relevant}.
The other terms in (\ref{deltaWightman9}) are evaluated in the next subsection and 
after being combined they are shown to become negligible.
We consider the case $x^0 >y^0$ in the following.
The calculations are performed in parallel  to those in  Appendix \ref{Sec-App-out-of-eq-Wightman-diag},
so we do not repeat detailed calculations.

We  suppose that the reference time $t$ and the arguments  $x^0,y^0$ in  $\Delta G(x^0,y^0)$
satisfy the conditions,
 $(X_{xy}-t) \ltsim 1/\Gamma_{\ell \phi}$ and
 $s_{xy} \ltsim 1/\Gamma_{\ell \phi}$ since
such situations appear  in the calculation of the lepton asymmetry (\ref{dndt2}).
It is due to the fast damping of the self-energy $\Pi_\gtrless$.  
In performing the Fourier transformation, we drop higher order terms
with respect to $H/\Gamma$ compared to the leading terms of order ${\mathcal O}(H/\Gamma)$. 

For $V'_1$, the leading order contribution 
with respect to $H/\Gamma$ comes from $V'_{14}$. Performing $q_0$ integration, it  becomes 
\begin{align}
V'_{14} \simeq &\sum_{\epsilon} Z_{\epsilon} \Pi^{'(eq)ij}_{R}(\epsilon \omega_{ q}) Z_{\epsilon}(-i) \frac{\partial \omega_{j q}}{\partial t}\frac{\partial}{\partial \omega_{j q}} \left\{ \begin{matrix} 1-f^{\epsilon}_{j q} \\ -f^{\epsilon}_{j q} \end{matrix} \right\}\frac{+1/2}{(\Omega_{\epsilon i}-\Omega_{\epsilon j})^2} \ e^{-i\Omega_{\epsilon j }s_{xy}}\n
+&\sum_{\epsilon} Z_{\epsilon } \Pi^{'(eq)ij}_{R}(\epsilon \omega_{\bold q}) Z_{\epsilon }(-i)  \frac{\partial \omega_{i q}}{\partial t}\frac{\partial}{\partial \omega_{i q}} \left\{ \begin{matrix} 1-f^{\epsilon}_{i q} \\ -f^{\epsilon}_{i q} \end{matrix} \right\} \left[ \frac{-1/2}{(\Omega_{\epsilon i}-\Omega_{\epsilon j})^2} + \frac{+1/2}{(\Omega_{\epsilon i}-\Omega^{*}_{\epsilon j})^2} \right] \ e^{-i\Omega_{\epsilon i }s_{xy}} \n
\simeq &\sum_{\epsilon} Z_{\epsilon } \Pi^{'(eq)ij}_{R}(\epsilon \omega_{\bold q}) Z_{\epsilon }(-i)  \frac{\partial \omega_{ q}}{\partial t}\frac{\partial}{\partial \omega_{ q}} \left\{ \begin{matrix} 1-f^{\epsilon}_{ q} \\ -f^{\epsilon}_{ q} \end{matrix} \right\} \frac{+1/2}{(\Omega_{\epsilon i}-\Omega^{*}_{\epsilon j})^2} \ e^{-i\Omega_{\epsilon }s_{xy}} \ . \label{leading of V'14}
\end{align}
$V'_{14}$ have poles at $q_0=\Omega_{i}$ of $G_R^{d(eq)}$ and $q_0=\Omega_{j}$
of $\partial_t G_\rho^{d(eq)}= \partial_t (G_R^{d(eq)}-G_A^{d(eq)})$.
The first line is the contribution of the pole at $q_0=\Omega_{j}$ while the second term 
corresponds to the residue of the pole at $q_0=\Omega_{i}$.
We have dropped higher order terms, e.g., a residue at the pole of the self-energy $\Pi'$
or the distribution function $f$ since they are much more suppressed by a factor $\Gamma/T$. 
The second equality is obtained by identifying $\Omega_i$ and $\Omega_j$ in the leading 
approximation.

$V'_2$ and $V'_3$ are evaluated as
\begin{align}
V'_2 \simeq &\sum_{\epsilon} Z_{\epsilon } \Pi^{'(eq)ij}_{R}(\epsilon \omega_{q}) Z_{\epsilon }(-i)  \frac{\partial \omega_{j q}}{\partial t}\frac{\partial}{\partial \omega_{j q}} \left\{ \begin{matrix} 1-f^{\epsilon}_{j q} \\ -f^{\epsilon}_{j q} \end{matrix} \right\} \n
&\times \left[ \frac{+1/2}{(\Omega_{\epsilon i}-\Omega_{\epsilon j})^2} + \frac{-1}{\Omega_{\epsilon i}-\Omega_{\epsilon j}}\frac{i}{\Gamma_{jq}} \right] \ e^{-i\Omega_{\epsilon j }s_{xy}}\n
+&\sum_{\epsilon} Z_{\epsilon} \Pi^{'(eq)ij}_{R}(\epsilon \omega_{ q}) Z_{\epsilon }(-i) \frac{\partial \omega_{i q}}{\partial t} \frac{\partial}{\partial \omega_{i q}} \left\{ \begin{matrix} 1-f^{\epsilon}_{i q} \\ -f^{\epsilon}_{i q} \end{matrix} \right\} \n
&\times \left[ \frac{+1/2}{(\Omega_{\epsilon i}-\Omega_{\epsilon j})^2}\frac{i\Gamma _{jq}}{\Omega_{\epsilon i}-\Omega^{*}_{\epsilon j}} + \frac{-1/2}{(\Omega_{\epsilon i}-\Omega^{*}_{\epsilon j})^2} \frac{i\Gamma _{jq}}{\Omega_{\epsilon i}-\Omega_{\epsilon j}} \right] \ e^{-i\Omega_{\epsilon i }s_{xy}} \n
\simeq &\sum_{\epsilon} Z_{\epsilon} \Pi^{'(eq)ij}_{R}(\epsilon \omega_{ q}) Z_{\epsilon }(-i) \frac{\partial \omega_{q}}{\partial t} \frac{\partial}{\partial \omega_{ q}} \left\{ \begin{matrix} 1-f^{\epsilon}_{ q} \\ -f^{\epsilon}_{ q} \end{matrix} \right\} \n
&\times \frac{-i}{2}\left[ \frac{1}{\Gamma_j}\left( \frac{1}{\Omega_{\epsilon i}-\Omega^{*}_{\epsilon j}} +\frac{1}{\Omega_{\epsilon i}-\Omega_{\epsilon j}} \right) + \frac{1}{(\Omega_{\epsilon i}-\Omega^{*}_{\epsilon j})^2} \frac{\Gamma _{jq}}{\Omega_{\epsilon i}-\Omega_{\epsilon j}} \right] \ e^{-i\Omega_{\epsilon }s_{xy}} 
\end{align}
and
\begin{align}
V'_3 \simeq &\sum_{\epsilon} Z_{\epsilon } \Pi^{'(eq)ij}_{R}(\epsilon \omega_{q}) Z_{\epsilon }(-i) \frac{\partial T}{\partial t}\frac{\partial}{\partial T} \left\{ \begin{matrix} 1-f^{\epsilon}_{j q} \\ -f^{\epsilon}_{j q} \end{matrix} \right\} \n
&\times \left[ \frac{+1/2}{(\Omega_{\epsilon i}-\Omega_{\epsilon j})^2} + \frac{-1}{\Omega_{\epsilon i}-\Omega_{\epsilon j}}\frac{i}{\Gamma_{jq}} \right] \ e^{-i\Omega_{\epsilon j }s_{xy}}\n
+&\sum_{\epsilon} Z_{\epsilon} \Pi^{'(eq)ij}_{R}(\epsilon \omega_{ q}) Z_{\epsilon }(-i) \frac{\partial T}{\partial t}\frac{\partial}{\partial T} \left\{ \begin{matrix} 1-f^{\epsilon}_{i q} \\ -f^{\epsilon}_{i q} \end{matrix} \right\} \n
&\times \left[ \frac{+1/2}{(\Omega_{\epsilon i}-\Omega_{\epsilon j})^2}\frac{i\Gamma _{jq}}{\Omega_{\epsilon i}-\Omega^{*}_{\epsilon j}} + \frac{-1/2}{(\Omega_{\epsilon i}-\Omega^{*}_{\epsilon j})^2} \frac{i\Gamma _{jq}}{\Omega_{\epsilon i}-\Omega_{\epsilon j}} \right] \ e^{-i\Omega_{\epsilon i }s_{xy}} \n
\simeq &\sum_{\epsilon} Z_{\epsilon} \Pi^{'(eq)ij}_{R}(\epsilon \omega_{ q}) Z_{\epsilon }(-i) \frac{\partial T}{\partial t}\frac{\partial}{\partial T} \left\{ \begin{matrix} 1-f^{\epsilon}_{ q} \\ -f^{\epsilon}_{ q} \end{matrix} \right\} \n
&\times \frac{-i}{2}\left[ \frac{1}{\Gamma_j}\left( \frac{1}{\Omega_{\epsilon i}-\Omega^{*}_{\epsilon j}} +\frac{1}{\Omega_{\epsilon i}-\Omega_{\epsilon j}} \right) + \frac{1}{(\Omega_{\epsilon i}-\Omega^{*}_{\epsilon j})^2} \frac{\Gamma _{jq}}{\Omega_{\epsilon i}-\Omega_{\epsilon j}} \right] \ e^{-i\Omega_{\epsilon }s_{xy}} \ .
\end{align}
For $V'_4$, the leading contribution comes from $V'_{41}$.
It becomes
\begin{align}
V'_{41}\simeq  &\sum_{\epsilon} Z_{\epsilon } \Pi^{'(eq)ij}_{R}(\epsilon \omega_{ q}) Z_{\epsilon }(-i) \frac{\partial T}{\partial t}\frac{\partial}{\partial T} \left\{ \begin{matrix} 1-f^{\epsilon}_{j q} \\ -f^{\epsilon}_{j q} \end{matrix} \right\}\frac{-1/2}{(\Omega_{\epsilon i}-\Omega_{\epsilon j})^2} \ e^{-i\Omega_{\epsilon j }s_{xy}} \n
+& \sum_{\epsilon} Z_{\epsilon } \Pi^{'(eq)ij}_{R}(\epsilon \omega_{\bold q}) Z_{\epsilon }(-i)  \frac{\partial T}{\partial t}\frac{\partial}{\partial T} \left\{ \begin{matrix} 1-f^{\epsilon}_{i q} \\ -f^{\epsilon}_{i q} \end{matrix} \right\} \left[ \frac{+1/2}{(\Omega_{\epsilon i}-\Omega_{\epsilon j})^2} + \frac{-1/2}{(\Omega_{\epsilon i}-\Omega^{*}_{\epsilon j})^2} \right] \ e^{-i\Omega_{\epsilon i }s_{xy}} \n
\simeq& \sum_{\epsilon} Z_{\epsilon } \Pi^{'(eq)ij}_{R}(\epsilon \omega_{\bold q}) Z_{\epsilon }(-i)  \frac{\partial T}{\partial t}\frac{\partial}{\partial T} \left\{ \begin{matrix} 1-f^{\epsilon}_{i q} \\ -f^{\epsilon}_{i q} \end{matrix} \right\} \frac{-1/2}{(\Omega_{\epsilon i}-\Omega^{*}_{\epsilon j})^2} \ e^{-i\Omega_{\epsilon }s_{xy}} \ .
\end{align}

Similarly, we get the leading order terms from $V''$ as
\begin{align}
V''_{1} \simeq  &\sum_{\epsilon} Z_{\epsilon } \Pi^{'(eq)ij}_{A}(\epsilon \omega_{ q}) Z_{\epsilon }(-i) \frac{\partial \omega_{i q}}{\partial t}\frac{\partial}{\partial \omega_{i q}} \left\{ \begin{matrix} 1-f^{\epsilon}_{i q} \\ -f^{\epsilon}_{i q} \end{matrix} \right\}\frac{-1/2}{(\Omega_{\epsilon i}-\Omega^{*}_{\epsilon j})^2} \ e^{-i\Omega_{\epsilon i}s_{xy}} \ ,
\end{align}
\begin{align}
V''_{2} \simeq  &\sum_{\epsilon} Z_{\epsilon } \Pi^{'(eq)ij}_{A}(\epsilon \omega_{q}) Z_{\epsilon }(-i) \frac{\partial \omega_{i q}}{\partial t}\frac{\partial}{\partial \omega_{i q}} \left\{ \begin{matrix} 1-f^{\epsilon}_{i q} \\ -f^{\epsilon}_{i q} \end{matrix} \right\} \n
&\times \left[ \frac{+1/2}{(\Omega_{\epsilon i}-\Omega^{*}_{\epsilon j})^2} + \frac{+1}{\Omega_{\epsilon i}-\Omega^{*}_{\epsilon j}}\frac{i}{\Gamma_{iq}} \right] \ e^{-i\Omega_{\epsilon i}s_{xy}}\ ,
\end{align}
\begin{align}
V''_{3}\simeq  &\sum_{\epsilon} Z_{\epsilon } \Pi^{'(eq)ij}_{A}(\epsilon \omega_{ q}) Z_{\epsilon }(-i) \frac{\partial T}{\partial t}\frac{\partial}{\partial T} \left\{ \begin{matrix} 1-f^{\epsilon}_{i q} \\ -f^{\epsilon}_{i q} \end{matrix} \right\} \n
&\times \left[ \frac{+1/2}{(\Omega_{\epsilon i}-\Omega^{*}_{\epsilon j})^2} + \frac{+1}{\Omega_{\epsilon i}-\Omega^{*}_{\epsilon j}}\frac{i}{\Gamma_{iq}} \right]  \ e^{-i\Omega_{\epsilon i}s_{xy}} \ ,
\end{align}
\begin{align}
V''_{4} \simeq  &\sum_{\epsilon} Z_{\epsilon} \Pi^{'(eq)ij}_{A}(\epsilon \omega_{ q}) Z_{\epsilon }(-i) \frac{\partial T}{\partial t}\frac{\partial}{\partial T} \left\{ \begin{matrix} 1-f^{\epsilon}_{i q} \\ -f^{\epsilon}_{i q} \end{matrix} \right\}\frac{+1/2}{(\Omega_{\epsilon i}-\Omega^{*}_{\epsilon j})^2} \ e^{-i\Omega_{\epsilon i}s_{xy}} \ .
\end{align}

Finally, the leading order contributions in (\ref{V'''2}) comes from the second term of $V_2'''$.
It becomes
\begin{align}
V_{2}''' &\simeq \sum_{\epsilon} Z_{\epsilon } \Pi^{'(eq)ij}_{R}(\epsilon \omega_{\bold q}) Z_{\epsilon }(-i)  \frac{\partial T}{\partial t}\frac{\partial}{\partial T} \left\{ \begin{matrix} 1-f^{\epsilon}_{i q} \\ -f^{\epsilon}_{i q} \end{matrix} \right\} \frac{+1}{(\Omega_{\epsilon i}-\Omega^{*}_{\epsilon j})^2} \ e^{-i\Omega_{\epsilon i }s_{xy}} \n
&+\sum_{\epsilon} Z_{\epsilon} \Pi^{'(eq)ij}_{A}(\epsilon \omega_{ q}) Z_{\epsilon}(-i) \frac{\partial T}{\partial t}\frac{\partial}{\partial T} \left\{ \begin{matrix} 1-f^{\epsilon}_{i q} \\ -f^{\epsilon}_{i q} \end{matrix} \right\}\frac{-1}{(\Omega_{\epsilon i}-\Omega^{*}_{\epsilon j})^2} \ e^{-i\Omega_{\epsilon i}s_{xy}} \ . \label{leading of V'''2}
\end{align}

Summing up all these contributions (\ref{leading of V'14})$\sim$(\ref{leading of V'''2}), we get the final expression 
\begin{align}
\Delta G^{'ij}_{\wightman}&(x^0,y^0;{\bold q}) \n
 \simeq & \  \Theta (s_{xy}){\Bigg [}\sum_{\epsilon} Z_{\epsilon} \Pi^{'(eq)ij}_{R}(\epsilon \omega_{ q}) Z_{\epsilon}(-i)\Delta \left\{ \begin{matrix} 1-f^{\epsilon}_{j q} \\ -f^{\epsilon}_{j q} \end{matrix} \right\}\frac{i}{\Omega_{\epsilon i}-\Omega^{*}_{\epsilon j}} \ e^{-i\Omega_{\epsilon }s_{xy}}\notag \\
&\hspace{30pt}+\sum_{\epsilon} Z_{\epsilon } \Pi^{'(eq)ij}_{A}(\epsilon \omega_{ q}) Z_{\epsilon}(-i)\Delta \left\{ \begin{matrix} 1-f^{\epsilon}_{i q} \\ -f^{\epsilon}_{i q} \end{matrix} \right\}\frac{-i}{\Omega_{\epsilon i}-\Omega^{*}_{\epsilon j}} \ e^{-i\Omega_{\epsilon  }s_{xy}} {\Bigg ]} \notag \\ 
 +& \Theta (-s_{xy}){\Bigg [}\sum_{\epsilon} Z_{\epsilon} \Pi^{'(eq)ij}_{R}(\epsilon \omega_{ q}) Z_{\epsilon}(-i)\Delta \left\{ \begin{matrix} 1-f^{\epsilon}_{j q} \\ -f^{\epsilon}_{j q} \end{matrix} \right\}\frac{i}{\Omega_{\epsilon i}-\Omega^{*}_{\epsilon j}} \ e^{-i\Omega^{*}_{\epsilon }s_{xy}}\notag \\
&\hspace{40pt}+\sum_{\epsilon} Z_{\epsilon } \Pi^{'(eq)ij}_{A}(\epsilon \omega_{ q}) Z_{\epsilon}(-i)\Delta \left\{ \begin{matrix} 1-f^{\epsilon}_{i q} \\ -f^{\epsilon}_{i q} \end{matrix} \right\}\frac{-i}{\Omega_{\epsilon i}-\Omega^{*}_{\epsilon j}} \ e^{-i\Omega^{*}_{\epsilon  }s_{xy}} {\Bigg ]}  \ .
\label{DelG_{W}^{'}1} 
\end{align}

\subsection{Irrelevant contributions}
\label{Sec-irrelevant}
In this Appendix we will see that after being combined, 
the other terms in (\ref{deltaWightman9}) neglected in Appendices  \ref{Sec-FTrelevant}
become subdominant 
in the limit of $H \ll \Gamma$
and $(X_{xy}-t) \ltsim 1/\Gamma_{\ell \phi}$, $s_{xy} \ltsim 1/\Gamma_{\ell \phi}$.

The neglected terms are obtained by taking variations of the retarded or advanced-type
functions.  By using (\ref{defDeltildeG_{RA}^{d}App}) and (\ref{Useful22}), 
the first term in (\ref{deltaWightman9}) becomes
\begin{align}
-&\int du dv \ \Delta G^{d }_{R}(X_{xu};s_{xu};{\bold q}) \Pi^{'(eq)}_{R}(s_{uv};{\bold q}) G^{d (eq)}_{\wightman}(s_{vy};{\bold q}) \n
=&\int \frac{d q_0}{2\pi}(-i) \left\{ \begin{matrix} 1-f \\ -f \end{matrix} \right\}
{\Big [}
\frac{-i}{2}\left( \partial_{q_0} G^{d(eq)}_{R} \zeta_{R} G_{R}^{d(eq)}-G_{R}^{d(eq)} \zeta_{R} \partial_{q_0} G^{d(eq)}_{R} \right) \n 
& - (X_{xy}-t+\frac{s_{xy}}{2})\partial _t  G^{d(eq)}_{R}  -\frac{i}{2}\partial_{q_0} \partial _t  G^{d(eq)}_{R} 
{\Big ]}\Pi_{R}^{'(eq)}G_{\rho}^{d(eq)} e^{-iq_0 s_{xy}} \ . \label{1stJ1} 
\end{align}
The second term becomes
\begin{align}
-&\int  du dv \ 
 G^{d (eq)}_{R}(s_{xu};{\bold q}) \Delta \Pi^{'(eq)}_{R}(X_{uv}, s_{uv};{\bold q}) G^{d(eq)}_{\wightman}(s_{vy};{\bold q}) \n
=&\int \frac{d q_0}{2\pi}(-i) \left\{ \begin{matrix} 1-f \\ -f \end{matrix} \right\}
{\Big [} 
-  G^{d(eq)}_{R} \partial _t \Pi_{R}^{'(eq)}  G_{\rho}^{d(eq)} (X_{xy}-t) \n
&- \frac{i}{2}\left( \partial_{q_0} G^{d(eq)}_{R} \partial _t \Pi_{R}^{'(eq)}G_{\rho}^{d(eq)}- G^{d(eq)}_{R} \partial _t \Pi_{R}^{'(eq)} \partial_{q_0}G_{\rho}^{d(eq)} \right)
{\Big ]}
e^{-iq_0 s_{xy}} \n
+&\int \frac{d q_0}{2\pi}(-i) \partial_{q_0} \left\{ \begin{matrix} 1-f \\
-f \end{matrix} \right\}\frac{+i}{2} G^{d(eq)}_{R} 
\partial _t  \Pi_{R}^{'(eq)}G_{\rho}^{d(eq)} e^{-iq_0 s_{xy}} \ . \label{2ndJ1}
\end{align}
The 5th and 6th term become
\begin{align}
-\int & du dv \ 
 G^{d (eq)}_{\wightman}(s_{xu};{\bold q}) \Delta \Pi^{'(eq)}_{A}(X_{uv}; s_{uv};{\bold q}) G^{d(eq)}_{A}(s_{vy};{\bold q}) \n
=&\int \frac{d q_0}{2\pi}(-i) \left\{ \begin{matrix} 1-f \\ -f \end{matrix} \right\}
{\Big [} 
-  G^{d(eq)}_{\rho} \partial _t \Pi_{R}^{'(eq)}  G_{A}^{d(eq)} (X_{xy}-t) \n
&- \frac{i}{2}\left( \partial_{q_0} G^{d(eq)}_{\rho} \partial _t \Pi_{A}^{'(eq)}G_{A}^{d(eq)}- G^{d(eq)}_{\rho} \partial _t \Pi_{A}^{'(eq)} \partial_{q_0}G_{A}^{d(eq)} \right)
{\Big ]}
e^{-iq_0 s_{xy}} \n
+&\int \frac{d q_0}{2\pi}(-i) \partial_{q_0} \left\{ \begin{matrix} 1-f \\
-f \end{matrix} \right\}\frac{-i}{2} G^{d(eq)}_{\rho} 
\partial _t  \Pi_{A}^{'(eq)}G_{A}^{d(eq)} e^{-iq_0 s_{xy}} \label{5thJ1}
\end{align}
and
\begin{align}
-&\int du dv \  G^{d (eq)}_{\wightman}(s_{xu};{\bold q}) \Pi^{'(eq)}_{A}(s_{uv};{\bold q}) \Delta G^{d}_{A}(X_{vy};s_{vy};{\bold q}) \n
=&\int \frac{d q_0}{2\pi}(-i) \left\{ \begin{matrix} 1-f \\ -f \end{matrix} \right\}
G_{\rho}^{d(eq)}\Pi_{A}^{'(eq)} \n
&\times {\Big [}
\frac{-i}{2}\left( \partial_{q_0} G^{d(eq)}_{A} \zeta_{A} G_{A}^{d(eq)}-G_{A}^{d(eq)} \zeta_{A} \partial_{q_0} G^{d(eq)}_{A} \right) \n 
&\ \ \  - (X_{xy}-t-\frac{s_{xy}}{2})\partial _t  G^{d(eq)}_{A}  +\frac{i}{2}\partial_{q_0} \partial _t  G^{d(eq)}_{A} 
{\Big ]} e^{-iq_0 s_{xy}} \ . \label{6thJ1}
\end{align}
The 7th and 9th term become
\begin{align}
-&\int du dv \ \Delta G^{d }_{R}(X_{xu},s_{xu};{\bold q}) \Pi^{'(eq)}_{\wightman}(s_{uv};{\bold q}) G^{d}_{A}(s_{vy};{\bold q})  \\
=&\int \frac{d q_0}{2\pi}(-i) \left\{ \begin{matrix} 1-f \\ -f \end{matrix} \right\}  e^{-iq_0 s_{xy}} \n
&\times {\Big [}
\frac{-i}{2}\left( \partial_{q_0} G^{d(eq)}_{R} \zeta_{R} G_{R}^{d(eq)}-G_{R}^{d(eq)} \zeta_{R} \partial_{q_0} G^{d(eq)}_{R} \right) \n
&\ \ \ - \frac{i}{2}  \partial_{q_0} \partial _t G_{R}^{d(eq)}
-  (X_{xy}-t+\frac{s_{xy}}{2})\partial _t  G_{R}^{d(eq)} 
{\Big ]} \Pi_{\rho}^{'(eq)} G_{A}^{d(eq)} \label{7thJ1}
\end{align}
and
\begin{align}
-&\int du dv \
 G^{d (eq)}_{R}(s_{xu};{\bold q})  \Pi^{'(eq)}_{\wightman}( s_{uv};{\bold q}) 
 \Delta G^{d}_{A}(X_{vy},s_{vy};{\bold q})  \notag \\
=
&\int \frac{d q_0}{2\pi}(-i) \left\{ \begin{matrix} 1-f \\ -f \end{matrix} \right\}
G_{R}^{d(eq)}\Pi_{\rho}^{'(eq)}  e^{-iq_0 s_{xy}} \n
&\times {\Big [}
\frac{-i}{2} \left(
\partial_{q_0} G^{d(eq)}_{A} \zeta_{A} G_{A}^{d(eq)}-G_{A}^{d(eq)} \zeta_{A} \partial_{q_0} G^{d(eq)}_{A} \right) \n
&\ \ \  + \frac{i}{2}  
\partial_{q_0} \partial _t G_{A}^{d(eq)}  
 -  \partial _t  G^{d(eq)}_{A}
(X_{xy}-t-\frac{s_{xy}}{2})
{\Big ]} \ . \label{9thJ1}
\end{align}

First let's combine terms containing  $\partial_{q_0} \partial_{t} G_{R,A,\rho}$.
Such terms appear in (\ref{V'12}), (\ref{V''1}), (\ref{1stJ1}), (\ref{6thJ1}), (\ref{7thJ1}), 
(\ref{9thJ1}) and  are shown to be combined to become
\begin{align}
\int \frac{d q_0}{2\pi}\frac{-1}{2} \left\{ \begin{matrix} 1-f \\
-f \end{matrix} \right\} {\Big [}&\left( \partial_{q_0} \partial _t G_{R}^{d(eq)}\Pi_{R}^{'(eq)}G^{d(eq)}_{R}  -G^{d(eq)}_{R} \Pi_{R}^{'(eq)}\partial_{q_0} \partial _t G_{R}^{d(eq)}\right) \n
-& \left( \partial_{q_0} \partial _t G_{A}^{d(eq)}\Pi_{A}^{'(eq)}G^{d(eq)}_{A}  -G^{d(eq)}_{A} \Pi_{A}^{'(eq)}\partial_{q_0} \partial _t G_{A}^{d(eq)}\right) e^{-iq_0 s_{xy}} \ . 
\label{sowhat}
\end{align}
Each of them is of order ${\mathcal O}(HT/\Gamma^2)$ but their difference becomes of order
${\mathcal O}(H/T) \ll H/\Gamma$ due to the cancellation between contributions from the pole $\Omega_{i}$ and $\Omega_{j}$.

Next let us see terms containing $(X_{xy}-t)$. Such terms appear in (\ref{V'13}),(\ref{V'42}), (\ref{V''1}),(\ref{V''4}),(\ref{V'''1}),(\ref{V'''2}),(\ref{1stJ1}),(\ref{6thJ1}),(\ref{7thJ1}).
They are combined to become
\be
\left. \partial_{X} G^{'(eq)}_{\wightman}(X;s_{xy}) \right|_{X=t} (X_{xy}-t) \ . \label{off-diagonal-local-temperature-dependence}
\label{GXxy}
\ee
The derivative $\partial_X$ acting on $G$ is ${\mathcal O}(HT/\Gamma)$ and $(X_{xy}-t) = {\mathcal O}(1/T)$.
Hence it seems that the term is of the same order ${\mathcal O}(H/\Gamma)$ as the leading order terms
in Appendices  \ref{Sec-FTrelevant}.
But there is further suppression because the off-diagonal Wightman function in thermal equilibrium
itself vanishes as $\Delta M/\Gamma_{\ell\phi} \sim \Gamma/\Gamma_{\ell \phi}$ as shown in (\ref{ODWinTEvanish}).
Hence (\ref{GXxy}) is smaller than $H/\Gamma$ by a factor $\Gamma/\Gamma_{\ell \phi} \sim \Gamma/T$,
and can be neglected.

The terms containing $s_{xy}$ in (\ref{V'13}),(\ref{V''1}),(\ref{1stJ1}),(\ref{6thJ1}),(\ref{7thJ1}) are combined to be
\begin{align}
\int \frac{d q_0}{2\pi} \left\{ \begin{matrix} 1-f \\ -f \end{matrix} \right\}  {\Big [} & \left( \partial _t  G_{R}^{d(eq)} \Pi_{R}^{'(eq)} G_{R}^{d(eq)} -  G_{R}^{d(eq)} \Pi_{R}^{'(eq)} \partial _t G_{R}^{d(eq)} \right) \n
-&\left( \partial _t  G_{A}^{d(eq)} \Pi_{A}^{'(eq)} G_{A}^{d(eq)} -  G_{A}^{d(eq)} \Pi_{A}^{'(eq)} \partial _t G_{A}^{d(eq)} \right) {\Big ]} \frac{i s_{xy}}{2} e^{-iq_0 s_{xy}} \ . \label{Gsxy}
\end{align}
Each of them is of order ${\mathcal O}(H/\Gamma)$ but their difference becomes of negligible order
${\mathcal O}(H/T) \ll H/\Gamma$ due to the cancellation between contributions from the pole $\Omega_{i}$ and $\Omega_{j}$.

(\ref{V'''1}) and the third line of (\ref{2ndJ1}) and (\ref{5thJ1}), which contain $\partial_{t} \Pi_{R,A}$, are combined to be
\begin{align}
\int \frac{d q_0}{2\pi} \frac{-1}{2} \left\{ \begin{matrix} 1-f \\ -f \end{matrix} \right\}  {\Big [} & \left( \partial _{q_0} G_{R}^{d(eq)} \partial _t  \Pi_{R}^{'(eq)} G_{R}^{d(eq)} -  G_{R}^{d(eq)} \partial _{q_0}  \Pi_{R}^{'(eq)} \partial _t G_{R}^{d(eq)} \right) \n
-&\left( \partial _{q_0}  G_{A}^{d(eq)} \partial _t  \Pi_{A}^{'(eq)} G_{A}^{d(eq)} -  G_{A}^{d(eq)} \partial _t  \Pi_{A}^{'(eq)} \partial _{q_0} G_{A}^{d(eq)} \right) {\Big ]}  e^{-iq_0 s_{xy}} \ . \label{partial_t Pi}
\end{align}
Again, each of them is of order ${\mathcal O}(H/\Gamma)$ but their difference turns out to be of negligible order
${\mathcal O}(H/T)$ because of the similar cancellations.

Finally the terms containing $\zeta $ in (\ref{V''1}),(\ref{1stJ1}),(\ref{6thJ1}),(\ref{7thJ1}),(\ref{9thJ1}), and (\ref{V'11}) are combined to be
\begin{align}
\int \frac{d q_0}{2\pi}&\frac{-1}{2} \left\{ \begin{matrix} 1-f \\ -f \end{matrix} \right\} e^{-iq_0 s_{xy}}  \n
\times  {\Big [} &
\left( \partial_{q_0} G^{d(eq)}_{R} \zeta_{R} G_{R}^{d(eq)}-G_{R}^{d(eq)} \zeta_{R} \partial_{q_0} G^{d(eq)}_{R} \right) \Pi_{R}^{'(eq)} G_{R}^{d(eq)} \n
-&G_{R}^{d(eq)} \Pi_{R}^{'(eq)} \left( G_{R}^{d(eq)} \zeta_{R} \partial_{q_0} G^{d(eq)}_{R} -\partial_{q_0} G^{d(eq)}_{R} \zeta_{R} G_{R}^{d(eq)} \right)  \n 
-&\left( \partial_{q_0} G^{d(eq)}_{A} \zeta_{A} G_{A}^{d(eq)}-G_{A}^{d(eq)} \zeta_{A} \partial_{q_0} G^{d(eq)}_{A} \right) \Pi_{A}^{'(eq)} G_{A}^{d(eq)} \n 
+&G_{A}^{d(eq)} \Pi_{A}^{'(eq)} \left( G_{A}^{d(eq)} \zeta_{A} \partial_{q_0} G^{d(eq)}_{A} -\partial_{q_0} G^{d(eq)}_{A} \zeta_{A} G_{A}^{d(eq)} \right) {\Big ]} \ . \label{GzetaGvanish}
\end{align}
Each of them is of order ${\mathcal O}(HT/\Gamma^2)$ but each line becomes of order ${\mathcal O}(H/\Gamma)$.
Moreover, because of the cancellation between the first and second line, or the third and fourth line,
it turns out to be of order ${\mathcal O}(H/T)$ so that they are negligible again.

\section{Another derivation of $\Delta G^{\prime}_{\gtrless}$ }
\label{AnotherDerivation}
In this appendix we give another, quick and heuristic, derivation of $\Delta G^{\prime}_{\gtrless}$.
The derivation use some of the results justified in the systematic derivation adopted in 
the paper.
First we assume that the deviations of the Wightman functions from the thermal value at time $t$
is given by the following form:
\begin{align}
\Delta G^{ij}_{\wightman}(X_{xy},s_{xy}=0;{\bold q})
=&\sum_{\epsilon}\frac{\epsilon}{2\omega_{q}}(\Slash{q}_{\epsilon}+M){\big \{ } \Delta {\mathcal A}^{ij}_{\wightman}+\Delta \dot{\mathcal A}^{ij}_{\wightman} (X_{xy}-t)   \n
&\hspace{75pt}+\epsilon \Delta {\mathcal B}^{ij}_{\wightman}+\epsilon \Delta \dot{\mathcal B}^{ij}_{\wightman} (X_{xy}-t) +\cdots{\big \}} \ .
\label{defAB} 
\end{align}
$\Delta {\mathcal A}, \Delta {\mathcal B}$ are terms which remain at $X_{xy}=t$, and $\mathcal A$ and $\mathcal B$ are introduced
to represent $\epsilon$ dependence of the sum.
Here we take the leading order with respect to $(X_{xy}-t) H \sim H/\Gamma$.
Both of $\Delta {\mathcal A}$ and $\Delta {\mathcal B}$ have no spinor indices.
\subsection{Solving KB equation for $G_\gtrless^{(eq)ij}$}
For the diagonal component, (\ref{DelG_{W}^{d}a}) shows that
\begin{align}
\Delta{\mathcal A}^{d}_{\wightman}=0\ ,\ \ \Delta{\mathcal B}^{d}_{\wightman}=\frac{d_t f^{(eq)}_{q}}{\Gamma_{q}} =\Delta \left\{ \begin{matrix} 1-f_{q} \\ -f_{q} \end{matrix} \right\} \ , \label{DelAB^{d}}
\end{align}
\begin{align}
\Delta \dot{\mathcal A}^{d}_{\wightman}=0\ ,\ \ \Delta \dot{\mathcal B}^{d}_{\wightman}=- d_t f^{(eq)}_{q} =-\Gamma_{q} \Delta{\mathcal B}^{d}_{\wightman} \ . \label{dotDelAB^{d}}
\end{align}

The Wightman functions in the thermal equilibrium at $t$ are given in 
(\ref{G_{W}^{d(eq)}1}) for the diagonal component and (\ref{G_{W}^{'(eq)}2}) for the
off-diagonal component. Hence they are similarly written in terms of ${\mathcal A}$ and ${\mathcal B}$ as
\begin{align}
G^{(eq)ij}_{\wightman}(X_{xy},s_{xy}=0;{\bold q})=\sum_{\epsilon}\frac{\epsilon}{2\omega_{q}}(\Slash{q}_{\epsilon}+M){\big \{ }&{\mathcal A}^{(eq)ij}_{\wightman}+\epsilon {\mathcal B}^{(eq)ij}_{\wightman}{\big \}} \notag
\end{align}
where
\begin{align}
{\mathcal A}^{d(eq)}_{>}=-{\mathcal A}^{d(eq)}_{<}=\frac{1}{2}\ ,\ \ {\mathcal B}^{d(eq)}_{>}={\mathcal B}^{d(eq)}_{<}=\frac{1}{2}(1-2f^{(eq)}_{q}) \label{AB^{d(eq)}}
\end{align}
for the diagonal component and
\begin{align}
{\mathcal A}^{'(eq)}_{\wightman}={\mathcal B}^{'(eq)}_{\wightman}=0 \label{AB^{'(eq)}}
\end{align}
for the off-diagonal component.
\subsection{KB equation for $\Delta G_\gtrless^{ij}$}
In the following we obtain the deviation of 
the off-diagonal component of the Wightman functions $\Delta{\mathcal A}^{'}_{\wightman}$
directly  by solving the KB equations using the above information.
The KB equations for the off-diagonal Wightman functions are given by
\begin{align}
i \gamma^0 &\partial_{x^0} G^{ij}_{\wightman}(x^0 ,y^0;{\bold q})  -{\Big \{ }\frac{{\boldsymbol \gamma}\cdot {\bold q}}{a(x^0)}+M_i {\Big \}}G^{ij}_{\wightman}(x^0 ,y^0;{\bold q}) \notag \\
&=\int dz^0 \ \Pi^{ik}_{R}(x^0 ,z^0;{\bold q}) G^{kj}_{\wightman}(z^0,y^0;{\bold q}) + \int dz^0 \ \Pi^{ik}_{\wightman}(x^0 ,z^0;{\bold q}) G^{kj}_{A}(z^0,y^0;{\bold q}) 
\end{align}
or
\begin{align}
-i \gamma^0 &\partial_{y^0} G^{ij}_{\wightman}(x^0 ,y^0;{\bold q}) -G^{ij}_{\wightman}(x^0 ,y^0;{\bold q}){\Big \{ }\frac{{\boldsymbol \gamma}\cdot {\bold q}}{a(y^0)}+M_{j}{\Big \}} \notag \\
&=\int dz^0 \ G^{ik}_{\wightman}(x^0 ,z^0;{\bold q}) \Pi^{kj}_{A}(z^0,y^0;{\bold q}) + \int dz^0 \ G^{ik}_{R}(x^0 ,z^0;{\bold q}) \Pi^{kj}_{\wightman}(z^0,y^0;{\bold q})
\end{align}
Setting $x^0 =y^0 =t$ and take a difference of these two equations.
Summing over the spinor indices, we have
\begin{align}
i&\partial_{X} G^{ij}_{\wightman V^0}(X;{\bold q}){\bigg |}_{ X=t} -(M_i -M_j ) G^{ij}_{\wightman S}(X=t;{\bold q})\notag \\
&=\int dz^0 \ \frac{1}{4}{\rm tr}\left\{ \Pi^{ik}_{R}(t ,z^0;{\bold q}) G^{kj}_{\wightman}(z^0,t;{\bold q})-G^{ik}_{\wightman}(t ,z^0;{\bold q}) \Pi^{kj}_{A}(z^0,t;{\bold q})\right\} \n
&+ \int dz^0 \ \frac{1}{4}{\rm tr}\left\{ \Pi^{ik}_{\wightman}(t ,z^0;{\bold q}) G^{kj}_{A}(z^0,t;{\bold q})- G^{ik}_{R}(t ,z^0;{\bold q}) \Pi^{kj}_{\wightman}(z^0,t;{\bold q})\right\} \ .
\label{trKB} 
\end{align}
On the other hand, multiplying $\gamma_0$ and then summing over the spinor indices, we have
\begin{align}
i&\partial_{X} G^{ij}_{\wightman S}(X;{\bold q}){\bigg |}_{X=t} -\frac{2}{4}{\rm tr}\left\{ \gamma^0 \frac{{\boldsymbol \gamma}\cdot {\bold q}}{a(t)} G^{ij}_{\wightman}(X=t;{\bold q})\right\}-(M_i -M_j )G^{ij}_{\wightman V^0}(X=t;{\bold q}) \notag \\
&=\int dz^0 \ \frac{1}{4}{\rm tr}\left\{ \gamma^{0}\Pi^{ik}_{R}(t,z^0;{\bold q}) G^{kj}_{\wightman}(z^0,t;{\bold q})-\gamma^{0} G^{ik}_{\wightman}(t,z^0;{\bold q}) \Pi^{kj}_{A}(z^0,t;{\bold q})\right\} 
\n
&+ \int dz^0 \ \frac{1}{4}{\rm tr}\left\{ \gamma^{0} \Pi^{ik}_{\wightman}(t,z^0;{\bold q}) G^{kj}_{A}(z^0,t;{\bold q})-\gamma^{0} G^{ik}_{R}(t ,z^0;{\bold q}) \Pi^{kj}_{\wightman}(z^0,t;{\bold q})\right\} 
\label{trgamma^{0}KB} 
\end{align}
where
\begin{align}
G^{ij}_{\wightman S}(X;{\bold q}) \equiv \frac{1}{4} {\rm tr}\{ G^{ij}_{\wightman}(X,s=0 ;{\bold q}) \} \ ,\ \ G^{ij}_{\wightman V^{\mu}}(X;{\bold q}) \equiv \frac{1}{4} {\rm tr}\{ \gamma^{\mu} G^{ij}_{\wightman}(X,s=0 ;{\bold q}) \} \ .
\end{align}
We are now interested in the deviation from the thermal values at time $t$.
The equations (\ref{trKB}) and (\ref{trgamma^{0}KB}) are rewritten as
\begin{align}
i&\Delta \dot{\mathcal A}^{ij}_{\wightman} -(M_i -M_j ) \frac{M}{\omega_{q}}\Delta {\mathcal B}^{ij}_{\wightman }\notag \\
&=\int dz^0 \ \frac{1}{4}{\rm tr}\left\{ \Pi^{(eq)ik}_{R}(t ,z^0;{\bold q}) \Delta {\mathcal G}^{kj}_{\wightman}(z^0,t;{\bold q})-\Delta {\mathcal G}^{ik}_{\wightman}(t ,z^0;{\bold q}) \Pi^{(eq)kj}_{A}(z^0,t;{\bold q})\right\} \notag \\
&= \sum_{\epsilon} \frac{\epsilon}{2\omega_{q}} \eta^{\mu \nu}q_{\epsilon \nu}\left\{ \Pi^{(eq)ik}_{R V^{\mu}}(\epsilon \omega_{q},{\bold q}) (\Delta {\mathcal A}^{kj}_{\wightman}+\epsilon \Delta {\mathcal B}^{kj}_{\wightman})-\Pi^{(eq)kj}_{A V^{\mu}}(\epsilon \omega_{q},{\bold q}) (\Delta {\mathcal A}^{ik}_{\wightman}+\epsilon \Delta {\mathcal B}^{ik}_{\wightman}) \right\}\notag \\
&= \frac{1}{2\omega_{q}} \left\{ q\cdot \Pi^{(eq)ik}_{\rho}( \omega_{q},{\bold q}) \Delta {\mathcal A}^{kj}_{\wightman}+q\cdot\Pi^{(eq)kj}_{\rho }( \omega_{q},{\bold q}) \Delta {\mathcal A}^{ik}_{\wightman} \right\}\notag \\
&\hspace{30pt} + \frac{1}{\omega_{q}} \left\{ q\cdot\Pi^{(eq)ik}_{h }( \omega_{q},{\bold q}) \Delta {\mathcal B}^{kj}_{\wightman}-q\cdot\Pi^{(eq)kj}_{h }(\omega_{q},{\bold q}) \Delta {\mathcal B}^{ik}_{\wightman} \right\} 
\label{Aequation}
\end{align}
and
\begin{align}
i&\frac{M}{\omega_{q}}\Delta \dot{\mathcal B}^{ij}_{\wightman} -\frac{2}{4}{\rm tr}\left\{ \gamma^0 \frac{{\boldsymbol \gamma}\cdot {\bold q}}{a(t)} \Delta G^{ij}_{\wightman}(X=t;{\bold q})\right\}-(M_i -M_j )\Delta {\mathcal A}^{ij}_{\wightman}\notag \\
&=\int dz^0 \ \frac{1}{4}{\rm tr}\left\{ \gamma^{0} \Pi^{(eq)ik}_{R}(t,z^0;{\bold q}) \Delta {\mathcal G}^{kj}_{\wightman}(z^0,t;{\bold q})-\gamma^{0} \Delta{\mathcal G}^{ik}_{\wightman}(t,z^0;{\bold q}) \Pi^{(eq)kj}_{A}(z^0,t;{\bold q})\right\} \notag \\
&= \sum_{\epsilon} \frac{\epsilon M}{2\omega_{q}} \left\{ \Pi^{(eq)ik}_{R V^{0}}(\epsilon \omega_{q},{\bold q}) (\Delta {\mathcal A}^{kj}_{\wightman}+\epsilon \Delta {\mathcal B}^{kj}_{\wightman})-\Pi^{(eq)kj}_{A V^{0}}(\epsilon \omega_{q},{\bold q}) (\Delta {\mathcal A}^{ik}_{\wightman}+\epsilon \Delta {\mathcal B}^{ik}_{\wightman}) \right\}\notag \\
&= \frac{M}{2\omega_{q}} \left\{ \Pi^{(eq)ik}_{\rho V^{0}}( \omega_{q},{\bold q}) \Delta {\mathcal B}^{kj}_{\wightman}+\Pi^{(eq)kj}_{\rho  V^{0}}( \omega_{q},{\bold q}) \Delta {\mathcal B}^{ik}_{\wightman} \right\}\notag \\
&\hspace{30pt} + \frac{M}{\omega_{q}} \left\{ \Pi^{(eq)ik}_{h  V^{0}}( \omega_{q},{\bold q}) \Delta {\mathcal A}^{kj}_{\wightman}- \Pi^{(eq)kj}_{h  V^{0}}(\omega_{q},{\bold q}) \Delta {\mathcal A}^{ik}_{\wightman} \right\} \ .
\label{Bequation}
\end{align}
The first line of each equation is nothing but the l.h.s. of 
(\ref{trKB}) and (\ref{trgamma^{0}KB}) written in term of the definitions in 
(\ref{defAB}). The second lines of them are the r.h.s. of 
(\ref{trKB}) and (\ref{trgamma^{0}KB}) in which small deviations from the thermal values
are considered. The terms represent the dominant contributions and 
terms like $\Delta \Pi _{R/A}$, $\Delta \Pi _{\wightman}$ and $\Delta G_{R/A}$ are dropped.
This is justified because of the large damping factor $\Pi(t,z^0)\sim e^{-|t-z^0|\Gamma_{\ell \phi}/2}$
of the self-energies. 
In the second equalities , we performed time integrations and 
taking the trace with respect to indices of spinor.
In the third equalities, we used 
 (\ref{sign-dependence of pi_{rho}}) and (\ref{sign-dependence of pi_{h}}).
\subsection{Diagonal component $\Delta G_\gtrless^{d ii}$}
Let us first look at the diagonal component.
We use the simple expression of the self-energy (\ref{SEsimplified})
by neglecting the medium effects and in the weak coupling limit. 
Then we have
\be
\Pi^{d(eq)}_{\rho V^{0}}( \omega_{q},{\bold q})
&=& (\omega_{q}/M^2)  q\cdot \Pi^{(eq)ik}_{\rho}( \omega_{q},{\bold q}) \n
&=& -i(\omega_{q}/M)  \Gamma \ =\ -i(\omega_{q}^2/M^2)  \Gamma_{q} \ .
\ee
With this relation, (\ref{Aequation}) and (\ref{Bequation}) are simplified to be
\begin{align}
i&\Delta \dot{\mathcal A}^{d}_{\wightman} = -i \Gamma_{q}  \Delta {\mathcal A}^{d}_{\wightman}\label{KBDelA^{d}} 
\end{align}
and
\begin{align}
i&\frac{M}{\omega_{q}}\Delta \dot{\mathcal B}^{d}_{\wightman} -\frac{2}{4}{\rm tr}\left\{ \gamma^0 \frac{{\boldsymbol \gamma}\cdot {\bold q}}{a(t)} \Delta G^{d}_{\wightman}(X=t;{\bold q})\right\} = -i\frac{\omega_{q}}{M}\Gamma_{q}   \Delta {\mathcal B}^{d}_{\wightman}  \ .
\label{KBDelB^{d}}  
\end{align}
(\ref{dotDelAB^{d}}) indeed satisfies  (\ref{KBDelA^{d}}).
The second term of the l.h.s. of (\ref{KBDelB^{d}})
vanishes in the leading order approximation (\ref{defAB}), but using
 the next to leading order approximation of $\Delta G^{d}$,
 the second term becomes  
 $-i\Gamma_{q} (|{\bold q}|^2 /a^2)/(M \omega_{q})$. 
 Then  eq. (\ref{KBDelB^{d}}) is satisfied.
 
\subsection{Off-diagonal component $\Delta G_\gtrless^{'ij}$}
Then we study the off-diagonal component.
Using $\Pi (\omega_q ,{\bold q}) \propto \Slash{q}$,
 (\ref{Aequation}) and (\ref{Bequation}) become
 \begin{align}
i&\Delta \dot{\mathcal A}^{'ij}_{\wightman} -(M_i -M_j ) \frac{M}{\omega_{q}}\Delta {\mathcal B}^{'ij}_{\wightman }\notag \\
&= -i\frac{M}{2\omega_{q}} \left\{ \Gamma_{i} +\Gamma_{j} \right\}\Delta {\mathcal A}^{'ij}_{\wightman} 
+ \frac{1}{2\omega_{q}} (q\cdot \Pi^{'(eq)ij}_{\rho}( \omega_{q},{\bold q}))\left\{  \Delta {\mathcal A}^{d jj}_{\wightman}+\Delta {\mathcal A}^{dii}_{\wightman} \right\}\notag \\
& + \frac{1}{\omega_{q}} (q\cdot\Pi^{'(eq)ij}_{h }( \omega_{q},{\bold q}))\left\{  \Delta {\mathcal B}^{djj}_{\wightman}- \Delta {\mathcal B}^{dii}_{\wightman} \right\}
\label{KBDelA^{'}}  
\end{align}
and
\begin{align}
i&\frac{M}{\omega_{q}}\Delta \dot{\mathcal B}^{'ij}_{\wightman} -\frac{2}{4}{\rm tr}\left\{ \gamma^0 \frac{{\boldsymbol \gamma}\cdot {\bold q}}{a(t)} \Delta G^{'ij}_{\wightman}(X=t;{\bold q})\right\}-(M_i -M_j )\Delta {\mathcal A}^{'ij}_{\wightman}\notag \\
&= -i\frac{1}{2} \left\{ \Gamma_{i} +\Gamma_{j} \right\}\Delta {\mathcal B}^{'ij}_{\wightman} 
+ \frac{1}{2M} (q\cdot \Pi^{'(eq)ij}_{\rho}( \omega_{q},{\bold q}))\left\{  \Delta {\mathcal B}^{djj}_{\wightman}+ \Delta {\mathcal B}^{dii}_{\wightman} \right\}\notag \\
& + \frac{1}{M} (q\cdot \Pi^{'(eq)ij}_{h }( \omega_{q},{\bold q}))\left\{  \Delta {\mathcal A}^{djj}_{\wightman}- \Delta {\mathcal A}^{dii}_{\wightman} \right\}
\label{KBDelB^{'}}  \ .
\end{align}
Here we absorbed the real part of the self-energy 
$\Pi_{h}$ into the mass term in the l.h.s. by the mass renormalisation.

For the off-diagonal component, we can expect that
\begin{align}
\Delta \dot{\mathcal A}^{'}_{\wightman}=\Delta \dot{\mathcal B}^{'}_{\wightman}=0 \ . \label{DeldotAB^{'}}
\end{align}
This comes from the relation (\ref{ODWinTEvanish}) or equivalently (\ref{AB^{'(eq)}}).
Since it vanishes in the  thermal equilibrium, its variation due to the change of the 
local temperature is also expected to vanish in the leading order approximation.
Indeed it  is confirmed by using (\ref{off-diagonal-local-temperature-dependence}).
On the contrary, since the equilibrium diagonal Wightman function survives in the same limit,
its variation (or $\Delta \dot{\mathcal B}^{d}\ne 0$) does not vanish either.
Furthermore, the second term in the l.h.s. of (\ref{KBDelB^{'}}) is also expected to
give no leading contribution like the first term $\Delta \dot{\mathcal B}^{'}_{\wightman}$.

Using the above arguments,  the equations (\ref{KBDelA^{'}}) and (\ref{KBDelB^{'}})
are simplified as equations to determine  
$\Delta {\mathcal A}^{'}_{\wightman}$ and $\Delta {\mathcal B}^{'}_{\wightman}$
in terms of 
$\Delta {\mathcal A}^{d}_{\wightman}$ and $\Delta {\mathcal B}^{d}_{\wightman}$,
and they are solved as
\begin{align}
\left( \begin{matrix} \Delta {\mathcal A}^{'ij}_{\wightman} \\ \Delta {\mathcal B}^{'ij}_{\wightman} \end{matrix} \right)=&\frac{-1}{(\Delta M_{ij}^2)^2+(M\Gamma_{i}+M\Gamma_{j})^2} \left( \begin{matrix} i(M\Gamma_{i}+M\Gamma_{j}) & \Delta M_{ij}^{2} \\ \Delta M_{ij}^{2} & i(M\Gamma_{i}+M\Gamma_{j}) \end{matrix} \right) \notag \\
& \times \left( \begin{matrix} 2(q\cdot\Pi^{'(eq)ij}_{h }( \omega_{q},{\bold q}))\left\{  \Delta {\mathcal B}^{djj}_{\wightman}- \Delta {\mathcal B}^{dii}_{\wightman} \right\} \\ (q\cdot \Pi^{'(eq)ij}_{\rho}( \omega_{q},{\bold q}))\left\{  \Delta {\mathcal B}^{djj}_{\wightman}+ \Delta {\mathcal B}^{dii}_{\wightman} \right\} \end{matrix} \right) \ .\label{DelAB^{'}}
\end{align}
The regulator $M_{i}\Gamma_{i}+M_{j}\Gamma_{j}$ controls the 
enhancement of the solutions for $\Delta G_\gtrless^{\prime}$.
This expression corresponds to (\ref{DelG_{W}^{'}2}).
\subsection{$\Delta G_\gtrless^{'}$ based on a wrong assumption $G_\gtrless^{'} \neq 0$}
Finally in this appendix, we discuss how we will obtain an erroneous answer with 
the regulator of the type $M_{i}\Gamma_{i}-M_{j}\Gamma_{j}$.
Let us assume (which turns out to be wrong) that the off-diagonal component did not 
vanish and is given by
\begin{align}
i&\Delta \dot{\mathcal A}^{''}_{\wightman} = -i \overline{\Gamma}_{q}  \Delta {\mathcal A}^{''}_{\wightman} \label{erroneousA}
\end{align}
and
\begin{align}
i&\frac{M}{\omega_{q}}\Delta \dot{\mathcal B}^{''}_{\wightman} -\frac{2}{4}{\rm tr}\left\{ \gamma^0 \frac{{\boldsymbol \gamma}\cdot {\bold q}}{a(t)} \Delta G^{''}_{\wightman}(X=t;{\bold q})\right\} = -i\frac{\omega_{q}}{M}\overline{\Gamma}_{q}   \Delta {\mathcal B}^{''}_{\wightman} \ . \label{erroneousB}
\end{align}
Here $\overline{\Gamma}_{q}= \overline{\Gamma}M/\omega_{q}$
is of the same order as 
 $\Gamma_{iq}=\Gamma_{i}M/\omega_{q}$ ($i=1,2$).
These are similar to  the correct relations for the diagonal components, (\ref{KBDelA^{d}}) and (\ref{KBDelB^{d}}).

The above equations 
(\ref{erroneousA}) and (\ref{erroneousB}) 
are based on a correct-looking assumption that
the deviations of the off-diagonal Wightman functions out of equilibrium are
obtained by taking a variation of the the equilibrium value with respect to the local temperature.
In other words, it is assumed that 
there exists an  "off-diagonal distribution function $f_{q}^{'(eq)}$''
which does not vanish at  $s_{xy}=x-y=0$ 
and its deviation from the equilibrium value satisfies the relation $\Delta f^{'}_{q}=-d_{t}f^{'(eq)}_{q}/\overline{\Gamma}_{q}$. (As a matter of fact, such a function does not exist.)  

Under such incorrect assumptions, 
 additional terms change the l.h.s of (\ref{KBDelA^{'}}) and (\ref{KBDelB^{'}}),
and the regulator is modified to be
\begin{align}
\Gamma_i + \Gamma_j \to \Gamma_i + \Gamma_j -2\overline{\Gamma} \sim \Gamma_i -\Gamma_j \ .
\end{align}
This is the way we could obtain an erroneous enhancement factor.

\section{Effects of backreactoin}
\label{Sec-App-calwashout}
Backreactions of the generated lepton asymmetry to the RH Wightman functions 
are given by inserting  (\ref{chempi}) into 
(\ref{intKBeqG_{W}^{d}}) and (\ref{intKBeqG_{W}^{'}1}):
\begin{align}
\Delta_\mu G_{\wightman}^{dii}(s_{xy};{\bold q})
=&-\int \frac{dq_0}{2\pi} e^{-iq_{0}(x^0 -y^0)}G_{R/A}^{d(eq)ii}(q)
\Delta_\mu \Pi^{d ii}_{R/A}(q)G_{R/A}^{d(eq)ii}(q) \ , \label{G_{W}^{d}(mu)}
\end{align}
\begin{align}
\Delta_\mu G_{\wightman}^{'ij}(s_{xy};{\bold q})
=&-\int \frac{dq_0}{2\pi} e^{-iq_{0}(x^0 -y^0)}G_{R}^{'(eq)ij}(q)
\Delta_\mu \Pi^{djj}_{\wightman}(q)G_{A}^{d(eq)jj}(q)  \notag \\
&-\int \frac{dq_0}{2\pi} e^{-iq_{0}(x^0 -y^0)}G_{R}^{d(eq)ii}(q)
\Delta_\mu \Pi^{dii}_{\wightman}(q)G_{A}^{d(eq)ij}(q) \label{G_{W}^{'}(mu)} \\
&-\int \frac{dq_0}{2\pi} e^{-iq_{0}(x^0 -y^0)}G_{R}^{d(eq)ii}(q)
\Delta_\mu \Pi^{'ij}_{\wightman}(q)G_{A}^{d(eq)jj}(q) \ . 
\end{align}
Here by using (\ref{Pi_{W}})
$\Delta_\mu \Pi_{\wightman}(q)$ becomes
\begin{align}
\Delta_\mu \Pi^{ij}_{\wightman}(q) =&\int ds\ e^{+iq_0 s} 
\Delta_\mu \Pi^{ij}_{\wightman}(t;s;{\bold q}) 
 \n
=& \sum_{\epsilon_{\ell},\epsilon_{\phi}}\int \frac{d^3 p}{(2\pi)^3 2\omega_{ p}}\frac{d^3 k}{(2\pi)^3 2\omega_{ k}}(-g_w)\ (2\pi)^3 \delta ^3(q-p-k)  \notag \\
&\hspace{35pt}\times \frac{\Gamma_{\ell \phi}}{(q_{0}-\epsilon_{\ell}\omega_{ p}-\epsilon_{\phi}\omega_{ k})^2 +\Gamma_{\ell \phi}^{2}/4} \ \Slash{p}_{\epsilon _{\ell}} \notag\\
&\hspace{35pt}\times {\bigg [}\frac{1}{2}\left\{ (h^{\dag}h)_{ij}\Delta {\mathcal D}^{\epsilon_{\ell} \epsilon_{\phi}}_{\wightman (p,k)}+(h^{\dag}h)^{*}_{ij}\Delta \overline{{\mathcal D}}^{\epsilon_{\ell} \epsilon_{\phi}}_{\wightman (p,k)} \right\} \notag \\
&\hspace{45pt}+\frac{1}{2}\gamma_{5} \left\{ (h^{\dag}h)_{ij}\Delta {\mathcal D}^{\epsilon_{\ell} \epsilon_{\phi}}_{\wightman (p,k)}-(h^{\dag}h)^{*}_{ij}\Delta \overline{{\mathcal D}}^{\epsilon_{\ell} \epsilon_{\phi}}_{\wightman (p,k)} \right\} {\bigg ]}
\label{Pi(mu)}
\end{align}
where
\begin{align}
\overline{{\mathcal D}}_{\wightman (p,k)}^{\epsilon_{\ell} \epsilon_{\phi}} \equiv \left\{ \begin{matrix} \epsilon_{\ell}\epsilon_{\phi}(1-f_{\overline{\ell} p}^{\epsilon_{\ell}})(1+f_{\overline{\phi} k}^{\epsilon_{\phi}})  \\ \epsilon_{\ell}\epsilon_{\phi}(-f_{\overline{\ell} p}^{\epsilon_{\ell}})(+f_{\overline{\phi} k}^{\epsilon_{\phi}})  \end{matrix} \right. = {\mathcal D}_{\wightmaninverse (p,k)}^{-\epsilon_{\ell} -\epsilon_{\phi}}
\end{align}
and
\begin{align}
\Delta {\mathcal D}^{\epsilon_{\ell} \epsilon_{\phi}}_{\wightman (p,k)}\equiv {\mathcal D}^{\epsilon_{\ell} \epsilon_{\phi}}_{\wightman (p,k)} -{\mathcal D}^{\epsilon_{\ell} \epsilon_{\phi} (eq)}_{\wightman (p,k)}\ , \ \ \ \Delta \overline{\mathcal D}^{\epsilon_{\ell} \epsilon_{\phi}}_{\wightman (p,k)}\equiv \overline{\mathcal D}^{\epsilon_{\ell} \epsilon_{\phi}}_{\wightman (p,k)} -{\mathcal D}^{\epsilon_{\ell} \epsilon_{\phi} (eq)}_{\wightman (p,k)} \ .
\end{align}
In the leading order approximation of small deviations from the thermal equalibrium,
the square bracket in  (\ref{Pi(mu)}) becomes
\begin{align}
\frac{1}{2}{\bigg [}i\Im (h^{\dag}h)_{ij}+\Re (h^{\dag}h)_{ij} \gamma_{5}{\bigg ]}\left\{ \Delta {\mathcal D}^{\epsilon_{\ell} \epsilon_{\phi}}_{\wightman (p,k)}-\Delta \overline{{\mathcal D}}^{\epsilon_{\ell} \epsilon_{\phi}}_{\wightman (p,k)} \right\} \ .
\end{align}

Let us write (\ref{G_{W}^{d}(mu)}) and (\ref{G_{W}^{'}(mu)}) as
\begin{align}
\Delta_\mu G^{ij}_{\wightman}(s;{\bold q})=
\sum_{\epsilon}
\Delta_\mu Z^{ij}_{\wightman \epsilon}({\bold q}) \  e^{-i\Omega_{\epsilon}s} \ .
\end{align}
We have adopted the approximation $e^{-i\Omega_i s}\simeq e^{-i\Omega_j s}\simeq e^{-i\Omega s}$ because we are especially interested in the case $1/\Gamma_{\ell \phi}\gtsim s>0$. 
Also we have dropped rapidly damping contributions $\sim e^{-s\Gamma_{\ell\phi}/2}$ from the pole of self-energy which are the higher order in $\Gamma /\Gamma_{\ell\phi}$.
Plugging this form into (\ref{dndt2}) with the equilibrium value of $\pi _{\wightman}$ and performing the time integration, we get
\begin{align}
&\left.\frac{d n_{L}}{dt}+3H n_{L}\right|_{\Delta _{\mu} G_{\wightman}} \notag \\
&= \int\frac{d^3 q}{(2\pi)^3}\sum_{\epsilon}\int \frac{dq_0}{2\pi} 2\Re {\Bigg [}(h^{\dag}h)_{ji} \frac{i}{q_0 -\Omega_{\epsilon}}{\bigg [}{\rm tr}{\Big \{ }P_{\rm R} 
\Delta_\mu Z_{<\epsilon}^{ij}({\bold q} )P_{\rm L}\pi^{(eq)}_{>}(q_0 ,{\bold q}){\Big \} }\notag \\
&\hspace{130pt}-{\rm tr}{\Big \{ }P_{\rm R} 
\Delta_\mu Z_{>\epsilon}^{ij}({\bold q} )P_{\rm L}\pi^{(eq)}_{<}(q_0 ,{\bold q}){\Big \} }{\bigg ]}{\Bigg ]} \ . \label{dndtG(mu)}
\end{align}
Focusing on $q_0$ integration and using the KMS condition for the self-energy,
we get
\begin{align}
\int & \frac{dq_0}{2\pi} \frac{i}{q_0 -\Omega_{\epsilon}} \pi^{(eq)}_{\wightman}(q_0 ,{\bold q}) =\int \frac{dq_0}{2\pi} \frac{i}{q_0 -\Omega_{\epsilon}} (-i)\left\{ \begin{matrix} 1-f(q_0) \\ -f(q_0) \end{matrix} \right\}\pi^{(eq)}_{\rho}(q_0 ,{\bold q}) \n
\simeq & \sum_{\epsilon_{\ell},\epsilon_{\phi}}\int \frac{d^3 p}{(2\pi)^3 2\omega_{ p}}\frac{d^3 k}{(2\pi)^3 2\omega_{ k}}(-g_w)\ (2\pi)^3 \delta ^3(q-p-k) \times (-1) \n
&\times \frac{- {\mathcal D}_{\rho (p,k)}^{\epsilon_{\ell} \epsilon_{\phi}(eq)} }{\epsilon \omega_{q}-\epsilon_{\ell}\omega_{ p}-\epsilon_{\phi}\omega_{ k}-i\Gamma_{\ell \phi}/2} \ \Slash{p}_{\epsilon _{\ell}} (-i)\left. \left\{ \begin{matrix} 1-f(q_0) \\ -f(q_0) \end{matrix} \right\} \right|_{q_0 =\epsilon_{\ell}\omega_{ p}+\epsilon_{\phi}\omega_{ k}+i\Gamma_{\ell \phi}/2} \ .
\end{align}
In the second line, we have neglected the poles of $f(q_0)$ in contour integral through the upper $q_0$ plane.
Considering the weak couplong limit in which $\Gamma_{\ell \phi}\ll T$, the imaginary part of the distribution function is negligible.
Moreover, in this limit, we can confirm that the real part of $1/(\epsilon \omega-\epsilon_{\ell}\omega_{ p}-\epsilon_{\phi}\omega_{ k}-i\Gamma_{\ell \phi}/2)$ does not contribute in (\ref{dndtG(mu)}) by the straightforward calculation.
the remaining imaginary part goes to the Dirac delta function $\pi \delta (\epsilon \omega-\epsilon_{\ell}\omega_{ p}-\epsilon_{\phi}\omega_{ k})$.
This means that, in (\ref{dndtG(mu)}), the approximation $f(\epsilon_{\ell}\omega_{ p}+\epsilon_{\phi}\omega_{ k}) \simeq f(\epsilon \omega_{q})$ is justified and the distribution function is allowed to come out from the inside of $p$ and $k$ integration.
This approximation is equivalent to the replacement:
\begin{align}
\int \frac{dq_0}{2\pi} \frac{i}{q_0 -\Omega_{\epsilon}} \pi^{(eq)}_{\wightman}(q_0 ,{\bold q}) \rightarrow (-i) \left\{ \begin{matrix} 1-f(\epsilon \omega_{q}) \\ -f(\epsilon \omega_{q}) \end{matrix} \right\} \frac{-1}{2} \pi^{(eq)}_{\rho}(\epsilon \omega_{q} ,{\bold q}) 
\end{align}
in (\ref{dndtG(mu)}).
Using the expression $\pi_{\rho}(\epsilon \omega_{ q})=-g_w i \epsilon \Slash{q}_{\epsilon} /(16\pi)$, \ $\pi_{h}(\epsilon \omega_{ q})=0$, It turns out that only the off-diagonal components of $\Delta _{\mu} G_{\wightman}$ contribute to (\ref{dndtG(mu)}).
After  simple calculations, we get the backreactions (\ref{C^{0}BR}) and (\ref{C^{1}BR}).

\section{Separation of $\Delta G_\gtrless'$ into ``on-shell'' and ``off-shell'' }
\label{SecAppOnOffShell}
In this appendix we give detailed calculations of the separation of various Green functions 
into ``on-shell" and ``off-shell" contributions
discussed in Section \ref{Sec-Interpretation}. 
Since   the leading order terms in $G_\gtrless^{'ij}$ contain
derivatives of the distribution functions, 
we consider only terms including  $\partial_{\omega} f$ and $\partial_{t} f$.
\subsection{$G_{R/A}^{'(eq)}$}
We first extract  terms oscillating with frequencies
$\Omega_1$ and $\Omega_2$ in $G_{R/A}^{'(eq)}$.
For the equilibrium Green functions,
the extraction can be easily carried out 
by looking at their time representations, (\ref{G_{R}^{'(eq)}}) and (\ref{G_{A}^{'(eq)}}):
\begin{align}
G_{R}^{'(eq)ij}(x^0,y^0;{\bold q})&=[G_{R}^{'(eq)ij}(x^0,y^0;{\bold q})]_{i}+[G_{R}^{'(eq)ij}(x^0,y^0;{\bold q})]_{j} \notag \\
[G_{R}^{'(eq)ij}(x^0,y^0;{\bold q})]_{i}&\equiv \Theta (s_{xy}) \sum_{\epsilon} Z_{\epsilon} \Pi^{'(eq)ij}_{R}(\epsilon \omega_{q}) Z_{\epsilon }\frac{-i}{\Omega_{\epsilon i}-\Omega_{\epsilon j}} e^{-i\Omega _{\epsilon i}(x^0 -y^0)} \notag \\
[G_{R}^{'(eq)ij}(x^0,y^0;{\bold q})]_{j}&\equiv \Theta (s_{xy}) \sum_{\epsilon} Z_{\epsilon} \Pi^{'(eq)ij}_{R}(\epsilon \omega_{q}) Z_{\epsilon }\frac{+i}{\Omega_{\epsilon i}-\Omega_{\epsilon j}} e^{-i\Omega _{\epsilon j}(x^0 -y^0)} \label{offshellG_{R}} 
\end{align}
and
\begin{align}
G_{A}^{'(eq)ij}(x^0,y^0;{\bold q})&=[G_{A}^{'(eq)ij}(x^0,y^0;{\bold q})]_{i}+[G_{A}^{'(eq)ij}(x^0,y^0;{\bold q})]_{j} \notag \\
[G_{A}^{'(eq)ij}(x^0,y^0;{\bold q})]_{i}&\equiv \Theta (-s_{xy}) \sum_{\epsilon} Z_{\epsilon} \Pi^{'(eq)ij}_{R}(\epsilon \omega_{q}) Z_{\epsilon }\frac{+i}{\Omega^{*}_{\epsilon i}-\Omega^{*}_{\epsilon j}} e^{-i\Omega^{*} _{\epsilon i}(x^0 -y^0)} \notag \\
[G_{R}^{'(eq)ij}(x^0,y^0;{\bold q})]_{j}&\equiv \Theta (-s_{xy}) \sum_{\epsilon} Z_{\epsilon} \Pi^{'(eq)ij}_{R}(\epsilon \omega_{q}) Z_{\epsilon }\frac{-i}{\Omega^{*}_{\epsilon i}-\Omega^{*}_{\epsilon j}} e^{-i\Omega^{*} _{\epsilon j}(x^0 -y^0)} \ . \label{offshellG_{A}} 
\end{align}
\subsection{$\Delta G_{R/A}^{'}$}
For the out-of-equilibrium Green functions $\Delta G_{R/A}^{'}$, the separation is a bit more involved.
Let us start with the expression (\ref{DelG_{RA}^{'}}) of the off-diagonal retarded propagator:
\be
&&\Delta G_{R}^{'ij}(x^0,y^0;{\bold q}) = W_1 +W_2 +W_3 \ , \n
&&W_1 \equiv -\int du^0 dv^0 G^{d(eq)ii}_{R}(x^0,u^0;{\bold q})\Pi^{'(eq)ij}_{R}(u^0,v^0;{\bold q}) \Delta G^{djj}_{R}(v^0,y^0;{\bold q}) \ , \n
&&W_2 \equiv -\int du^0 dv^0 \Delta G^{dii}_{R}(x^0,u^0;{\bold q})\Pi^{'(eq)ij}_{R}(u^0,v^0;{\bold q})  G^{d(eq)jj}_{R}(v^0,y^0;{\bold q}) \ , \n
&&W_3 \equiv -\int du^0 dv^0 G^{d(eq)ii}_{R}(x^0,u^0;{\bold q}) \Delta \Pi^{'(eq)ij}_{R}(u^0,v^0;{\bold q}) G^{d(eq)jj}_{R}(v^0,y^0;{\bold q}) \ . \n
\label{W_123}
\ee
By using identities in Appendix \ref{App-Useful}, we can calculate these integrals.
$W_1$ becomes
\begin{align}
W_1&=\int \frac{dq_0}{2\pi}\frac{-i}{2}G_{R}^{d(eq)ii}\Pi_{R}^{'(eq)ij}(\partial_{q_0}G_{R}^{d(eq)}\zeta_{R}G_{R}^{d(eq)}-G_{R}^{d(eq)}\zeta_{R} \partial_{q_0}G_{R}^{d(eq)})^{jj} e^{-iq_0 s_{xy}}\notag \\
&+\int \frac{dq_0}{2\pi}\frac{+i}{2}G_{R}^{d(eq)ii}\Pi_{R}^{'(eq)ij}\partial_{q_0}\partial_{t}G_{R}^{d(eq)jj} e^{-iq_0 s_{xy}} \notag \\
&+\int \frac{dq_0}{2\pi}(-1)G_{R}^{d(eq)ii}\Pi_{R}^{'(eq)ij}\partial_{t}G_{R}^{d(eq)jj} \left( X_{xy}-t-\frac{s_{xy}}{2} \right)e^{-iq_0 s_{xy}} \ . \label{W_1i}
\end{align}
The leading order terms containing waves with frequency $\Omega_i$ is given by
taking the residue of the  $\Omega_i$ in $G_{R}^{d(eq)ii}$ and we have
\begin{align}
[W_1]_i\simeq \left. \sum_{\epsilon} Z_{\epsilon}\Pi^{'(eq)ij}_{R}(\epsilon \omega_{q}) Z_{\epsilon} \frac{\epsilon}{(\Omega_{\epsilon i}-\Omega_{\epsilon j})^2}\frac{\partial \omega_{q}}{ \partial t}\left(i\partial_{Q_1}-t -\frac{i}{2}\partial_{\Omega_{\epsilon i}} \right) e^{-i\Omega_{\epsilon i} s_{xy} -iQ_1 X_{xy}} \right|_{Q_1 =0} \ . \label{W_1i}
\end{align}
Here we used the identity (\ref{Usefulint}).
To extract a leading order contribution with frequency $\Omega_{j}$,
it's convenient to rewrite $W_1$ into the following form:
\begin{align}
W_1&=\int \frac{dq_0}{2\pi}\frac{-i}{2}G_{R}^{d(eq)ii}\Pi_{R}^{'(eq)ij}(\partial_{q_0}G_{R}^{d(eq)}\zeta_{R}G_{R}^{d(eq)}-G_{R}^{d(eq)}\zeta_{R} \partial_{q_0}G_{R}^{d(eq)})^{jj} e^{-iq_0 s_{xy}} \n
&+\int \frac{dq_0}{2\pi}\frac{-i}{2}\partial_{q_0} \left( G_{R}^{d(eq)ii}\Pi_{R}^{'(eq)ij} \right) \partial_{t}G_{R}^{d(eq)jj} e^{-iq_0 s_{xy}} \notag \\
&+\int \frac{dq_0}{2\pi}(-1)G_{R}^{d(eq)ii}\Pi_{R}^{'(eq)ij}\partial_{t}G_{R}^{d(eq)jj} \left( X_{xy}-t \right)e^{-iq_0 s_{xy}} \ . \label{W_1j}
\end{align}
It is equal to (\ref{W_1i}) up to a total derivative.
Taking the residue of the pole $\Omega_j$ in $\Delta G_{R}^{djj}$, we get 
the leading order contribution with frequency $\Omega_j$ as
\begin{align}
[W_1]_j \simeq&\left. \sum_{\epsilon} Z_{\epsilon}\Pi^{'(eq)ij}_{R}(\epsilon \omega_{q}) Z_{\epsilon} \frac{-\epsilon}{(\Omega_{\epsilon i}-\Omega_{\epsilon j})^2}\frac{\partial \omega_{q}}{ \partial t}\left(i\partial_{Q_1}-t +\frac{i}{2}\partial_{\Omega_{\epsilon i}} \right) e^{-i\Omega_{\epsilon i} s_{xy} -iQ_1 X_{xy}} \right|_{Q_1 =0} \notag \\
+&\left. \sum_{\epsilon} Z_{\epsilon}\Pi^{'(eq)ij}_{R}(\epsilon \omega_{q}) Z_{\epsilon} \frac{+i\epsilon}{\Omega_{\epsilon i}-\Omega_{\epsilon j}}\frac{\partial \omega_{q}}{ \partial t}\left(i\partial_{Q_1}-t \right)\frac{i}{2}\partial_{\Omega_{\epsilon i}} e^{-i\Omega_{\epsilon i} s_{xy} -iQ_1 X_{xy}} \right|_{Q_1 =0} \ .
\end{align}

Similarly,
we can extract the following leading order contributions from $W_2$:
\begin{align}
[W_2]_j \simeq \left. \sum_{\epsilon} Z_{\epsilon}\Pi^{'(eq)ij}_{R}(\epsilon \omega_{q}) Z_{\epsilon} \frac{\epsilon}{(\Omega_{\epsilon i}-\Omega_{\epsilon j})^2}\frac{\partial \omega_{q}}{ \partial t}\left(i\partial_{Q_1}-t +\frac{i}{2}\partial_{\Omega_{\epsilon i}} \right) e^{-i\Omega_{\epsilon i} s_{xy} -iQ_1 X_{xy}} \right|_{Q_1 =0}
\end{align}
and
\begin{align}
[W_2]_i \simeq& \left. \sum_{\epsilon} Z_{\epsilon}\Pi^{'(eq)ij}_{R}(\epsilon \omega_{q}) Z_{\epsilon} \frac{-\epsilon}{(\Omega_{\epsilon i}-\Omega_{\epsilon j})^2}\frac{\partial \omega_{q}}{ \partial t}\left(i\partial_{Q_1}-t -\frac{i}{2}\partial_{\Omega_{\epsilon i}} \right) e^{-i\Omega_{\epsilon i} s_{xy} -iQ_1 X_{xy}} \right|_{Q_1 =0} \n
+&\left. \sum_{\epsilon} Z_{\epsilon}\Pi^{'(eq)ij}_{R}(\epsilon \omega_{q}) Z_{\epsilon} \frac{-i\epsilon}{\Omega_{\epsilon i}-\Omega_{\epsilon j}}\frac{\partial \omega_{q}}{ \partial t}\left(i\partial_{Q_1}-t \right)\frac{i}{2}\partial_{\Omega_{\epsilon i}} e^{-i\Omega_{\epsilon i} s_{xy} -iQ_1 X_{xy}} \right|_{Q_1 =0} \ .
\end{align}

$W_3$ in (\ref{W_123}) is calculated as
\begin{align}
W_3&=\int \frac{dq_0}{2\pi}\frac{-i}{2}\left( \partial_{q_0} G_{R}^{d(eq)ii}\partial_t \Pi_{R}^{'(eq)ij} G_{R}^{d(eq)}-G_{R}^{d(eq)ii}\partial_t \Pi_{R}^{'(eq)ij}\partial_{q_0}G_{R}^{d(eq)} \right) e^{-iq_0 s_{xy}}\notag \\
&\ +\int \frac{dq_0}{2\pi}(-1)G_{R}^{d(eq)ii}\partial_{t} \Pi_{R}^{'(eq)ij}G_{R}^{d(eq)jj} \left( X_{xy}-t \right)e^{-iq_0 s_{xy}} \ .
\end{align}
It turns out that all terms are suppressed by a factor $\Gamma/T$ and 
 there are no leading order contributions from $W_3$:
\begin{align}
[W_3]_i = [W_3]_j =0 \ .
\end{align}

Summing up $W_i$ ($i=1,2,3$), the retarded Green function
$\Delta G_R'$ is separated  in the leading order approximation as
\begin{align}
\Delta G_{R}^{'ij}(x^0,y^0;{\bold q}) \simeq &[\Delta G_{R}^{'ij}(x^0,y^0;{\bold q})]_{i}+[\Delta G_{R}^{'ij}(x^0,y^0;{\bold q})]_{j} \ , \notag \\
[\Delta G_{R}^{'ij}(x^0,y^0;{\bold q})]_{i}\equiv & \Theta (s_{xy})  \sum_{\epsilon} Z_{\epsilon}\Pi^{'(eq)ij}_{R}(\epsilon \omega_{q}) Z_{\epsilon} \frac{-i\epsilon}{\Omega_{\epsilon i}-\Omega_{\epsilon j}} \notag \\
&\times \left. \frac{\partial \omega_{q}}{ \partial t} \left(i\partial_{Q_1}-t \right)\frac{i}{2}\partial_{\Omega_{\epsilon i}} e^{-i\Omega_{\epsilon i} s_{xy} -iQ_1 X_{xy}} \right|_{Q_1 =0} \ , \notag \\
[\Delta G_{R}^{'ij}(x^0,y^0;{\bold q})]_{j}\equiv &\Theta (s_{xy})  \sum_{\epsilon} Z_{\epsilon}\Pi^{'(eq)ij}_{R}(\epsilon \omega_{q}) Z_{\epsilon} \frac{+i\epsilon}{\Omega_{\epsilon i}-\Omega_{\epsilon j}} \notag \\
&\times \left. \frac{\partial \omega_{q}}{ \partial t} \left(i\partial_{Q_1}-t \right)\frac{i}{2}\partial_{\Omega_{\epsilon i}} e^{-i\Omega_{\epsilon i} s_{xy} -iQ_1 X_{xy}} \right|_{Q_1 =0} \ . \label{offshellG_{R}2} 
\end{align}

\subsection{Useful identities}
In order to perform the following calculations, we first introduce two useful identities
\begin{align}
\int \frac{dq}{2\pi}&\frac{dq_2}{2\pi} F(\Omega_{\epsilon}, q, q_2)\int^{x} du \int dv e^{-i\Omega_{\epsilon} (x-u) -iQ_1 (x+u)/2 -iq(u-v) -iXQ -iq_2 (v-y) -iQ_2 (v+y)/2}\n
=& \int \frac{dq}{2\pi} F(\Omega_{\epsilon},q+Q/2 +Q_2 /2 ,q) \frac{+i}{q-\Omega_{\epsilon} +Q+ (Q_1 +Q_2)/2}  \n
&\hspace{30pt}\times e^{-i(q+Q/2 +Q_1 /2)s_{xy} -i(Q_1 +Q_2 +Q)X_{xy}}  \label{intexp3}
\end{align}
and
\begin{align}
\int \frac{dq}{2\pi}&\frac{dq_1}{2\pi} F(q_1 ,q, \Omega^{*}_{\epsilon})\int du \int^{y} dv e^{-iq_1 (x-u) -iQ_1 (x+u)/2 -iq(u-v) -iXQ -i\Omega^{*}_{\epsilon} (v-y) -iQ_2 (v+y)/2}\n
=& \int \frac{dq}{2\pi} F(q,q-Q/2 -Q_{1}/2,\Omega^{*}_{\epsilon})\frac{-i}{q-\Omega^{*}_{\epsilon} -Q- (Q_1 +Q_2)/2}  \n
&\hspace{30pt}\times e^{-i(q-Q/2 -Q_2 /2)s_{xy} -i(Q_1 +Q_2 +Q)X_{xy}} \ . \label{intexp2}
\end{align}
Note that the complex frequency $\Omega$ is introduced in the above identities
unlike  (\ref{Usefulint}).
In the following, we will use these identities together with (\ref{Usefulint}).

\subsection{$G_\gtrless^{'}$}
By using the decomposition  (\ref{offshellG_{R}}) and (\ref{offshellG_{A}}), we separate
$G_\gtrless{'(eq)}$ into  ``on-shell" and``off-shell" terms as
\begin{align}
G_{\wightman {\bold q}}^{'(eq)ij}=&[G_{\wightman {\bold q}}^{'(eq)ij}]_{\mbox{on-shell}} +[G_{\wightman {\bold q}}^{'(eq)ij}]_{\mbox{off-shell}} \ , \notag \\
[G_{\wightman {\bold q}}^{'(eq)ij}]_{\mbox{on-shell}} \equiv & - [G^{'(eq)ij}_{R {\bold q}}]_{j} * \Pi^{d(eq)jj}_{\wightman {\bold q}} * G^{d(eq)jj}_{A {\bold q}}\notag \\
&-  G^{d(eq)ii}_{R {\bold q}}* \Pi^{d(eq)ii}_{\wightman {\bold q}} *[G^{'(eq)ij}_{A {\bold q}}]_{i} \ , \notag \\
[G_{\wightman {\bold q}}^{'(eq)ij}]_{\mbox{off-shell}} \equiv & - [G^{'(eq)ij}_{R {\bold q}}]_{i} * \Pi^{d(eq)jj}_{\wightman {\bold q}} * G^{d(eq)jj}_{A {\bold q}}\notag \\
&-  G^{d(eq)ii}_{R {\bold q}}* \Pi^{d(eq)ii}_{\wightman {\bold q}} *[G^{'(eq)ij}_{A {\bold q}}]_{j}\notag \\
&-  G^{d(eq)ii}_{R {\bold q}}* \Pi^{'(eq)ij}_{\wightman {\bold q}} *G^{d(eq)jj}_{A {\bold q}} \ .\label{G^{'(eq)}On-shell-Off-shell}
\end{align}
In the ``on-shell" terms, the same mass eigenstate $i$ or $j$ propagates\footnote{Since $\Pi'$ is
flavor off-diagonal, differences between the flavor eigenstates and the mass eigenstates in $G_{R/A}$ are
higher orders with respect to $(h^\dagger h)'/(h^\dagger h)^d$.}.
On the other hand, the ``off-shell" terms contain both of the mass eigenstates $i$ and $j$ simultaneously. 
\subsection{On-shell part of $G_\gtrless^{'(eq)}$}
Plugging the decomposition (\ref{offshellG_{R}}) and (\ref{offshellG_{A}}) into above 
and using the identities (\ref{intexp3}) and (\ref{intexp2}), 
 we obtain the following on-shell contributions:
\begin{align}
-&{\big [} G^{'(eq)ij}_{R\ {\bold q}} {\big ]}_{j}
*\Pi^{d(eq)jj}_{\wightman \ {\bold q}} * G^{d(eq)jj}_{A\ {\bold q}} 
 \n
&\simeq \sum_{\epsilon} Z_{\epsilon } \Pi^{'(eq)ij}_{R}(\epsilon \omega_{ q}) Z_{\epsilon}\frac{-i}{\Omega_{\epsilon i}-\Omega_{\epsilon j}}\frac{i\Gamma_{jq}}{\Omega_{\epsilon j}-\Omega^{*}_{\epsilon j}}(-i)\left\{ \begin{matrix} 1-f^{\epsilon}_{iq} \\
-f^{\epsilon}_{iq} \end{matrix} \right\} e^{-i\Omega _{\epsilon i}(x^0 -y^0)} \n
&= \sum_{\epsilon} Z_{\epsilon } \Pi^{'(eq)ij}_{R}(\epsilon \omega_{ q}) Z_{\epsilon}\frac{+i}{\Omega_{\epsilon i}-\Omega_{\epsilon j}}(-i)\left\{ \begin{matrix} 1-f^{\epsilon}_{iq} \\
-f^{\epsilon}_{iq} \end{matrix} \right\} e^{-i\Omega _{\epsilon i}(x^0 -y^0)} 
\label{onshell(eq)1}
\end{align}
and 
\begin{align}
-&G^{d(eq)ii}_{R\ {\bold q}} *\Pi^{d(eq)ii}_{\wightman \ {\bold q}} * 
{\big [}  G^{'(eq)ij}_{A\ {\bold q}} {\big ]}_{i} \n
&\simeq \sum_{\epsilon} Z_{\epsilon} \Pi^{'(eq)ij}_{A}(\epsilon \omega_{ q}) Z_{\epsilon}\frac{-i}{\Omega^{*}_{\epsilon i}-\Omega^{*}_{\epsilon j}}\frac{-i\Gamma_{iq}}{\Omega_{\epsilon i}-\Omega^{*}_{\epsilon i}} (-i)\left\{ \begin{matrix} 1-f^{\epsilon}_{iq} \\
-f^{\epsilon}_{iq} \end{matrix} \right\} e^{-i\Omega _{\epsilon i}(x^0 -y^0)} \n
&= \sum_{\epsilon} Z_{\epsilon} \Pi^{'(eq)ij}_{A}(\epsilon \omega_{ q}) Z_{\epsilon}\frac{-i}{\Omega^{*}_{\epsilon i}-\Omega^{*}_{\epsilon j}} (-i)\left\{ \begin{matrix} 1-f^{\epsilon}_{iq} \\
-f^{\epsilon}_{iq} \end{matrix} \right\} e^{-i\Omega _{\epsilon i}(x^0 -y^0)} \ .
\label{onshell(eq)2} 
\end{align}

Summing the above two on-shell contributions,
we have eq. (\ref{onshell(eq)}).

\subsection{Off-shell part of $G_\gtrless^{'(eq)}$}
The first off-shell contribution comes from the
$i$-th propagation in  $G_R^{'ij}$ of the first term of  (\ref{intKBeqG_{W}^{'}1}), 
and becomes
(supposing $x^0 > y^0$)
\begin{align}
-&{\big [} G^{'(eq)ij}_{R\ {\bold q}} {\big ]}_{i}
*\Pi^{d(eq)jj}_{\wightman \ {\bold q}} * G^{d(eq)jj}_{A\ {\bold q}} 
 \n
&\simeq \sum_{\epsilon} Z_{\epsilon } \Pi^{'(eq)ij}_{R}(\epsilon \omega_{ q}) Z_{\epsilon}\frac{+i}{\Omega_{\epsilon i}-\Omega_{\epsilon j}}\frac{i\Gamma_{jq}}{\Omega_{\epsilon i}-\Omega^{*}_{\epsilon j}}(-i)\left\{ \begin{matrix} 1-f^{\epsilon}_{iq} \\
-f^{\epsilon}_{iq} \end{matrix} \right\} e^{-i\Omega _{\epsilon i}(x^0 -y^0)} \ .
\label{offshell(eq)1}
\end{align}
The $j$-th propagation of the second term in (\ref{intKBeqG_{W}^{'}1}) gives the off-shell 
contribution and becomes (for $x^0 > y^0$)\footnote{
Note that the frequency of the wave function is given by $\Omega _{\epsilon i}$.
It is because, for  $x^0 > y^0$, the pole of the $j$-th eigenstate of $G_A^{'ij}$ does not contribute
to the Cauchy integral because the pole is in the lower complex plane. 
}
\begin{align}
-&G^{d(eq)ii}_{R\ {\bold q}} *\Pi^{d(eq)ii}_{\wightman \ {\bold q}} * 
{\big [}  G^{'(eq)ij}_{A\ {\bold q}} {\big ]}_{j} \n
&\simeq \sum_{\epsilon} Z_{\epsilon} \Pi^{'(eq)ij}_{A}(\epsilon \omega_{ q}) Z_{\epsilon}\frac{+i}{\Omega^{*}_{\epsilon i}-\Omega^{*}_{\epsilon j}}\frac{-i\Gamma_{iq}}{\Omega_{\epsilon i}-\Omega^{*}_{\epsilon j}} (-i)\left\{ \begin{matrix} 1-f^{\epsilon}_{iq} \\
-f^{\epsilon}_{iq} \end{matrix} \right\} e^{-i\Omega _{\epsilon i}(x^0 -y^0)} \ .
\label{offshell(eq)2} 
\end{align}

From the last line of (\ref{G^{'(eq)}On-shell-Off-shell}), 
we get the off-shell contribution by
using the identity (\ref{Usefulint}) (for $x^0 > y^0$):
\begin{align}
-&G^{d(eq)ii}_{R\ {\bold q}} *\Pi^{'(eq)ij}_{\wightman \ {\bold q}} * G^{d(eq)jj}_{A\ {\bold q}}
\n
&\simeq \sum_{\epsilon} Z_{\epsilon} \Pi^{'(eq)ij}_{\rho}(\epsilon \omega_{ q}) Z_{\epsilon}\frac{-i}{\Omega_{\epsilon i}-\Omega^{*}_{\epsilon j}}(-i)\left\{ \begin{matrix} 1-f^{\epsilon}_{iq} \\
-f^{\epsilon}_{iq} \end{matrix} \right\} e^{-i\Omega _{\epsilon i}(x^0 -y^0)} \ .
 \label{offshell(eq)3} 
\end{align}

Note that the enhancement factors of these off-shell contributions have a factor
$1/(\Omega_i-\Omega_j^*)$.
Summing the on-shell (\ref{onshell(eq)}) and off-shell contributions, we can of course recover the 
full result  (\ref{G_{W}^{'(eq)}1}).
\subsection{On-shell part of $\Delta G_\gtrless^{'ij}$}
\label{App-onshellDeltaG}
Finally, we move on to $\Delta G_\gtrless^{\prime ij}$.
Taking a variation of (\ref{intKBeqG_{W}^{'}1}),
we get the following 7 terms:
\begin{align}
\Delta G^{' ij}_{\wightman } 
=& - \Delta G^{'ij}_{R}* \Pi^{d(eq)jj}_{\wightman} *G^{d(eq)jj}_{A}
- G^{'(eq)ij}_{R}* \Delta \Pi^{d(eq)jj}_{\wightman} *G^{d(eq)jj}_{A}\n
& - G^{'(eq)ij}_{R}* \Pi^{d(eq)jj}_{\wightman} *\Delta G^{djj}_{A} - \Delta G^{dii}_{R}* \Pi^{d(eq)ii}_{\wightman } *G^{'(eq)ij}_{A}\n
& - G^{d(eq)ii}_{R}* \Delta \Pi^{d(eq)ii}_{\wightman} *G^{'(eq)ij}_{A}
- G^{d(eq)ii}_{R}* \Pi^{d(eq)ii}_{\wightman } *\Delta G^{'ij}_{A} \n
& - \Delta \left\{ G^{dii}_{R}* \Pi^{'ij}_{\wightman} *G^{djj}_{A} \right\} \ .
\label{deltaWightman9-2}
\end{align}

Let us apply the decomposition (\ref{offshellG_{R}2}) into the first term.
Using the identity (\ref{intexp3}) and performing 
$q_0$ integration to pick up the pole $q_0 = \Omega_{\epsilon i,j}-Q_1 /2$ in $[\Delta G^{'ij}_{R}]_{i,j}$,
$f(q_0)$ is replaced by $f(\Omega_{\epsilon i,j}-Q_1 /2)$.
The derivatives with respect to $Q_1$ or $\Omega_{\epsilon i,j}$ act on it in the leading order
approximation. As a result we have
\begin{align}
-&[\Delta G^{'ij}_{R {\bold q}} *\Pi^{d(eq)jj}_{\wightman {\bold q}}*G^{d(eq)jj}_{A {\bold q}}]_{\mbox{on-shell}} \equiv -[\Delta G^{'ij}_{R {\bold q}}]_j *\Pi^{d(eq)jj}_{\wightman {\bold q}}*G^{d(eq)jj}_{A {\bold q}} \notag \\
&\simeq\sum_{\epsilon} Z_{\epsilon} \Pi^{'(eq)ij}_{R}(\epsilon \omega_{ q}) Z_{\epsilon}(-i) \frac{\partial \omega_{j q}}{\partial t}\frac{\partial}{\partial \omega_{j q}} \left\{ \begin{matrix} 1-f^{\epsilon}_{j q} \\ -f^{\epsilon}_{j q} \end{matrix} \right\}\frac{1}{(\Omega_{\epsilon j}-\Omega^{*}_{\epsilon j})^2}\frac{+i \Gamma_{jq}}{\Omega_{\epsilon i}-\Omega_{\epsilon j}} \ e^{-i\Omega_{\epsilon j}s_{xy}} \notag \\
&=\sum_{\epsilon} Z_{\epsilon} \Pi^{'(eq)ij}_{R}(\epsilon \omega_{ q}) Z_{\epsilon}(-i) \frac{\partial \omega_{j q}}{\partial t}\frac{\partial}{\partial \omega_{jq}} \left\{ \begin{matrix} 1-f^{\epsilon}_{j q} \\ -f^{\epsilon}_{j q} \end{matrix} \right\}\frac{-1}{\Gamma_{jq}}\frac{i}{\Omega_{\epsilon i}-\Omega_{\epsilon j}} \ e^{-i\Omega_{\epsilon j}s_{xy}} \ . \label{offdiagonal7-1-on}
\end{align}

For the second  term in (\ref{deltaWightman9-2}),
we apply the decomposition of $G_R^{\prime}$ in  (\ref{offshellG_{R}}). 
After using (\ref{intexp3}),
we get
\begin{align}
-&[ G^{'(eq)ij}_{R {\bold q}} *\Delta \Pi^{d(eq)jj}_{\wightman {\bold q}}*G^{d(eq)jj}_{A {\bold q}} ]_{\mbox{on-shell}} \equiv -[ G^{'(eq)ij}_{R {\bold q}}]_j *\Delta \Pi^{d(eq)jj}_{\wightman {\bold q}}*G^{d(eq)jj}_{A {\bold q}} \notag \\
&\simeq\sum_{\epsilon} Z_{\epsilon} \Pi^{'(eq)ij}_{R}(\epsilon \omega_{ q}) Z_{\epsilon}(-i) \frac{\partial T}{\partial t}\frac{\partial}{\partial T} \left\{ \begin{matrix} 1-f^{\epsilon}_{j q} \\ -f^{\epsilon}_{j q} \end{matrix} \right\}\frac{1}{(\Omega_{\epsilon j}-\Omega^{*}_{\epsilon j})^2}\frac{+i \Gamma_{jq}}{\Omega_{\epsilon i}-\Omega_{\epsilon j}} \ e^{-i\Omega_{\epsilon j}s_{xy}} \notag \\
&=\sum_{\epsilon} Z_{\epsilon} \Pi^{'(eq)ij}_{R}(\epsilon \omega_{ q}) Z_{\epsilon}(-i) \frac{\partial T}{\partial t}\frac{\partial}{\partial T} \left\{ \begin{matrix} 1-f^{\epsilon}_{j q} \\ -f^{\epsilon}_{j q} \end{matrix} \right\}\frac{-1}{\Gamma_{jq}}\frac{i}{\Omega_{\epsilon i}-\Omega_{\epsilon j}} \ e^{-i\Omega_{\epsilon j}s_{xy}} \  .\label{offdiagonal7-2-on}
\end{align}

Summing up the these two on-shell contributions,
 (\ref{offdiagonal7-1-on}) and (\ref{offdiagonal7-2-on}), we get
\begin{align}
-&\Delta  \left\{{\big [} G^{'ij}_{R\ {\bold q}} {\big ]}_j *\Pi^{djj}_{\wightman \ {\bold q}} * G^{djj}_{A\ {\bold q}} \right\} \n
&\simeq \sum_{\epsilon} Z_{\epsilon } \Pi^{'(eq)ij}_{R}(\epsilon \omega_{q}) Z_{\epsilon}(-i) \ d_t \left\{ \begin{matrix} 1-f^{\epsilon}_{j q} \\ -f^{\epsilon}_{j q} \end{matrix} \right\} \frac{+1}{(\Omega_{\epsilon j}-\Omega^{*}_{\epsilon j})^2} \frac{+i \Gamma_{jq} }{\Omega_{\epsilon i}-\Omega_{\epsilon j}} \ e^{-i\Omega_{\epsilon j }(x^0 -y^0)} \n
&= \sum_{\epsilon} Z_{\epsilon } \Pi^{'(eq)ij}_{R}(\epsilon \omega_{q}) Z_{\epsilon}(-i) \ d_t \left\{ \begin{matrix} 1-f^{\epsilon}_{j q} \\ -f^{\epsilon}_{j q} \end{matrix} \right\} \frac{-1}{\Gamma_{jq}} \frac{i}{\Omega_{\epsilon i}-\Omega_{\epsilon j}} \ e^{-i\Omega_{\epsilon j }(x^0 -y^0)} .
 \label{Delonshell1}  
\end{align}

For the 4th term in (\ref{deltaWightman9-2}),
we apply the decomposition of $G_A^{'(eq)}$ (\ref{offshellG_{A}}).
Using (\ref{intexp2}), we get
\begin{align}
-&[\Delta G^{dii}_{R {\bold q}} *\Pi^{d(eq)ii}_{\wightman {\bold q}}* G^{'(eq)ij}_{A {\bold q}}]_{\mbox{on-shell}} \equiv -\Delta G^{dii}_{R {\bold q}} *\Pi^{d(eq)ii}_{\wightman {\bold q}}*[G^{'(eq)ij}_{A {\bold q}}]_i \notag \\
&\simeq\sum_{\epsilon} Z_{\epsilon} \Pi^{'(eq)ij}_{A}(\epsilon \omega_{ q}) Z_{\epsilon}(-i) \frac{\partial \omega_{i q}}{\partial t}\frac{\partial}{\partial \omega_{i q}} \left\{ \begin{matrix} 1-f^{\epsilon}_{i q} \\ -f^{\epsilon}_{i q} \end{matrix} \right\}\frac{1}{(\Omega_{\epsilon i}-\Omega^{*}_{\epsilon i})^2}\frac{-i \Gamma_{iq}}{\Omega^{*}_{\epsilon i}-\Omega^{*}_{\epsilon j}} \ e^{-i\Omega_{\epsilon i}s_{xy}} \notag \\
&=\sum_{\epsilon} Z_{\epsilon} \Pi^{'(eq)ij}_{A}(\epsilon \omega_{ q}) Z_{\epsilon}(-i) \frac{\partial \omega_{i q}}{\partial t}\frac{\partial}{\partial \omega_{i q}} \left\{ \begin{matrix} 1-f^{\epsilon}_{i q} \\ -f^{\epsilon}_{i q} \end{matrix} \right\}\frac{-1}{\Gamma_{iq}}\frac{-i}{\Omega^{*}_{\epsilon i}-\Omega^{*}_{\epsilon j}} \ e^{-i\Omega_{\epsilon i}s_{xy}} \ . \label{offdiagonal7-4-on}
\end{align}

On-shell contribution from the 5th term in (\ref{deltaWightman9-2}) is similarly calculated as
\begin{align}
-&[G^{d(eq)ii}_{R {\bold q}} *\Delta \Pi^{d(eq)ii}_{\wightman {\bold q}}*G^{'(eq)ij}_{A {\bold q}}]_{\mbox{on-shell}} \equiv -G^{d(eq)ii}_{R {\bold q}} *\Delta \Pi^{d(eq)ii}_{\wightman {\bold q}}*[G^{'(eq)ij}_{A {\bold q}}]_i \notag \\
&\simeq \sum_{\epsilon} Z_{\epsilon} \Pi^{'(eq)ij}_{A}(\epsilon \omega_{ q}) Z_{\epsilon}(-i) \frac{\partial T}{\partial t}\frac{\partial}{\partial T} \left\{ \begin{matrix} 1-f^{\epsilon}_{i q} \\ -f^{\epsilon}_{i q} \end{matrix} \right\}\frac{1}{(\Omega_{\epsilon j}-\Omega^{*}_{\epsilon j})^2}\frac{-i \Gamma_{iq}}{\Omega^{*}_{\epsilon i}-\Omega^{*}_{\epsilon j}} \ e^{-i\Omega_{\epsilon i}s_{xy}} \notag \\
&=\sum_{\epsilon} Z_{\epsilon} \Pi^{'(eq)ij}_{A}(\epsilon \omega_{ q}) Z_{\epsilon}(-i) \frac{\partial T}{\partial t}\frac{\partial}{\partial T} \left\{ \begin{matrix} 1-f^{\epsilon}_{i q} \\ -f^{\epsilon}_{i q} \end{matrix} \right\}\frac{-1}{\Gamma_{iq}}\frac{-i}{\Omega^{*}_{\epsilon i}-\Omega^{*}_{\epsilon j}} \ e^{-i\Omega_{\epsilon i}s_{xy}} \ . \label{offdiagonal7-5-on}
\end{align}

Summing  us another set of these two on-shell contributions,
(\ref{offdiagonal7-4-on}) and (\ref{offdiagonal7-5-on}), we get  
\begin{align}
-&\Delta \left\{G^{d(eq)ii}_{R\ {\bold q}} *\Pi^{d(eq)ii}_{\wightman \ {\bold q}} * 
{\big [}G^{'(eq)ij}_{A\ {\bold q}} {\big ]}_i \right\} 
\n
&\simeq \sum_{\epsilon} Z_{\epsilon} \Pi^{'(eq)ij}_{A}(\epsilon \omega_{q}) Z_{\epsilon}(-i) \ d_t \left\{ \begin{matrix} 1-f^{\epsilon}_{i q} \\ -f^{\epsilon}_{i q} \end{matrix} \right\} \frac{-1}{(\Omega_{\epsilon i}-\Omega^{*}_{\epsilon i})^2} \frac{+i \Gamma_{iq} }{\Omega^{*}_{\epsilon i}-\Omega^{*}_{\epsilon j}} \ e^{-i\Omega_{\epsilon i }(x^0 -y^0)} \n
&= \sum_{\epsilon} Z_{\epsilon} \Pi^{'(eq)ij}_{A}(\epsilon \omega_{q}) Z_{\epsilon}(-i) \ d_t \left\{ \begin{matrix} 1-f^{\epsilon}_{i q} \\ -f^{\epsilon}_{i q} \end{matrix} \right\} \frac{-1}{\Gamma_{iq}} \frac{-i }{\Omega^{*}_{\epsilon i}-\Omega^{*}_{\epsilon j}} \ e^{-i\Omega_{\epsilon i }(x^0 -y^0)} .
\label{Delonshell2} 
\end{align}

To summarize, after summing all on-shell contributions,   (\ref{Delonshell1} ) and (\ref{Delonshell2}),
we have (for $x^0 >y^0$)
\begin{align}
{\big [}
\Delta G^{'ij}_{\wightman}&(x^0,y^0;{\bold q})  {\big ]}_{\mbox{on-shell}}
\n
\simeq &\sum_{\epsilon} Z_{\epsilon} \Pi^{'(eq)ij}_{R}(\epsilon \omega_{ q}) Z_{\epsilon}(-i)\Delta \left\{ \begin{matrix} 1-f^{\epsilon}_{j q} \\ -f^{\epsilon}_{j q} \end{matrix} \right\}\frac{i}{\Omega_{\epsilon i}-\Omega_{\epsilon j}} \ e^{-i\Omega_{\epsilon j}s_{xy}}\notag \\
&+\sum_{\epsilon} Z_{\epsilon } \Pi^{'(eq)ij}_{A}(\epsilon \omega_{ q}) Z_{\epsilon }(-i)\Delta \left\{ \begin{matrix} 1-f^{\epsilon}_{iq} \\ -f^{\epsilon}_{iq} \end{matrix} \right\}\frac{-i}{\Omega^{*}_{\epsilon i}-\Omega^{*}_{\epsilon j}} \ e^{-i\Omega_{\epsilon i}s_{xy}}.
 \label{Delonshell} 
\end{align}

\subsection{Off-shell part of $\Delta G_\gtrless^{'ij}$}
Let us also calculate off-shell contributions to $\Delta G_\gtrless^{'ij}$ for completeness.

The off-shell contribution of the first term in (\ref{deltaWightman9-2}) is given by
\begin{align}
-&[\Delta G^{'ij}_{R {\bold q}} *\Pi^{d(eq)jj}_{\wightman {\bold q}}*G^{d(eq)jj}_{A {\bold q}} ]_{\mbox{off-shell}} \equiv -[\Delta G^{'ij}_{R {\bold q}}]_i *\Pi^{d(eq)jj}_{\wightman {\bold q}}*G^{d(eq)jj}_{A {\bold q}} \notag \\
&\simeq\sum_{\epsilon} Z_{\epsilon} \Pi^{'(eq)ij}_{R}(\epsilon \omega_{ q}) Z_{\epsilon}(-i) \frac{\partial \omega_{i q}}{\partial t}\frac{\partial}{\partial \omega_{i q}} \left\{ \begin{matrix} 1-f^{\epsilon}_{i q} \\ -f^{\epsilon}_{i q} \end{matrix} \right\}\frac{1}{(\Omega_{\epsilon i}-\Omega^{*}_{\epsilon j})^2}\frac{-i \Gamma_{jq}}{\Omega_{\epsilon i}-\Omega_{\epsilon j}} \ e^{-i\Omega_{\epsilon i}s_{xy}} \ . \label{offdiagonal7-1-off}
\end{align}

From the second term in (\ref{deltaWightman9-2}), we have the following off-shell contribution:
\begin{align}
-&[ G^{'(eq)ij}_{R {\bold q}} *\Delta \Pi^{d(eq)jj}_{\wightman {\bold q}}*G^{d(eq)jj}_{A {\bold q}} ]_{\mbox{off-shell}} \equiv -[ G^{'(eq)ij}_{R {\bold q}}]_i *\Delta \Pi^{d(eq)jj}_{\wightman {\bold q}}*G^{d(eq)jj}_{A {\bold q}} \notag \\
&\simeq\sum_{\epsilon} Z_{\epsilon} \Pi^{'(eq)ij}_{R}(\epsilon \omega_{ q}) Z_{\epsilon}(-i) \frac{\partial T}{\partial t}\frac{\partial}{\partial T} \left\{ \begin{matrix} 1-f^{\epsilon}_{i q} \\ -f^{\epsilon}_{i q} \end{matrix} \right\}\frac{1}{(\Omega_{\epsilon i}-\Omega^{*}_{\epsilon j})^2}\frac{-i \Gamma_{jq}}{\Omega_{\epsilon i}-\Omega_{\epsilon j}} \ e^{-i\Omega_{\epsilon i}s_{xy}} \ . \label{offdiagonal7-2-off}
\end{align}

Summing (\ref{offdiagonal7-1-off}) and (\ref{offdiagonal7-2-off}),
we have (for $x^0 > y^0$)
\begin{align}
-&\Delta  \left\{{\big [} G^{'ij}_{R\ {\bold q}} {\big ]}_i *\Pi^{djj}_{\wightman \ {\bold q}} * G^{djj}_{A\ {\bold q}} \right\} \n
&\simeq\sum_{\epsilon} Z_{\epsilon } \Pi^{'(eq)ij}_{R}(\epsilon \omega_{q}) Z_{\epsilon}(-i) \ d_t \left\{ \begin{matrix} 1-f^{\epsilon}_{i q} \\ -f^{\epsilon}_{i q} \end{matrix} \right\} \frac{-1}{(\Omega_{\epsilon i}-\Omega^{*}_{\epsilon j})^2} \frac{+i \Gamma_{jq} }{\Omega_{\epsilon i}-\Omega_{\epsilon j}} \ e^{-i\Omega_{\epsilon i }(x^0 -y^0)} \ .
 \label{Deloffshell1}  
\end{align}

The third term in (\ref{deltaWightman9-2}) has no leading order contribution:
\begin{align}
-[ G^{'(eq)ij}_{R {\bold q}}]_i * \Pi^{d(eq)jj}_{\wightman {\bold q}}*\Delta G^{djj}_{A {\bold q}}=-[ G^{'(eq)ij}_{R {\bold q}}]_j * \Pi^{d(eq)jj}_{\wightman {\bold q}}*\Delta G^{djj}_{A {\bold q}}=0 \ . \label{offdiagonal7-3}
\end{align}
To see this, perform the time and frequency integration.
Using (\ref{intexp3}) with $Q=0$, we have a factor $f(q_0+Q_2/2) =f(\Omega_{\epsilon})$ 
by picking up the pole at $\Omega_{\epsilon} -Q_2 /2$.
The derivative with respect to $\Omega$ would come from the double poles of $\Delta G_A$
 (\ref{defDeltildeG_{RA}^{d}App}), but
since we are interested in the region $x^0>y^0$, these poles on the upper complex plane
do not contribute. Therefore no derivatives of the distribution function appear.
Similarly the 6th term in (\ref{deltaWightman9-2})  also vanishes in the leading order approximation:
\begin{align}
- G^{d(eq)ii}_{R {\bold q}} * \Pi^{d(eq)ii}_{\wightman {\bold q}}*[\Delta G^{'ij}_{A {\bold q}}]_j=- G^{d(eq)ii}_{R {\bold q}} * \Pi^{d(eq)ii}_{\wightman {\bold q}}*[\Delta G^{'ij}_{A {\bold q}}]_i=0 \ . \label{offdiagonal7-6}
\end{align}

The 4th term in (\ref{deltaWightman9-2})  give the following off-shell contribution:
\begin{align}
-&[ \Delta G^{dii}_{R {\bold q}} *\Pi^{d(eq)ii}_{\wightman {\bold q}}* G^{'(eq)ij}_{A {\bold q}}]_{\mbox{off-shell}} \equiv -\Delta G^{dii}_{R {\bold q}} *\Pi^{d(eq)ii}_{\wightman {\bold q}}*[G^{'(eq)ij}_{A {\bold q}}]_j \notag \\
&\simeq\sum_{\epsilon} Z_{\epsilon} \Pi^{'(eq)ij}_{A}(\epsilon \omega_{ q}) Z_{\epsilon}(-i) \frac{\partial \omega_{i q}}{\partial t}\frac{\partial}{\partial \omega_{i q}} \left\{ \begin{matrix} 1-f^{\epsilon}_{i q} \\ -f^{\epsilon}_{i q} \end{matrix} \right\}\frac{1}{(\Omega_{\epsilon i}-\Omega^{*}_{\epsilon j})^2}\frac{+i \Gamma_{iq}}{\Omega^{*}_{\epsilon i}-\Omega^{*}_{\epsilon j}} \ e^{-i\Omega_{\epsilon i}s_{xy}} \ .
\label{offdiagonal7-4-off}
\end{align}

The 5th term in (\ref{deltaWightman9-2}) gives the off-shell contribution as
\begin{align}
-& [G^{d(eq)ii}_{R {\bold q}} *\Delta \Pi^{d(eq)ii}_{\wightman {\bold q}}* G^{'(eq)ij}_{A {\bold q}}]_{\mbox{off-shell}} \equiv - G^{d(eq)ii}_{R {\bold q}} *\Delta \Pi^{d(eq)ii}_{\wightman {\bold q}}*[G^{'(eq)ij}_{A {\bold q}}]_j \notag \\
&\simeq\sum_{\epsilon} Z_{\epsilon} \Pi^{'(eq)ij}_{A}(\epsilon \omega_{ q}) Z_{\epsilon}(-i) \frac{\partial T}{\partial t}\frac{\partial}{\partial T} \left\{ \begin{matrix} 1-f^{\epsilon}_{i q} \\ -f^{\epsilon}_{i q} \end{matrix} \right\}\frac{1}{(\Omega_{\epsilon i}-\Omega^{*}_{\epsilon j})^2}\frac{+i \Gamma_{iq}}{\Omega^{*}_{\epsilon i}-\Omega^{*}_{\epsilon j}} \ e^{-i\Omega_{\epsilon i}s_{xy}} \ . \label{offdiagonal7-5-off}
\end{align}

Summing  (\ref{offdiagonal7-4-off}) and (\ref{offdiagonal7-5-off}),
we have
\begin{align}
-&\Delta \left\{G^{d(eq)ii}_{R\ {\bold q}} *\Pi^{d(eq)ii}_{\wightman \ {\bold q}} * 
{\big [}G^{'(eq)ij}_{A\ {\bold q}} {\big ]}_j \right\} 
\n
&\simeq \sum_{\epsilon} Z_{\epsilon} \Pi^{'(eq)ij}_{A}(\epsilon \omega_{q}) Z_{\epsilon}(-i) \ d_t \left\{ \begin{matrix} 1-f^{\epsilon}_{i q} \\ -f^{\epsilon}_{i q} \end{matrix} \right\} \frac{+1}{(\Omega_{\epsilon i}-\Omega^{*}_{\epsilon j})^2} \frac{+i \Gamma_{iq} }{\Omega^{*}_{\epsilon i}-\Omega^{*}_{\epsilon j}} \ e^{-i\Omega_{\epsilon i }(x^0 -y^0)} \ .
\label{Deloffshell2} 
\end{align}

The last line in (\ref{deltaWightman9-2}) gives off-shell contributions.
They are composed of (\ref{V'''}),(\ref{7thJ1}) and (\ref{9thJ1}).
Using (\ref{Usefulint}),
we get the following leading order contributions:
\begin{align}
{\rm (\ref{V'''})} \simeq &\sum_{\epsilon} Z_{\epsilon } \Pi^{'(eq)ij}_{R}(\epsilon \omega_{ q}) Z_{\epsilon }(-i)  \frac{\partial T}{\partial t}\frac{\partial}{\partial T} \left\{ \begin{matrix} 1-f^{\epsilon}_{i q} \\ -f^{\epsilon}_{i q} \end{matrix} \right\} \frac{+1}{(\Omega_{\epsilon i}-\Omega^{*}_{\epsilon j})^2} \ e^{-i\Omega_{\epsilon i }s_{xy}} \n
+&\sum_{\epsilon} Z_{\epsilon } \Pi^{'(eq)ij}_{A}(\epsilon \omega_{q}) Z_{\epsilon }(-i) \frac{\partial T}{\partial t}\frac{\partial}{\partial T} \left\{ \begin{matrix} 1-f^{\epsilon}_{i q} \\ -f^{\epsilon}_{i q} \end{matrix} \right\}\frac{-1}{(\Omega_{\epsilon i}-\Omega^{*}_{\epsilon j})^2} \ e^{-i\Omega_{\epsilon i}s_{xy}} \ ,
\end{align}
\begin{align}
{\rm (\ref{7thJ1}) + (\ref{9thJ1})} \simeq &\sum_{\epsilon} Z_{\epsilon } \Pi^{'(eq)ij}_{R}(\epsilon \omega_{ q}) Z_{\epsilon }(-i)  \frac{\partial \omega_{i q}}{\partial t} \frac{\partial}{\partial \omega_{i q}} \left\{ \begin{matrix} 1-f^{\epsilon}_{i q} \\ -f^{\epsilon}_{i q} \end{matrix} \right\} \frac{+1}{(\Omega_{\epsilon i}-\Omega^{*}_{\epsilon j})^2} \ e^{-i\Omega_{\epsilon i }s_{xy}} \notag \\
+&\sum_{\epsilon} Z_{\epsilon } \Pi^{'(eq)ij}_{A}(\epsilon \omega_{q}) Z_{\epsilon }(-i) \frac{\partial \omega_{i q}}{\partial t}\frac{\partial}{\partial \omega_{i q}} \left\{ \begin{matrix} 1-f^{\epsilon}_{i q} \\ -f^{\epsilon}_{i q} \end{matrix} \right\}\frac{-1}{(\Omega_{\epsilon i}-\Omega^{*}_{\epsilon j})^2} \ e^{-i\Omega_{\epsilon i}s_{xy}} \ .
\end{align}
Summing up these, we have
\begin{align}
-&\Delta \left\{ G^{d(eq)ii}_{R\ {\bold q}} *\Pi^{'(eq)ij}_{\wightman \ {\bold q}} * G^{d(eq)jj}_{A\ {\bold q}} \right\} 
\n
&\simeq \sum_{\epsilon} Z_{\epsilon} \Pi^{'(eq)ij}_{R}(\epsilon \omega_{q}) Z_{\epsilon}(-i)\ d_t \left\{ \begin{matrix} 1-f^{\epsilon}_{i q} \\ -f^{\epsilon}_{i q} \end{matrix} \right\} \frac{+1}{(\Omega_{\epsilon i}-\Omega^{*}_{\epsilon j})^2} \ e^{-i\Omega_{\epsilon i }(x^0 -y^0)}\notag \\
&+\sum_{\epsilon} Z_{\epsilon} \Pi^{'(eq)ij}_{A}(\epsilon \omega_{q}) Z_{\epsilon}(-i) \ d_t \left\{ \begin{matrix} 1-f^{\epsilon}_{i q} \\ -f^{\epsilon}_{i q} \end{matrix} \right\} \frac{-1}{(\Omega_{\epsilon i}-\Omega^{*}_{\epsilon j})^2}  \ e^{-i\Omega_{\epsilon i }(x^0 -y^0)} \ .
\label{Deloffshell3} 
\end{align}

Of course, if we sum up all the on-shell and off-shell contributions,
(\ref{Delonshell}) and (\ref{Deloffshell1}), (\ref{Deloffshell2}), (\ref{Deloffshell3}),
we can recover the full result
(\ref{DelG_{W}^{'}1}).


\end{document}